\documentclass[]{emulateapj}
\usepackage{amsmath}
\usepackage{natbib}
\usepackage{txfonts}
\usepackage{graphicx,color}
\usepackage{hyperref}
\usepackage{endnotes}
\usepackage{color}

\shorttitle{FARGO3D}
\shortauthors{P. Ben\'\i tez-Llambay \& F.\ S.\ Masset}

\begin{document}

\title{FARGO3D: A new GPU-oriented MHD code}

\author{Pablo Ben\'itez-Llambay} 

\affil{Instituto de Astronom\'\i a Te\'orica y Experimental,
  Observatorio Astron\'onomico, Universidad Nacional de
  C\'ordoba. Laprida 854, X5000BGR, C\'ordoba, Argentina}
\email{pbllambay@oac.unc.edu.ar} \author{Fr\'ed\'eric S. Masset}

\affil{Instituto de Ciencias F\'\i sicas, Universidad Nacional
  Aut\'onoma de M\'exico (UNAM), Apdo. Postal 48-3,62251-Cuernavaca,
  Morelos, Mexico} \email{masset@icf.unam.mx}

\begin{abstract}

  We present the FARGO3D code, recently publicly released. It is a
  magnetohydrodynamics code developed with special emphasis on
  protoplanetary disks physics and planet-disk interactions, and
  parallelized with MPI.  The hydrodynamics algorithms are based on
  finite difference upwind, dimensionally split methods. The
  magnetohydrodynamics algorithms consist of the constrained transport
  method to preserve the divergence-free property of the magnetic
  field to machine accuracy, coupled to a method of characteristics
  for the evaluation of electromotive forces and Lorentz forces.
  Orbital advection is implemented, and an N-body solver is included
  to simulate planets or stars interacting with the gas. We present
  our implementation in detail and present a number of widely known
  tests for comparison purposes.  One strength of FARGO3D is that it
  can run on both \emph{Graphical Processing Units} (GPUs) or
  \emph{Central Processing unit} (CPUs), achieving large speed up with
  respect to CPU cores.  We describe our implementation choices, which
  allow a user with no prior knowledge of GPU programming to develop
  new routines for the CPU, and have them translated automatically for
  the GPU.

\end{abstract}


\section{Introduction}
\label{sect:introduction}

Theories of planet-disk interactions have considerably evolved in the
last two decades.  A significant part of their progress is
attributable to numerical experiments.  In many of these, the role
played by the different parameters of the problem has been deciphered
by systematic, computationally expensive explorations of the parameter
space.  These could be achieved in part thanks to the steadily
increasing power of computational resources, and in part by the
development of fast algorithms adapted to the specifics of gas motion
in thin, nearly Keplerian disks, known as orbital advection algorithms
\citep{Masset.2000,2008ApJS..177..373J,2010ApJS..189..142S}.  Although
two-dimensional (2D) calculations are still a valuable tool in
tackling specific problems linked to planet-disk interactions, most of
the latest results in this field have been obtained through
three-dimensional calculations.  In problems involving magnetic field
tied to the matter, the prevalence of the magneto-rotational
instability, which requires resolving the vertical dimension across
the disk, implies that most studies of planet-disks interactions must
be tackled through expensive, three dimensional calculations, with
only a few notable exceptions.

The development of the CUDA language, aimed at tapping the huge
computational resources of \emph{Graphical Processing Units} (GPUs)
for general purpose processing, commonly known as GPGPU computing,
allows the development of astrophysical computational codes which can
tackle expensive problems at relatively moderate cost.  We can adopt
as a rule of thumb that the same code runs between one and two orders
of magnitude faster on a given high profile GPU than on a high profile
CPU core of same generation. This ratio turns out to be of same order
of magnitude as the vertical number of zones in many three-dimensional
(3D) studies. To put it simply, GPUs put 3D calculations at the cost
of 2D calculations on a CPU core.

With this in mind, we have developed a new code that solves the
equations of hydrodynamics (HD) or magnetohydrodynamics (MHD) on a
mesh (either Cartesian, cylindrical or spherical) with special
emphasis on describing protoplanetary disks and their interactions
with forming planets.

While there is a plethora of astrophysical fluid dynamics codes, there
are a lot fewer that can run on GPUs.  One of them is the public
Adaptive Mesh Refinement (AMR) code ENZO \citep{2014ApJS..211...19B},
which contains hard coded CUDA versions of the PPM and MHD
solvers. Like ENZO, the GAMER code \citep{2010ApJS..186..457S} is
another AMR code essentially oriented toward cosmological simulations,
which contains a hydrodynamic solver coded in CUDA.  Another one is
the hydrodynamical code CHOLLA \citep{2015ApJS..217...24S}, which is
based on high order Godunov's methods, and contains hard coded CUDA
kernels.  Another one is the PEnGUIn code \citep{2014ApJ...782...88F},
which is a Lagrangian, dimensionally split, shock capturing hydrocode
that uses the Parabolic Piecewise Method \citep[PPM,][]{cw84}. It is
written in CUDA and~C.  As memory is a significant concern in
cosmological simulations, and since the standard RAM available on a
cluster node is generally much larger than the RAM on board of the
GPUs, ENZO and GAMER constantly transfer back and forth data between
the CPU and GPU to benefit at the same time from the large
memory of the host (CPU) and the large computational throughput of the
GPU. Since these transfers constitute a bottleneck for effective GPU
performance, the GAMER code uses sophisticated asynchronous data
transfers in order to hide their latency.  Our needs are very
different: in the domain of predilection of FARGO3D, that of
protoplanetary disks and their interactions with embedded planets,
speed is generally much more a concern than memory. We therefore
require that our whole simulation fits on the memory of the GPU(s), in
order to avoid data transfer. All our routines, including those
dedicated to boundary conditions, run on the GPU. Consequently, we
have striven to obtain the smallest possible memory footprint per
cell.

When the HD or MHD algorithms are directly
coded in CUDA, a user who wishes to amend the core routines, or who
wishes to incorporate new physics into them must be proficient in
CUDA.  We have adopted a radically different approach: we have
developed all our code for CPU, and enforced the use of strict syntax
rules in all computationally expensive routines which expose their
parallelism, so that they can be translated automatically during
compilation to CUDA code.  This allows users who have no prior
knowledge of GPU programming to easily modify the code for their own
needs.  This also made the development of the code much faster and
reliable.  The scope of this paper is a comprehensive description of
the algorithms used in our code; we do not present the syntax rules
that we have developed to enable the automatic translation, which can
be found in the online manual. We nonetheless describe the conceptual
ideas of this process in section~\ref{sec:building-gpus}.

The code that we have developed is based on ideas similar to those
used in the ZEUS code \citep{Stone.Norman.1992.a}. In this code, the
velocities are not centered on the cells but staggered on their
faces. This allows us to calculate easily the fluxes of mass, momenta and
specific energy on the cell edges, by the use of upwind methods.

ZEUS-like codes have oftentimes been referred to as finite difference
codes, as opposed to the so-called finite volume codes. These, like
those based on Godunov's method that have emerged in astrophysics in
the past two decades, in which Riemann's problem is used to evaluate
fluxes at the faces between cells, automatically ensure the
conservation of the physical quantities for which a flux can be
evaluated.  This could suggest that finite difference codes like the
one we present in this work do not have such conservation
properties. In fact, a significant fraction of the sub steps performed
over a full update correspond to finite volume operations. In
particular, we shall see that FARGO3D conserves mass and momenta to
machine accuracy.

We summarize hereafter the salient points of our implementation
choices:

\begin{enumerate}
\item Our method is dimensionally split, thereby requiring as few as
  possible temporary arrays. Our code has probably the smallest
  possible memory imprint, which is well suited to GPUs, which have
  moderate amounts of random access memory (RAM) compared to CPUs.
\item Staggered mesh codes, contrary to Godunov's methods, do not have
  issues with steady flows with sources terms. This bears some
  importance in many situations in protoplanetary disks, for which the
  vertical and rotational equilibrium of the unperturbed disk, as well
  as that of the envelope that appears around embedded planets, must
  be accurately captured by the method.
\item The cost of a full time step with a staggered mesh code is
  significantly smaller than that of a Godunov's method.
\item Codes based on Godunov's method use generally fluxes of total
  energy, thus enforcing the conservation of energy to machine
  accuracy. On the other hand, staggered mesh codes consider the
  fluxes of internal energy, and do not fulfill the conservation of
  energy to machine accuracy.  Although this at first sight may seem a
  disadvantage, it is not so: in protoplanetary disks, which usually
  have a large Mach number, the kinetic energy is two to three orders
  of magnitude larger than the internal energy.  Any truncation error
  affecting the kinetic energy is forcibly transferred to the internal
  energy budget, which compounds the relative error.  This problem is
  known as the high Mach number problem
  \citep{1993ApJ...414....1R,2004NewA....9..443T}.  Many planet-disk
  interaction problems require an accurate advection of entropy in the
  planet's coorbital region.  It is therefore desirable to deal with
  the internal energy separately.
\item Inclusion of new processes generally requires significantly more
  work with Godunov's method, such as rewriting or significantly
  editing the Riemann solver, whereas it is easier to incorporate new
  or different physics in staggered mesh codes through the operator
  splitting technique.  Although we have striven to make the process
  of translation to GPUs as transparent and automatic as possible, the
  use of simple routines ensures that our code can be
  straightforwardly adapted to the user's needs, without requiring
  proficiency in GPU programming.
\item The predecessor of our new code, FARGO, has been intensively
  used in the field of planet-disk interactions, and proven to perform
  well over the whole range of planetary masses, from deeply embedded,
  low mass objects, which require finely tuned rotational equilibrium
  and accurate entropy and vortensity advection, to giant planets,
  which clear a gap and excite strong shocks in their immediate
  vicinity.  Although Godunov's codes are better at handling shocks
  than staggered mesh codes, the latter have been found to yield
  similar results to the former in problems involving gap clearing and
  shocks excitation \citep{valborro06}.
\end{enumerate}

As does its predecessor, the FARGO3D code includes orbital advection.
Initially, FARGO (\emph{Fast Advection in Rotating Gaseous Objects})
was an orbital advection algorithm \citep{Masset.2000}.  This acronym
was then used to name the code originally based on this algorithm and
publicly released in 2005.  The orbital advection has been extended to
MHD in the present implementation, using the upstream averaged
electric field method of \citet{2010ApJS..189..142S}.

One of the main novelties introduced in FARGO3D is that it can be run
effortlessly on GPU platforms with a high computational
throughput. This, together with the fact that it features orbital
advection in its HD and MHD versions,
makes it a tool of choice to study protoplanetary disk dynamics and
planet-disk interactions, which are extremely demanding in terms of
computational power.

This paper is organized as follows: in section~\ref{sec:overview} we
present the main characteristics of the code, list the equations that
it solves and introduce our notation.  In section~\ref{sect:methods}
we present a flow chart of the code and describe step-by-step the
numerical methods used to solve the governing equations.  In
section~\ref{sec:impl-cons} we discuss some implementation details, in
particular our choice to export computationally expensive routines to
the GPU.  In section~\ref{sec:tests} we present a set of tests on
different problems in HD and MHD. In section~\ref{sec:discussion}, we
provide some discussion on the impact of orbital advection on the
properties of the code, both in HD and MHD, and we also present
potential pitfalls of single precision calculations in the context of
astrophysical disks. Finally, we discuss ongoing and future
developments in section~\ref{sec:perspectives}.  Throughout the
manuscript the reader will find many superscripts referring to end
notes.  These refer to the specifics of our implementation and should
significantly speed up the learning process of the reader interested
in modifying the code.

\section{Overview}
\label{sec:overview}
\subsection{Governing equations}
\label{sec:governing-equations}

FARGO3D solves the equations of HD or MHD on an Eulerian mesh, which
can be either Cartesian, cylindrical or spherical. As its name
indicates, it is designed to solve three-dimensional problems, but it
can also be used in lower dimension (one or two-dimensions).  The
continuity equation is:
\begin{equation}
  \frac{\partial \rho}{\partial t} + \nabla.\left(\rho \vec{v}\right) = 0,
  \label{eq:continuity}
\end{equation}
where $\rho$ is the volumic density, and $\vec{v}$ is the velocity of
the fluid with respect to the mesh.  The Navier-Stokes equation reads:
\begin{eqnarray}
  \label{eq:euler}
\rho \left(\frac{\partial \vec{v}}{\partial t} + \vec{v}.\nabla\vec{v}\right) &=& -\nabla P + \frac{1}{\mu_0} \left( \nabla \times \vec{B} \right) \times \vec{B} + \nabla . \vec{T} + \vec{F_{ext}}\\
&&  -[2\vec{\Omega}\times\vec{v}+\vec{\Omega}\times\left(\vec{\Omega}\times\vec{r}\right)+\dot{\vec{\Omega}}\times\vec{r}]\rho\nonumber,
\end{eqnarray}
where $P$ is the pressure, $\vec{B}$ the magnetic field, and
$\vec{F}_{ext}$ is any external force (e.g. gravity force), and where
the second term of the right hand side is dealt with only if MHD is
included.  The last term of Eq.~\eqref{eq:euler} accounts for a
possible rotation of the mesh about the vertical axis, at a rate
$\vec{\Omega}$ which can vary with time.  The third term features the
stress tensor $\vec{T}$ , which has the following expression:
\begin{equation}
  \vec{T} = \rho \nu \left[ \nabla \vec{v} + \left(\nabla \vec{v} \right)^T - \frac{2}{3}\left(\nabla.\vec{v}\right) \vec{I} \right],
\label{eq:stresstensor}
\end{equation}
where $\nu$ is the kinematic viscosity and $\vec{I}$ is the unit tensor of same rank as the tensor $\nabla{\vec{v}}$.
For the energy equation, we use as discussed in section~\ref{sect:introduction} a non-conservative form using the volumic internal energy $e$:
\begin{equation}
  \partial_te + \nabla\cdot\left(e\vec{v}\right)= -P\nabla.\vec{v}.
  \label{eq:energy}
\end{equation}
Finally, when MHD is included, we solve the induction equation, which
reads:
\begin{equation}
  \frac{\partial \vec{B}}{\partial t} = \nabla \times \left(\vec{v}\times\vec{B}-\eta\nabla\times\vec{B}\right),
\label{eq:induction}
\end{equation}
where $\eta$ is the Ohmic diffusivity.  The Eqs.~\eqref{eq:continuity}
to~\eqref{eq:induction} are closed using an equation of state, giving
a relation between pressure or the internal energy.  In the public
release\endnote{Homepage: {\tt http://fargo.in2p3.fr}}, two forms for
the equation of state are provided:
\begin{equation}
  P = c_s^2 \rho,
  \label{eq:eos1}
\end{equation}
and
\begin{equation}
  P = \left(\gamma-1\right)e,
  \label{eq:eos2}
\end{equation}
where respectively $c_s$ is the isothermal sound speed and $\gamma$
the ratio of specific heats at constant pressure and volume.  The
first form given by Eq.~\eqref{eq:eos1} is often called {\it (locally)
  isothermal equation of state}. Its field $c_s(\vec{r})$ is constant
in time and given by the initial conditions.  When this equation is
used, Eq.~\eqref{eq:energy} is decoupled from the others and does not
have to be solved.  The second form, often called {\it adiabatic} or
{\it ideal equation of state}, is the relation between the internal
energy and pressure.  This form is commonly used when
Eq.~\eqref{eq:energy} has to be solved.

\subsection{Coordinate names}
The three different directions are generically named $X$, $Y$ and $Z$
in our implementation, even if the mesh geometry is non-Cartesian.
Tab.~\ref{tab:corresp} lists the correspondence between these names
and the corresponding coordinate in the different geometries.

In Cartesian coordinates one may use the shearing sheet\endnote{This
  is activated by the use of the \texttt{SHEARINGBOX} macro-variable
  in the option (\texttt{.opt}) file.} formalism to describe a sheared
flow in a rotating frame, subject to the standard expansion of the
effective (centrifugal plus central) potential.  This will trigger the
use of a different momentum component along the $X$ direction (see
Tab.~\ref{tab:momenta}) and the inclusion of specific source terms
(see section~\ref{sec:source-step}).  The frame rotating rate is not
allowed to vary in the shearing sheet case\endnote{Note that owing to
  our convention that orbital advection is performed along the $X$
  axis, the $X$ and $Y$ axis are swapped in our shearing sheet
  description with respect to standard notation.  Also, either our
  $(X,Y,Z)$ base is no longer direct, or the rotation is directed
  toward decreasing $X$, which yields opposite sign for the Coriolis
  force.}.

\begin{table*}
\footnotesize
  \centering 
\begin{tabular}{l || c c c}
  Geometry & $X$ & $Y$ & $Z$\\
  \hline 
  Shearing sheet & along advection & along gradient of unperturbed velocity & perpendicular to the previous ones \\  
  Cylindrical & azimuth ($\phi$) & radius ($r$) & $Z$ \\ 
  Spherical & azimuth ($\phi$) &  radius ($r$) & colatitude ($\theta$)\\ 
\end{tabular} 
\caption{Correspondence between names $X$, $Y$, $Z$  and the coordinate name in different geometries.}
\label{tab:corresp}
\end{table*}

\subsection{Centering and notation}
\label{subsect:staggmesh}
We store the face-centered values of vectorial quantities (velocity
and magnetic field components) while we store the cell-centered values
of scalar quantities (internal energy and density).

\begin{figure}
  \centering 
  \includegraphics[width=\columnwidth]{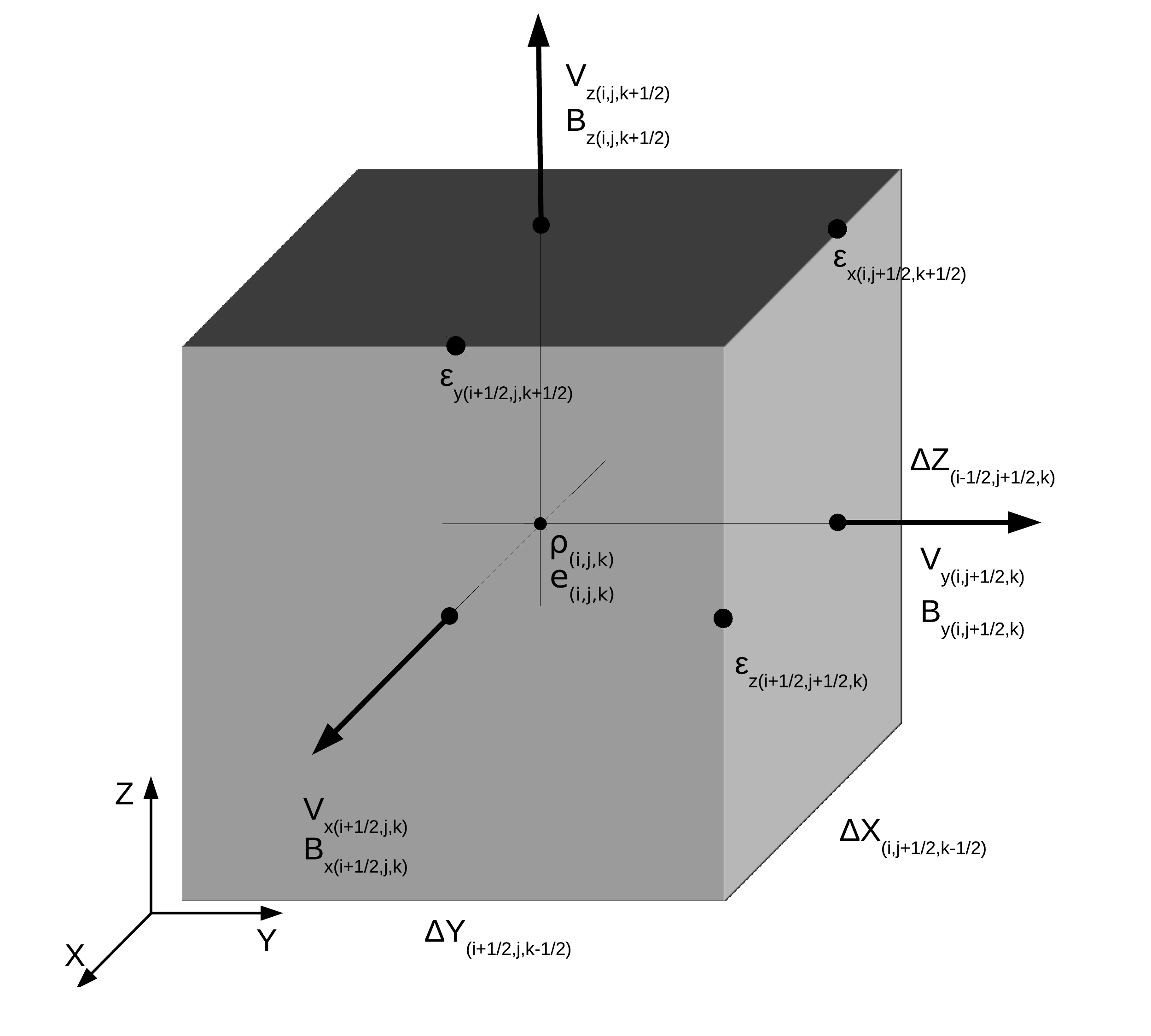}
  \caption{\label{fig:cell} Sketch of a cell in FARGO3D.  Vectorial 
    quantities ($v_x$, $v_y$ $v_z$, $B_x$, $B_y$ $B_z$) are located on 
    the faces, while scalar ones ($e$, $\rho$) are cell centered. 
    Electromotive forces $\varepsilon$ are defined at middle of the 
    edges (see section~\ref{sec:magnetohydrodynamic}).  Also shown are 
    the edge lengths ($\Delta X$, $\Delta Y$, $\Delta Z$), which 
    generally depend on $j$ and $k$.}
\end{figure}

In figure \ref{fig:cell} we display a sketch of a single cell, showing
the centering of different quantities.  We adopt half integer indices
for edges, while integer indices correspond to cell centers. The rule
we follow is: if a quantity $Q$ is defined at the center of cell
$i,j,k$, its value there reads $Q_{i,j,k}$, while if it is defined at
the center of the interface between cell $i,j,k$ and cell $i,j+1,k$,
its value reads $Q_{i,j+\frac{1}{2},k}$.

\section{Methods}
\label{sect:methods}
FARGO3D solves the hydrodynamical equations with a time-explicit 
method, using operator splitting and upwind techniques on an Eulerian 
mesh.  Some differential equations are discretized as finite 
differences, whereas others are solved using finite volume methods. 
To update the magnetic field with the induction equation 
\ref{eq:induction}, we use the {\it Method of Characteristics\it}
\citep[MOC,][]{Stone.Norman.1992.b} and to preserve the free divergence 
property of the magnetic field we use the {\it constrained transport}
method \citep[CT,][]{Evans.Hawley.1988}. 

\begin{figure*}
  \centering 
  \includegraphics[width=1.3\columnwidth]{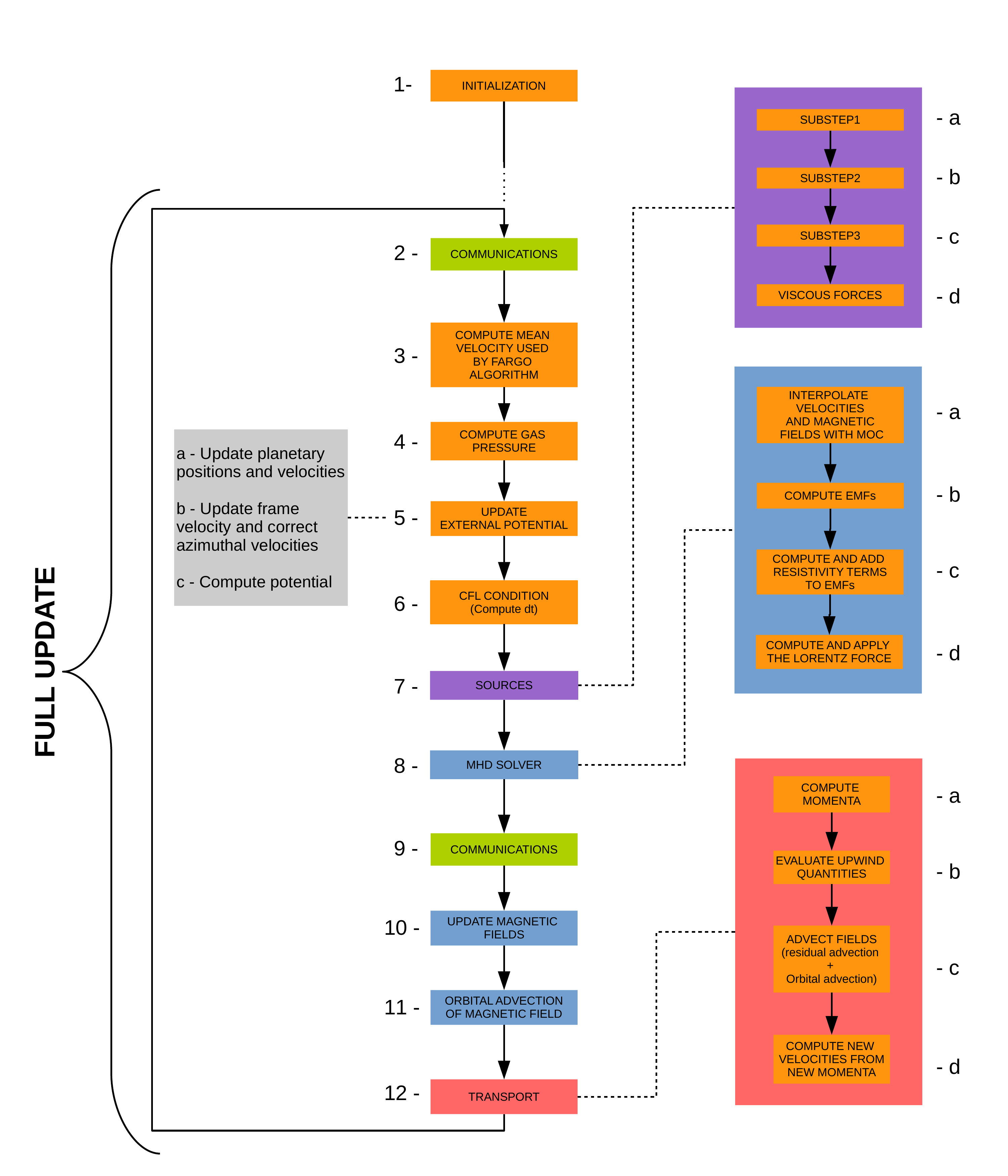}
  \caption{\label{fig:flowchart} Flow chart of the operations
    performed in succession in FARGO3D during a full update. Boxes
    with blue background are specific to the MHD case. The
    corresponding substeps are run only when the code is compiled for
    MHD.  Communications (cells~2 and~9) are discussed in
    section~\ref{sec:gener-cons}. The evaluation of the mean velocity
    at cell~3 is discussed in sections \ref{subsect:fargo} and
    section~\ref{sec:amendm-cfl-cond}. The evaluation of the gas
    pressure at cell~4 is done either according to Eq.~\eqref{eq:eos1}
    or according to \eqref{eq:eos2}. Calculations relative to the
    external potential (cell~5) are discussed in
    section~\ref{sec:orbit-integr-plan}. The time step evaluation of
    cell~6 is detailed in section~\ref{subsect:time-step}. The detail
    of the different source steps of box~7 are given in
    section~\ref{sec:source-step}, and in
    appendix~\ref{sec:visc-stress-tens} for the viscous
    forces. Algorithms specific to MHD, corresponding
    to boxes~8, 10 and~11, are described in
    sections~\ref{sec:magnetohydrodynamic}. Finally, algorithms
    specific to the transport step (box~12) are described in
    section~\ref{subsect:transport}.}
\end{figure*}

Fig.~\ref{fig:flowchart} shows a flow chart corresponding to a
complete time step (or full update). The governing equations are
solved by performing in succession the different substeps shown in
this figure. In the following two subsections we explain how we split
the equations of HD into a source and transport step, corresponding to
boxes~7 and~12 in Fig.~\ref{fig:flowchart}.  The practical application
of these steps will be discussed in further detail in
Section~\ref{sec:source-step} and~\ref{subsect:transport}. Our
implementation of the CFL condition is presented in
Section~\ref{subsect:time-step}. Additional terms required by MHD are
discussed in section~\ref{sec:magnetohydrodynamic}, and the FARGO
algorithm for MHD is discussed in Section~\ref{subsect:fargo-mhd}. We
conclude the section with a discussion of the orbital integrator for
planets, in Section~\ref{sec:orbit-integr-plan}.

\subsection{Operator splitting technique}

\label{subsect:opsplitting}
As is done frequently in (magneto)hydrodynamical solvers, the update
of the different fields is done in a step-by-step manner, along the
lines of the \emph{operator splitting technique}
\citep[e.g.][]{Stone.Norman.1992.a}.  Consider a problem defined by:
\begin{equation}
   \frac{\partial U}{\partial t} + \mathcal{A}(U) = 0,  \ \ \ \ U(0) = U_0.
  \label{eq:split1}
\end{equation}
If we can decompose $\mathcal{A}$ such that $\mathcal{A} =
\mathcal{A}_1+\mathcal{A}_2$, the solution of Eq.~\eqref{eq:split1}
could be obtained as the linear combination of the solutions of:
\begin{eqnarray}
   \frac{\partial U_1}{\partial t} + \mathcal{A}_1(U_1) &=& 0, \\
   \frac{\partial U_2}{\partial t} + \mathcal{A}_2(U_2) &=& 0.
  \label{eq:split2}
\end{eqnarray}
In a finite differences scheme, the approximate solution can be
written in the form:
\begin{eqnarray}
  \frac{U^1-U^0}{\Delta t} = - A_1(U^0), \\
  \frac{U^2-U^1}{\Delta t} = - A_2(U^1).
  \label{eq:split3}
\end{eqnarray}
We note that in hydrodynamics, differential equations have the general
form\footnote{If $Q$ is a vectorial quantity, Eq.~\eqref{eq:q} is
  still valid component wise.}:

\begin{equation}
  \frac{\partial Q}{\partial t} + \nabla.\left(Q\vec{v}\right) = \mathcal{S}(Q,\vec{v},t)  \label{eq:q}
\end{equation}
where $Q$ can be any quantity (such as a momentum component), and
$\mathcal{S}$ are source or sink terms of $Q$ (for instance a body
force).  In that case, splitting Eq.~\eqref{eq:q} is straightforward:
\begin{eqnarray}
  A_1 &=& -\mathcal{S} \\
  A_2 &=& \nabla.\left(Q\vec{v}\right)
\end{eqnarray}
The technique of operator splitting therefore divides the problem into
two partial ones, the {\it Source Step} and the {\it Transport Step}
\citep{Stone.Norman.1992.a}. The equations corresponding to the
partial problems read respectively:
\begin{equation}
  \label{eq:1}
  \partial_tQ=\mathcal{S}(Q,\vec{v},t)
\end{equation}
and
\begin{equation}
  \label{eq:2}
  \partial_tQ+\nabla.\left(Q\vec{v}\right)=0
\end{equation}
A complete time step is therefore composed of a partial update from
$Q(t_0) = Q^0 \rightarrow Q^1$ by the source step described by
Eq.~\eqref{eq:1}, and a subsequent update from $Q^1 \rightarrow
Q^2=Q(t_0+\Delta t)$ by the transport step of Eq.~\eqref{eq:2}. In
addition, a substep dedicated to the interaction of the flow with the
magnetic field may be added, and will be considered in
section~\ref{sec:magnetohydrodynamic}.

\subsection{Splitting source and transport in practice}
\label{sec:splitt-source-transp}
We entertain here the equations of HD in spherical coordinates, the
frame being in rotation about the polar axis with an angular velocity
$\Omega_f(t)$.  The reason for the possible time dependence of the
frame rotation rate is that in many circumstances it is useful that
the $x$ axis (direction of azimuth's origin) tracks either a planet or
its guiding-center, which yields a non constant rotation rate if the
planet is on a non-circular orbit in the first case, or if it is
migrating in the second case, or if the orbit is inclined in either
case.  The cases of Cartesian and cylindrical coordinates can be
straightforwardly worked out, and are summed up at the end of this
section.

The equations that govern the flow are the continuity equation
(Eq.~\ref{eq:continuity}), where the divergence operator, in spherical
coordinates, is:
\begin{equation}
  \label{eq:3}
  \nabla . \vec{u} \equiv\frac{1}{r^2}\partial_r(r^2u_r)+\frac{1}{r\sin\theta}\partial_\phi u_\phi  
  +\frac {1}{r\sin\theta}\partial_\theta(u_\theta\sin\theta),
\end{equation} 
the Euler equation (Eq.~\ref{eq:euler}), which can be recast
respectively in radius, azimuth and colatitude as (the color coding of
the terms will be explained later):
\begin{equation}
  \label{eq:4}
  D_tv_r={\color{blue}\frac{v_\theta^2+(v_\phi+r\sin\theta\Omega_f)^2}{r}-\frac{\partial_rP}{\rho}-\partial_r\Phi},
\end{equation}
\begin{eqnarray}
  D_tv_\phi&=&\label{eq:5}
  -r\sin\theta\dot\Omega_f-\frac{v_rv_\phi+v_\theta v_\phi\cot\theta}{r}\\\nonumber
  &&-2\Omega_f(v_r\sin\theta+v_\theta\cos\theta) {\color{blue}-\frac{\partial_\phi P}{r\sin\theta\rho}-\frac{\partial_\phi\Phi}{r\sin\theta}}  
\end{eqnarray}
and
\begin{eqnarray}
  \label{eq:6}
  D_tv_\theta =
  -\frac{v_rv_\theta}{r}{\color{blue}+\frac{(v_\phi+r\sin\theta\Omega_f)^2}{r}\cot\theta
    -\frac{\partial_\theta P}{r\rho}-\frac{\partial_\theta\Phi}{r}},
\end{eqnarray}
where the Lagrangian derivative $D_t$ is defined by:
\begin{equation}
  \label{eq:7}
  D_tQ=\partial_tQ+v_r\partial_rQ+\frac{v_\phi}{r\sin\theta}\partial_\phi Q+\frac{v_\theta}{r}\partial_\theta Q
\end{equation}
for any scalar quantity $Q$, and the energy
equation~\eqref{eq:energy}:
\begin{equation}
  \label{eq:8}
  \partial_te+\nabla.(e\vec{v})={\color{blue}-P\nabla.\vec{v}}.
\end{equation}
Using Eqs.~\eqref{eq:continuity} and~\eqref{eq:4}, we can write:
\begin{equation}
  \label{eq:9}
  \partial_t(\rho v_r)+\nabla.(\rho v_r\vec{v})=\rho\frac{v_\theta^2+(v^t_\phi)^2}{r}-\partial_rP-\rho\partial_r\Phi,
\end{equation}
where, for the sake of brevity, we use
$v^t_\phi=v_\phi+r\sin\theta\Omega_f$, the azimuthal velocity in the
inertial frame.  Similarly, using the specific angular momentum
$j=r\sin\theta v_\phi+r^2\sin^2\theta\Omega_f$, one may recast
Eq.~\eqref{eq:5} as:
\begin{equation}
  \label{eq:10}
  \partial_t(\rho j)+\nabla.(\rho j\vec{v})=\rho D_tj = -\partial_\phi P-\rho\partial_\phi\Phi.
\end{equation}
Finally, using Eqs~\eqref{eq:continuity} and~\eqref{eq:6}, we can
write:
\begin{equation}
  \label{eq:11}
  \partial_t(\rho rv_\theta)+\nabla.(\rho r v_\theta\vec{v})=\rho(v^t_\phi)^2\cot\theta+\partial_\theta P+\rho\partial_\theta\Phi.
\end{equation}
This equation on the ``meridional momentum'' has also been considered
by \citet{2009A&A...506..971K} in their implementation of the NIRVANA
code.  As we have seen in section~\ref{subsect:opsplitting}, a
transport substep is a combination of routines that solves
Eq.~\eqref{eq:2} on the mesh, for an arbitrary quantity $Q$.  We see
from the above that, by transporting respectively the variables
$\rho$, $\rho v_r$, $\rho j$, $\rho r v_\theta$ and $e$ (when we want
to solve the energy equation), we take into account all the terms in
black in Eqs.~\eqref{eq:4}, \eqref{eq:5} and~\eqref{eq:6}.  The terms
in blue, which also appear on the right hand side of
Eqs.~\eqref{eq:9}, \eqref{eq:10} and~\eqref{eq:11} (scaled by $\rho$
and geometric factors), are dealt with outside of the transport steps,
in the source substep, which is the subject of the next section.  The
name source step is somewhat misleading as many source terms of
Eqs.~\eqref{eq:4}, \eqref{eq:5} and~\eqref{eq:6} are dealt with during
the conservative update of the transport step.  In particular, the
Coriolis force and all source terms involving a product of different
velocity components are embedded in the transport step.  This is of
particular importance in the context of planets in
disks. \citet{1998A&A...338L..37K} has shown, by considering a
gap-opening planet in a disk, that failing to deal with Coriolis
forces in a conservative manner can lead to incorrect gap surface
density profiles.  Tab.~\ref{tab:momenta} generalizes the above
discussion to Cartesian and cylindrical geometries, and shows the
quantities to transport in each case. The Cartesian case is trivial
and does not allow for a rotation of the frame about the vertical
axis.

\begin{table*}
  \centering 
\begin{tabular}{l || c c c}
Geometry & X-momentum & Y-momentum & Z-momentum\\
\hline 
Cartesian & $\rho v_x$ & $\rho v_y$ & $\rho v_z$ \\  
Shearing sheet & $\rho (v_x+2\Omega_fy)$ & $\rho v_y$ & $\rho v_z$ \\  
Cylindrical & $\rho (r v_\phi+r^2\Omega_f)$ & $\rho v_r$ & $\rho v_z$ \\ 
Spherical & $\rho (r v_\phi \sin\theta+r^2\sin^2\theta\Omega_f)$ & $\rho v_r$ & $\rho r v_\theta$ \\ 
\end{tabular} 
\caption{Quantities transported in $X$, $Y$ and $Z$ to ensure a
  formulation as conservative as possible, for the different
  geometries implemented. Note that FARGO3D's notation differs from
  the usual notation for the shearing sheet by a swap of the $X$ and
  $Y$ coordinates.}\label{tab:momenta}
\end{table*}

\subsection{Source step}
\label{sec:source-step}
In the source step, described by Eq.~\eqref{eq:1}, all terms that are
not included in the transport step must be considered.  Specifying to
the case of spherical geometry contemplated in the previous section,
these are the terms that appear in blue in Eqs~\eqref{eq:4},
\eqref{eq:5}, \eqref{eq:6} and \eqref{eq:8} if the energy equation is
solved.  They correspond respectively to the centrifugal force, the
pressure gradient, the body forces and the work done by pressure
forces.  All other terms in Eqs.~\eqref{eq:4}, \eqref{eq:5}
and~\eqref{eq:6} are included in the conservative update of the
transport step.

In addition, as we shall see in section~\ref{sec:magnetohydrodynamic},
additional source terms arise from the Lorentz force in the MHD case.

We follow the procedure of \citet{Stone.Norman.1992.a}, separating the
source step into three sub-steps:

\begin{itemize}
\item Sub-step~1: we update\endnote{The source code corresponding to
    this step can be found in the files \texttt{substep1\_x.c},
    \texttt{substep1\_y.c} and \texttt{substep1\_z.c}.} the velocity
  field by pressure gradients and gravitational forces. This
  corresponds to the terms in blue in Eqs.~\eqref{eq:4}
  to~\eqref{eq:6}, and to the cell~7a of the flow chart of
  Fig.~\ref{fig:flowchart}.
\item Sub-step~2: we add an artificial von Neumann-Richtmyer viscosity
  and corresponding heating terms\endnote{The source code
    corresponding to this step can be found in the files
    \texttt{substep2\_a.c} and \texttt{substep2\_b.c}. The artificial
    viscous pressure is calculated in the first file, its gradient is
    then used in the second file. We had to split the operations to
    avoid race conditions, and to fully expose the parallelism of
    each substep. This kind of decomposition is common in our
    implementation.}, in the exact same manner as described by
  \citet{Stone.Norman.1992.a}. Instead of using a covariant form in
  curvilinear coordinates, component-wise expressions are employed in
  all cases. This corresponds to the cell~7b of the flow chart of
  Fig.~\ref{fig:flowchart}.
\item Sub-step~3: we add the work done by pressure forces\endnote{The
    source code corresponding to this step can be found in the file
    \texttt{substep3.c}.}, corresponding to the blue term of
  Eq.~\eqref{eq:8}, if the energy equation is solved. This corresponds
  to cell~7c of Fig.~\ref{fig:flowchart}.
\end{itemize}

Below we show the source terms that have to be included in Sub-step 1
for different geometries:

\begin{description}
\item[\bf Cartesian case]
\begin{eqnarray}
  \frac{\partial v_\chi}{\partial_t} = -\frac{\partial_\chi P}{\rho} - \partial_\chi\Phi,
\end{eqnarray}
where $\chi$ is an arbitrary direction. 
\item[\bf Shearing sheet]
\begin{eqnarray}
  \frac{\partial v_x}{\partial_t} &=& -\frac{\partial_x P}{\rho} - \partial_x\Phi\\
  \frac{\partial v_y}{\partial_t} &=& -\frac{\partial_y P}{\rho} - \partial_y\Phi+2\Omega_fv_x-4A\Omega_fy\\
  \frac{\partial v_z}{\partial_t} &=& -\frac{\partial_z P}{\rho} - \partial_z\Phi\\
\end{eqnarray}
\item[\bf Cylindrical case] 
\begin{eqnarray}
  \frac{\partial v_{\phi}}{\partial t} &=& -\frac
                                           1r\frac{\partial_\phi
                                           P}{\rho} -\frac 1r \partial_\phi\Phi    \\
  \frac{\partial v_r}{\partial t} &=& -\frac{\partial_r P}{\rho} - \partial_r\Phi + \frac{\left(v^t_\phi\right)^2}{r}  \\
  \frac{\partial v_{z}}{\partial t} &=& -\frac{\partial_z P}{\rho} - \partial_z\Phi 
\end{eqnarray}
\item[\bf Spherical case]
\begin{eqnarray}
  \frac{\partial v_\phi}{\partial t} &=& -\frac{1}{r\sin\theta}\frac{\partial_\phi P}{\rho} - \frac{1}{r\sin\theta}\partial_\phi\Phi \\
  \frac{\partial v_r}{\partial t} &=& -\frac{\partial_r P}{\rho} - \partial_r\Phi + \frac{\left(v^t_\phi\right)^2+v_\theta^2}{r} \\
  \frac{\partial v_\theta}{\partial t} &=& -\frac
                                           1r\frac{\partial_\theta
                                           P}{\rho} - \frac 1r\partial_\theta\Phi + \frac{\left(v^t_\phi\right)^2 \cot{\theta}}{r}
\end{eqnarray}
\end{description}

These source terms are applied directly to the velocity
components. The staggering of the velocity fields makes trivial the
evaluation of the pressure and potential gradients.  Geometrical
source terms (centrifugal forces) require in general averaging the
information of several cells.  In such cases we take the arithmetic
average of the velocities, then we evaluate the square, as shown in
the following example for the radial velocity in cylindrical geometry:

\begin{multline}
  \label{eq:12}
  \frac{{v^b_r}_{ij-\frac{1}{2} k}-{v^a_r}_{ij-\frac{1}{2} k}}{\Delta t}=\\\frac 1r\left[\frac14\left({v^t_\phi}_{i-\frac{1}{2} jk}+{v^t_\phi}_{i+\frac{1}{2} jk}+{v^t_\phi}_{i-\frac{1}{2} j-1k}+{v^t_\phi}_{i+\frac{1}{2} j-1k}\right)\right]^2,
\end{multline}
where the superscripts $a$ and $b$ denote the intermediate values at the
beginning and end of the substep, respectively.  Generalization to
other geometries or directions is straightforward.  The source term of
Eq.~\eqref{eq:8} is applied as described by Eq.~(40) of
\citet{Stone.Norman.1992.a}, valid for gamma-law gases.

The implementation of the viscous stress tensor is presented in
appendix~\ref{sec:visc-stress-tens}.

\subsection{Transport step}
\label{subsect:transport}
As we have seen in section~\ref{subsect:opsplitting}, the transport
step consists in solving Eq.~\eqref{eq:2}, which reads as a
conservation law for the quantity $Q$ ($Q$ being any of the quantities
in the list given in section~\ref{sec:splitt-source-transp}.)  The
integral form of Eq.~\eqref{eq:2} is, using the divergence theorem:
\begin{equation}
  \frac{\partial}{\partial t} \int\!\!\!\int\!\!\!\int_V Q dV + \int\!\!\!\int_{\partial V} Q\vec{v} . d\vec{S} = 0,
\label{eq:advection2}
\end{equation}
where we have assumed that the control volume $V$ has no explicit
temporal dependency.  Thus, time variations of $Q$ inside a control
volume are due exclusively to its flux across the boundary $\partial
V$ of the control volume.  If $Q$ is defined at a cell center, the
finite difference representation of Eq.~\eqref{eq:advection2} is:
\begin{eqnarray}
  \frac{Q^{n+1}_{ijk} - Q^{n}_{ijk}}{\Delta t}\mathcal{V} = &-& \left[ {\mathcal{F}_X}_{i+\frac{1}{2} jk} - {\mathcal{F}_X}_{i-1/2 jk}\right. \nonumber \\
  &+& {\mathcal{F}_Y}_{ij+1/2 k}-{\mathcal{F}_Y}_{ij-1/2 k}  \\
  &+& \left. {\mathcal{F}_Z}_{ijk+1/2 }-{\mathcal{F}_Z}_{ijk-1/2 }\right]^{n+1/2 } ,\nonumber 
  \label{eq:advection_finite}
\end{eqnarray}
where $\mathcal{V}$ is the volume of the cell and $\mathcal{F}$ is the
flux of $Q$ across the faces of the cubic cell.  Expressions of the
fluxes along other directions can be straightforwardly inferred from
Eq.~\eqref{eq:flux}.  For multidimensional advection, we reduce the
transport problem to several one-dimensional problems, each one
updating partially the field with the corresponding fluxes:
\begin{eqnarray}
  \left[Q^{n+a}_{ijk} - Q^{n}_{ijk}\right]\mathcal{V} = &- \Delta t\left[{\mathcal{F}_X}_{i+1/2jk } - {\mathcal{F}_X}_{i-1/2jk }\right]^{n+1/2 } 
  \label{eq:1D-advection-X}  \\
  \left[Q^{n+b}_{ijk} - Q^{n+a}_{ijk}\right]\mathcal{V} = &- \Delta t\left[ {\mathcal{F}_Y}_{ij+1/2k } - {\mathcal{F}_Y}_{ij-1/2k }\right]^{n+1/2 }
  \label{eq:1D-advection-Y} \\
   \left[Q^{n+1}_{ijk} - Q^{n+b}_{ijk}\right]\mathcal{V} = &- \Delta t\left[ {\mathcal{F}_Z}_{ijk+1/2 } - {\mathcal{F}_Z}_{ijk-1/2 }\right]^{n+1/2 }
  \label{eq:1D-advection-Z}
\end{eqnarray}
Note that the fluxes in Eq.~\eqref{eq:1D-advection-Y} are evaluated
using the quantity $Q$ at the intermediate stage $n+a$, and similarly
in Eq.~\eqref{eq:1D-advection-Z}, at the intermediate stage $n+b$,
thus in principle the solution depends on the order of evaluation,
which is a characteristic of numerical schemes based on dimensionally
split methods \citep{Stone.Norman.1992.a}.  Performing in succession
the operations described by Eqs.~\eqref{eq:1D-advection-X}
to~\eqref{eq:1D-advection-Z} amounts to take into account all terms in
black in the right hand side of Eqs.~\eqref{eq:4} to~\eqref{eq:6}
(except for the term in $\dot\Omega_f$). These operations
correspond to cell~12c of Fig.~\ref{fig:flowchart}.

We note that even though Eq.~\eqref{eq:10} embeds the first term of
the right hand side of Eq.~\eqref{eq:5}, the update of the frame
angular velocity (first part of cell~5b of Fig.~\ref{fig:flowchart}) is not
performed between the beginning and the end of the transport substep
(respectively cells~12a and~12d of Fig.~\ref{fig:flowchart}), so that
the new velocities do not reflect the variation of the frame rotation
rate. Instead, upon the update of the frame angular velocity, we must
explicitly correct the azimuthal velocities\endnote{Source code in
  \texttt{change\_frame.c}.}. In the spherical case, this is done as
follows:
\begin{equation}
  \label{eq:13}
  {v^c_\phi}_{ijk}=  {v^b_\phi}_{ijk}-(\Omega_f'-\Omega_f)r\sin\theta,
\end{equation}
where $\Omega_f$ is the old rotation rate of the frame, and
$\Omega_f'$ the new rate, and where the superscripts~$b$ and~$c$
denote respectively the value of the azimuthal velocity at the
beginning of cells~5b and~5c of Fig.~\ref{fig:flowchart}, which are
also the values at the beginning and end of the velocity correction
routine (second part of cell 5b of Fig.~\ref{fig:flowchart}).  The
geometrical terms of Eq.~\eqref{eq:13} are evaluated at the
center of the faces where $v_\phi$ is defined.  This substep trivially
conserves angular momentum to machine accuracy.

The flux evaluation is done through an upwind method to infer the
value of the quantity $Q^*$ at the center of the face at half time
step.  While there has been a trend in the past decade to turn to
Riemann's problem to evaluate the fluxes at the interface in many
astrophysical fluid dynamics codes, as initially devised by
\citet{godunov59}, we make use here of the staggering of the velocity
field to express simply the flux as:
\begin{equation}
  {\mathcal{F}_X}_{i+1/2 }^{n+1/2 } =  {v_X}_{i+1/2 }\left[
   Q^{*x}_{i+1/2 }
  \right]^{n+1/2 }\mathcal{S}_{i+1/2},
\label{eq:flux}
\end{equation}
where the subscript $X$ represent the direction normal to the face,
along which the flux is evaluated, and where $Q^{*x}_{i+1/2 }$ is the
$x$-interpolated value of the cell centered quantity $Q$ onto the face
$i+1/2 $, at the middle of the time step.  In Eq.~\eqref{eq:flux} we
have omitted the $j,k$ subscripts for the sake of brevity, since the
flux calculation only involves operations along one axis.  As in
\cite{Stone.Norman.1992.a}, we consider a unique characteristic speed
which is the velocity of the flow at the interface, at the beginning
of the time step, given directly by ${v_x^n}_{i+1/2 jk}$. This
characteristic is integrated back in time over $\Delta t/2$, and the
interpolated value $Q^*$ of the variable $Q$ under consideration is
sought at the location $x_*$ thus reached, which is:
\begin{equation}
  \label{eq:14}
  x_*=x_{i+1/2 }-{v_x^n}_{i+1/2 }\Delta t/2.
\end{equation}
The interpolation, regardless of its order, naturally contains a
test on the velocity sign, since $x_*$ falls in cell $i$ if
${v_x^n}_{i+1/2 }<0$, and in cell $i+1$ otherwise.
Eq.~\eqref{eq:flux} can finally be recast as:
\begin{equation}
  {\mathcal{F}_X}_{i+1/2 }^{n+1/2 } =
    {v_X}^{n}_{i+1/2 }\left(Q^{*x}\right)_{i+1/2 }\mathcal{S}_{i+1/2 }.
\label{flux_upwind}
\end{equation}
For the interpolated value of $Q$, we use a zone-wise linear
reconstruction using van Leer's slopes \citep{1977JCoPh..23..276V} for
most of our sub steps\endnote{The source code corresponding to the
  calculation of the slopes can be found in \texttt{vanleer\_x\_a.c},
  \texttt{vanleer\_y\_a.c} and \texttt{vanleer\_z\_a.c}.}. There is an
exception to this: when we perform the uniform residual step of
orbital advection (which we will present in
section~\ref{subsect:fargo}), we use a zone-wise parabolic
reconstruction of the field, using the so-called PPA or
\emph{Piecewise parabolic advection} algorithm\endnote{The succession
  of operations involved in the PPA reconstruction is invoked in the
  file \texttt{vanleerx\_ppa.c}, and the different operations called
  therein are found in the files \texttt{fargo\_ppa\_a.c},
  \texttt{fargo\_ppa\_b.c}, \texttt{fargo\_ppa\_c.c} and
  \texttt{fargo\_ppa\_d.c}. Again, the operation is fragmented as
  necessary to avoid race conditions, thereby exposing the parallelism
  of each substep.}, in a manner similar to what is done in the PLUTO
code~\citep{Mignone.2012}.  For the sake of completeness we give
hereafter the detail of the evaluation of the interpolated value using
van Leer's slopes:
\begin{equation}
Q^{*x}_{i+1/2 } = \left\{
\begin{array}{cl}
\label{eq:15}
  Q_i + a_{i}\left(\Delta X_{i} -
    {v_x}_{i+1/2 } \Delta t \right)/2 &\mbox{if } {v_x}_{i+1/2 } \geq 0 \\
  Q_{i+1} - a_{i+1}\left(\Delta X_{i} +
    {v_x}_{i+1/2 } \Delta t\right)/2 &\mbox{if } {v_x}_{i+1/2 } < 0,
\end{array}\right.
\end{equation}
where $a_{i}$ is van Leer's slope, given by:
\begin{equation}
a_{i} = \left\{
\begin{array}{cl}
  0 & \mbox{if~~}
  \Delta Q_{i+1/2 }\Delta Q_{i-1/2 } < 0 \\
  \displaystyle{2\frac{\Delta Q_{i+1/2 }\Delta Q_{i-1/2 }}{\Delta
    Q_{i+1/2 }+\Delta Q_{i-1/2 }}} &\mbox{otherwise}\\
\end{array}\right. \nonumber
\end{equation}
with $\Delta Q_{i+1/2 } = \left(Q_{i+1}-Q_{i}\right)/\Delta
X_{i}$. This evaluation of the interface value at half time step
corresponds to cell~12b of Fig.~\ref{fig:flowchart}. During this
substep, the interface values of density, energy and momenta are
evaluated at the cell interfaces. 

We use consistent transport
\citep{1980ApJ...239..968N,Stone.Norman.1992.a} for all quantities
other than mass\endnote{Consistent transport is achieved by a division
  by the density of the quantities to transport, prior to perform the
  upwind evaluation. This division is carried out by invoking
  \texttt{DivideByRho()} in \texttt{transport.c}.}.

\subsubsection{Momenta advection}
\label{sec:momenta-advection}
As discussed in section~\ref{sec:splitt-source-transp}, we must
transport a number of momenta-like quantities. These momenta do not
have a trivial definition on a staggered mesh, as they involve the
product of quantities which have different centering.  In contrast
with what is done in the ZEUS code \citep{Stone.Norman.1992.a}, where
the control volumes of staggered quantities are shifted and involve
zone centered fluxes along the staggered direction, we define two
flavors of each momentum, that we call the left and right momenta,
which we define respectively as:
\begin{eqnarray}
  \label{eq:16}
  \Pi_i^- &=& \rho_iv_{i-1/2}\\
\label{eq:17}
  \Pi_i^+ &=& \rho_iv_{i+1/2},
\end{eqnarray}
which we transport as any cell-centered quantity. Eq.~\eqref{eq:16}
and~\eqref{eq:17} involve the product of the cell-centered density
$\rho_i$ and respectively of the left (right) interface value of
velocity $v_{i-1/2}$ ($v_{i+1/2}$). These momenta are evaluated when
entering the set of transport sub steps, which corresponds to cell~12a
of Fig.~\ref{fig:flowchart}. Upon completion of the transport sub
steps, the new velocity is inferred from the new momenta and new
density as follows:
\begin{equation}
  \label{eq:18}
  v_{i-1/2}^{n+1}=\frac{{\Pi_{i-1}^+}^{n+1}+{\Pi_i^-}^{n+1}}{\rho_{i-1}^{n+1}+\rho_i^{n+1}},
\end{equation}
where we note that Eq.~\eqref{eq:18} applied to the old momenta and
density yields the old velocity. This transformation corresponds to
cell~12d of Fig.~\ref{fig:flowchart}. Eqs.~\eqref{eq:16}
and~\eqref{eq:18} are used in the Cartesian case. In other geometries
we amend them according to Tab.~\ref{tab:momenta}.  Our procedure
ensures the conservation to machine accuracy of both the left and
right momentum, and therefore of the discrete momentum defined as:
\begin{equation}
  \label{eq:19}
  \Pi_i = \frac12 (v_{i-1/2}+v_{i+1/2})\rho_i,
\end{equation}
which appears as the arithmetic mean of left and right momenta.  When
used with consistent transport, the technique of left and right
momenta advection is tantamount, for the transport of $\Pi_i$, to
evaluating the interpolated velocity $v^*$ at the left interface as
the arithmetic average of the zone centered interpolates $v^*_i$ and
$v^*_{i+1}$.  Its main difference with respect to the method used in
the ZEUS code is therefore that in the latter the mass fluxes are
averaged to yield zone centered mass fluxes, whereas in our method it
is the star values that are averaged in order to keep the mesh zones as
control volumes.  This avoids, in particular, the averaging of mass
fluxes, and makes the transport of momenta exactly consistent with the
transport of zone centered quantities.  Besides, it is important that
the control volume be the same for all variables when using the
orbital advection (aka FARGO) algorithm, which is the subject of
section~\ref{subsect:fargo}.

\subsubsection{Momentum conservation in our implementation}
\label{sec:pressure-source-term}
The previous section shows that momenta are conserved to machine
accuracy \emph{during the transport step}. In codes based on Godunov's
method, the pressure at the interface between zones is included in the
momenta fluxes, which ensures that momenta are conserved to machine
accuracy during a full update. In our case, however, the pressure
gradient is dealt with apart from the transport step, in substep~1
(see section~\ref{sec:source-step}), and it is necessary to address
separately the conservation of momenta under the action of this source
term.  Since it is a case of particular interest for FARGO3D, we
specialize our discussion to the case of the angular momentum, in
spherical coordinates, for a rotating frame. The generalization to
other components of momentum is straightforward.

During the pressure source step, the following transformation is
applied to the azimuthal velocity (note that all the discussion in
this section applies at fixed values of $r_j$ and $\sin\theta_k$, so
we drop for the rest of this section the $j$ and $k$ notation):
\begin{equation}
  \label{eq:20}
  \frac{v_{i-1/2}^{n+a}-v_{i-1/2}^{n}}{\Delta t} \equiv \frac{\Delta v_{i-1/2}}{\Delta t}=
  -2\frac{P_{i}-P_{i-1}}{\Delta x(\rho_i+\rho_{i-1})},
\end{equation}
where $P_i$ is the pressure defined in the center of cell $i$, $\Delta
t$ is the time step and $\Delta x=r\Delta\phi$ the cell width. During
the source step, the integrated zone centered momentum therefore
varies of the quantity:
\begin{equation}
  \label{eq:21}
  \sum_{i=0}^{N-1}\Delta \Pi_i\Delta x = 
  - r\sin\theta \Delta t
 \left(\sum_{i=0}^{N-1}\frac{P_i-P_{i-1}}{\rho_i+\rho_{i-1}}\rho_i+
    \frac{P_{i+1}-P_{i}}{\rho_{i+1}+\rho_{i}}\rho_i\right),
\end{equation}
where
\begin{equation}
  \label{eq:22}
  \Delta \Pi_i = \frac 12(\Delta \Pi_i^-+\Delta \Pi_i^+)
\end{equation}
is the variation of the zone centered momentum in zone $i$ during the
source step. We have:
\begin{eqnarray}
  \label{eq:23}
  \Delta \Pi_i^\pm &=& \Delta\left[\rho_i\left(rv^{i\pm 1/2}_\phi\sin\theta+r^2\sin^2\theta\Omega_f\right)\right]\\
                    &=& \rho_ir\sin\theta\Delta v^{i\pm 1/2}\nonumber.
\end{eqnarray}
Renumbering the second sum of the right hand side of Eq.~\eqref{eq:21}
from $1$ to $N$, we are exclusively left with edge terms:
\begin{equation}
  \label{eq:24}
  \sum_{i=0}^{N-1}\Delta \Pi_i\Delta x = r\sin\theta\Delta
  t\left(\frac{P_0\rho_{-1}+P_{-1}\rho_0}{\rho_0+\rho_{-1}}-\frac{P_{N-1}\rho_N+P_N\rho_{N-1}}{\rho_{N-1}+\rho_N}\right),
\end{equation}
which cancel each other if we account for the mesh periodicity in
azimuth: $\rho_{-1}=\rho_{N-1}$, $\rho_N=\rho_0$, and similar
relations for the pressure:
\begin{equation}
  \label{eq:25}
  \sum_{i=0}^{N-1}\Delta \Pi_i\Delta x = 0.
\end{equation}
The angular momentum is therefore conserved to machine accuracy,
regardless of whether the frame is rotating.  A corollary of this is
that shock jump conditions are satisfied in isothermal setups, even if
the pressure contains an artificial viscosity, and the production of
vortensity that occurs when a fluid parcel crosses a shock, which is
governed by the shock jump conditions, is also captured correctly by
the code, as was noted by \citet{2010MNRAS.405.1473L}.

The conservation of momentum under the pressure source term is due to
the fact that the value of the density evaluated at the zone interface
that we use in Eq.~\eqref{eq:20} is the arithmetic mean of the
adjacent zone centered values. This conservation property can be
generalized to any source term arising from the gradient of a zone
centered quantity divided by the density.

\subsection{Stability}
\label{subsect:time-step}
The integration time step $\Delta t$ over which a full cycle of
(magneto)hydrodynamical sub-steps is performed has to be limited in
order to ensure stability of the explicit methods we use.  This
condition is known as the Courant-Friedrichs-Levy (CFL) condition, or
the Courant condition.  Its broad physical meaning is that information
cannot travel over more than one cell per time step.  Although in
simple cases it is possible to determine exactly the stability
criterion with a von Neumann and Richtmyer's analysis, in general, as
is the case here, we resort to heuristic methods to work out the
maximum allowed time step. Following \citet{Stone.Norman.1992.a}, we
take as the maximum time step allowed\endnote{The argument of the min
  function of Eq.~\eqref{cfl2} is evaluated in file
  \texttt{cfl.c}. Its minimum value is then sought in the file
  \texttt{cfl\_b.c}, which also incorporates the shear time step limit
  of Eq.~\eqref{eq:38}, when needed.}:
\begin{equation}
  \Delta t = C\text{min}\left\{ \left(\sum_i \Delta t_i^{-2}\right)^{-1/2}\right\},
  \label{cfl2}
\end{equation}
where $C$ is a real parameter smaller than one, called the Courant
number.  Unless specified otherwise we use $C = 0.44$, a value that we
regard as a good compromise between speed and stability for our test
problems.  In Eq.~\eqref{cfl2}, the minimum is sought over the full
computational domain (excluding ghost zones), which corresponds to
cell~6 of the flow chart of Fig.~\ref{fig:flowchart}. The different
$\Delta t_i$ correspond to different processes that individually limit
the time step.  We draw hereafter the list of these individual time
steps.  The $j$ index below represents each direction of the mesh
($xyz$/$\phi r z$/ $\phi r \theta$) or a subset of those in lower
dimension.  In a multidimensional case we set $\Delta t_i$ as
$\text{min}_j\left\{ \Delta t_{i,j}\right\}$.

\begin{enumerate}
\item Sound or magnetosonic waves: $\Delta t_1 = \Delta_j/ C_w$, where
  $C_w$ is the maximum speed of the waves that can propagate in the
  medium.  In the MHD case, it is the fast magnetosonic wave, with
  expression $C_w^2=C_s^2+v_A^2$, where $C_s$ is the speed of sound
  and $v_A$ is the Alfv\'en velocity: $v_A^2=B^2/(\mu_0\rho)$. In the
  purely HD case $C_w$ is simply $C_s$.
\item Fluid motion: $\Delta t_2 = \Delta_j/|V_j|$.
\item Artificial viscosity: $\Delta t_3 = C_2|\Delta_j/\Delta v_j|$,
  where we choose the constant value $C_2=4\sqrt 2$. $(\Delta v_j)$ is
  the difference between successive values of $v_j$ along the $j$
  direction (e.g. $\Delta v_x = {v_x}_{i+1/2} - {v_x}_{i-1/2}$).
  \item Viscosity: $\Delta t_4 = \Delta_j^2/(4\nu)$, $\nu$ being the
    kinematic viscosity.
  \item Resistivity: $\Delta t_5 =\Delta_j^2/(4\eta)$, $\eta$ being
    the resistivity.
\end{enumerate}

\subsubsection{Orbital advection}
\label{subsect:fargo}

The transport step presented in section~\ref{subsect:transport} is
known to yield the following issues:
\begin{enumerate}
\item Large nearly uniform azimuthal velocities severely limit the
  time step (see section~\ref{subsect:time-step})
\item Truncation errors depend on the frame of reference
  \citep{Robertson.2009}.
\end{enumerate}
We present here a derivation which is similar to that of
\cite{Masset.2000}, except for the use of \emph{Piecewise Parabolic
  Advection} to perform the fractional shift.
The main idea of orbital advection techniques is to decompose, for
each ring of cells at a given radius and colatitude, the azimuthal
velocity into a large, uniform velocity and a residual, smaller
velocity (which is tantamount to working in a nearly corotating frame
in each ring):
\begin{equation}
  \label{eq:26}
  v= v^0+\delta v,
\end{equation}
where $v\equiv v_\phi$ in spherical and cylindrical geometries, and
$v_x$ in the Cartesian (shearing sheet) case.  The transport equation
in the azimuthal direction, which reads
\begin{equation}
  \label{eq:27}
  \partial_t Q+\partial_x(vQ)=0,
\end{equation}
where $\partial_x\equiv (1/r)\partial_\phi$ in cylindrical geometry
and $\partial_x\equiv (1/r\sin\theta)\partial_\phi$ in spherical
geometry.  Eq.~\eqref{eq:27} is solved in two steps, using the
operator splitting technique:
\begin{eqnarray}
  \partial_t Q + \partial_x (\delta v Q) &=& 0  \label{eq:fargo1} \\
  \partial_t Q +v^0\partial_x Q &=& 0 \label{eq:fargo2}
\end{eqnarray}
Eq.~(\ref{eq:fargo1}) is solved using the upwind method described in
section~\ref{subsect:transport}.  Eq.~(\ref{eq:fargo2}) amounts to a
shift of the initial profile:
\begin{equation}
  \label{eq:28}
  Q(x,t+\Delta t) = Q(x-v^0\Delta t,t).
\end{equation}
Since this shift does not necessarily represent an integer number of
zones, we further split Eq.~\eqref{eq:fargo2} into two sub steps.  We
decompose $v^0$ as:
\begin{equation}
  \label{eq:29}
  v^0=v_\mathrm{shift}+v_\mathrm{shift}^R,
\end{equation}
where
\begin{equation}
  v_\mathrm{shift}= N\frac{\Delta X}{\Delta t}
\end{equation}
with ${N=E\left(v^0\frac{\Delta t}{\Delta X}\right)}$, $E(X)$ being
the nearest integer to $X$. Using again the operator splitting
technique, Eq.~\eqref{eq:fargo2} is equivalent to solving in
succession\endnote{The integer shift is implemented in file
  \texttt{advect\_shift.c}, while the residual uniform shift uses the
  same function as standard advection, but with a different (uniform)
  velocity field. The invocation of the different sub steps is found
  in file \texttt{transport.c}. The user may switch between standard
  and orbital advection by commenting or commenting out the line
  \texttt{FARGO\_OPT += -DSTANDARD} in the option (\texttt{.opt})
  file.}:
\begin{equation}
  \label{eq:30}
  \partial_tQ+v_\mathrm{shift}\partial_xQ=0
\end{equation}
and
\begin{equation}
  \label{eq:31}
  \partial_tQ+v_\mathrm{shift}^R\partial_xQ=0.
\end{equation}
The solution of Eq.~\eqref{eq:30} is given by
\begin{eqnarray}
  \label{eq:32}
  Q_i^{n+1}&\equiv& Q(x_i,t+\Delta t) = Q(x-v_\mathrm{shift}\Delta t,t) \\\nonumber
  &=& Q(x_i-N\Delta X,t) = Q_{i-N}^n,\nonumber
\end{eqnarray}
which is simply a circular permutation of the zone values within the
ring under consideration. 

When using orbital advection, the
velocity to be used in Eq.~\eqref{eq:15} is the residual velocity
$\delta v$. Similarly, Eq.~\eqref{eq:31} can be
recast as:
\begin{equation}
  \label{eq:33}
  \partial_tQ+\partial_x(v_\mathrm{shift}^RQ)=0,
\end{equation}
and can therefore be dealt with using the same advection machinery. We
perform this fractional uniform advection using a higher order
interpolation (\emph{Piecewise Parabolic Advection}).
The interface values are given by:
\begin{equation}
Q^{*x}_{i+1/2 } = \left\{
\begin{array}{ll}
  Q_{R,i} + \xi\left(Q_i-Q_{R,i}\right) & \\
  \;\; +\xi\left(1-\xi\right)\left(2Q_i-Q_{R,i}-Q_{L,i}\right) & \mbox{if } {v_x}_{i+1/2 } \geq 0 \\
  Q_{L,i+1} + \xi\left(Q_{i+1}-Q_{L,i+1}\right) & \\
  \;\; +\xi\left(1-\xi\right)\left(2Q_{i+1}-Q_{R,i+1}-Q_{L,i+1}\right) &\mbox{if } {v_x}_{i+1/2 } < 0 \\
\end{array}\right. \nonumber
\end{equation}
where $\xi=v_{i+1/2 }\Delta t / \Delta x$, and $Q_{L/R}$ are the
monotonized left/right interface values for $Q$, obtained as follows.

From $\delta Q_i=(Q_{i+1}-Q_{i-1})/2$, we firstly define a monotonized
centered slope in each cell with:
\begin{equation}
  \label{eq:34}
  \delta_mQ_i= \left\{
    \begin{array}{l}
0  \mbox{~~~~~~~if~}(Q_{i+1}-Q_{i}) (Q_{i}-Q_{i-1})\le 0\\
\min(2|Q_i-Q_{i-1}|,2|Q_{i+1}-Q_i|,|\delta Q_i|) \\
\;\;\;\times\mathrm{sgn}(\delta Q_i)
       \mbox{~~~~~~~otherwise,}\\
    \end{array}\right.\nonumber
\end{equation}
This procedure corresponds to that of \citet{cw84} with a uniform cell
size. In a second stage, parabolic interpolations are used to
reconstruct the quantity at the interfaces with:
\begin{equation}
  \label{eq:35}
  Q^L_{i+1}=Q^R_i =Q_i+\frac 12(Q_{i+1}-Q_i)-\frac 16(\delta_mQ_{i+1}-\delta_mQ_i),
\end{equation}
and final tests are warranted to remove local extrema potentially
created by the above procedure, as follows:
\begin{equation}
  \label{eq:36}
\left\{
  \begin{array}{ll}
    Q^L_i=Q^R_i=Q_i&\mbox{if~}(Q_{i+1}-Q_{i}) (Q_{i}-Q_{i-1})\le 0\\
    Q^L_i=3Q_i-2Q^R_i&\mbox{if~}\Delta_iC_i>\Delta_i^2/6\\
    Q^R_i=3Q_i-2Q^L_i&\mbox{if~}-\Delta_i^2/6>\Delta_iC_i,\\
  \end{array}
\right.\nonumber
\end{equation}
where $\Delta_i = Q^R_i-Q^L_i$ and $C_i=Q_i-(Q^L_i+Q^R_i)/2$.  The use
of this method requires one additional layer of ghost or buffer
zones (extra zones outside the mesh that are used either to
synchronize the data between different computing processes, or to
prescribe boundary conditions) with respect to the use of van Leer's
slope presented in section~\ref{subsect:transport}.  We employ it
exclusively to perform the fractional uniform residual of the orbital
advection, as in \citet{Mignone.2012}.  Along the orbital direction,
no ghost zones are used, as the algebra on the corresponding zone
index ($i$) is hard coded to account for the mesh
periodicity\endnote{See the macrocommands \texttt{lxp} and
  \texttt{lxm} in \texttt{define.h}.}.  We also mention that we have
tried to use the steepened version of the PPA method, as described by
\citet{cw84} at Eqs.~(1.15) to~(1.17), using the same numerical
thresholds. This method is definitely not suitable for Keplerian
disks, where it is found to spuriously generate small scale vortices.

The circular permutation of Eq.~\eqref{eq:32} does not introduce
numerical errors, and $v_\mathrm{shift}$ does not have to be included
in the stability analysis: it does not contribute to the Courant
condition.  By construction of $N$, we have
$|v_\mathrm{shift}^R\Delta t| < 1/2 $, so that $v_\mathrm{shift}^R$
does not enter the Courant condition either.

The Courant condition is thus determined by a criterion similar to
that of Eq.~\eqref{cfl2}, in which we replace the azimuthal velocity
$v$ by $\delta v$.  So far we have not specified how the decomposition
of Eq.~\eqref{eq:26} is performed.  It should be done so that the
residual velocity $\delta v$ has the smallest possible
$\mathcal{L}_\infty$ norm, which is realized if one adopts:
\begin{equation}
  \label{eq:37}
  v^0_{jk}=\frac 12\left[\mathrm{max}(v_{ijk})+\mathrm{min}(v_{ijk})\right],
\end{equation}
where the minimum and maximum values are sought with $j$ and $k$ being
fixed, for all values of $i$.  Most implementations of the FARGO
algorithm consider instead the azimuthal average
\citep{Masset.2000,2009A&A...506..971K,Mignone.2012}.  We suggest that
some difference might be expected when the perturbed velocity is
significantly larger than the sound speed (so it is what primarily
limits the time step), which happens, in the context of planets in
disks, when giant planets are present.  Tests we have performed in
two-dimensional disks with an aspect ratio of $5$~\% and a Jupiter
mass planet have shown differences at the percent level between the
execution times of the two implementations, hence the azimuthal
average is good enough for most applications.

We conclude this section with the remark that we apply the orbital
advection algorithm to the same variables presented in
section~\ref{sec:splitt-source-transp}. The orbital advection, which
consists of a circular permutation of indices and conservative updates
involving azimuthal fluxes similar to those of
section~\ref{subsect:transport}, conserves therefore these fields to
machine accuracy. Just as its predecessor code (FARGO), the FARGO3D
code conserves mass and angular momentum to computer accuracy.

As pointed out in section~\ref{sec:momenta-advection}, the technique
of left and right momenta allows us to have the same control volume for
all the variables transported.  This is an important prerequisite to
implement orbital advection in staggered mesh codes.  With the
standard method of the ZEUS code, for instance, we would have to
consider rings staggered in radius for the transport of radial
momentum, so that the implementation of orbital advection would be far
more involved.  To the best of our knowledge, all implementations of
the FARGO algorithm in staggered mesh codes make use of the left and
right momenta technique.  Codes based on higher order Godunov methods
do not suffer from this problem, since all their hydrodynamics
variables are zone centered \citep[eg][]{Mignone.2012}.

Orbital advection must also be implemented to solve the induction
equation in the MHD case. This is presented in
section~\ref{subsect:fargo-mhd}.

\subsubsection{Amendments to the CFL condition}
\label{sec:amendm-cfl-cond}
The large, axisymmetric azimuthal velocity $v_0$ of Eq.~\eqref{eq:26}
does not enter into play in the stability criterion of
Eq.~\eqref{cfl2}.  Instead, one must substitute the total azimuthal
velocity by the residual velocity $\delta v$ in the CFL criterion (in
the second item of the list of section~\ref{subsect:time-step}). Since
the limit timestep must be evaluated before entering the different
(magneto-)hydrodynamical substeps, and since the evaluation of the
limit timestep requires the knowledge of the bulk azimuthal velocity
for orbital advection, the determination of this quantity is one of
the first tasks performed during a full update. It corresponds to
cell~3 of the flow chart of Fig.~\ref{fig:flowchart}. In geometrically
thin disks, which have an aspect ratio $h=c_s/v_K\ll 1$, the large
orbital velocity usually dominates largely the Courant condition.  By
getting rid of orbital motion in the latter, we can leverage the time
step by a factor $\sim h^{-1}$ for weakly perturbed flows. This ratio
typically amounts to an order of magnitude in protoplanetary disks.
In addition to relaxing this constraint on the time step, we introduce
a new time step limit linked to the shear, intended to prevent
radially neighboring zones from becoming disconnected after one time
step. This criterion reads:
\begin{equation}
  \label{eq:38}
  \Delta t_\mathrm{shear} < C_0\left(\frac{V_X(i,j,k)}{\Delta X_{j}}-\frac{V_X(i,j+1,k)}{\Delta X_{j+1}}\right)^{-1},
\end{equation}

where we impose that the content of radially neighboring zones are
offset at most a fraction $1-C_0$ of the azimuthal zone width after
advection during one time step along the orbital motion.  In practice,
this criterion is seldom useful.  The shear could become an issue in
the inner regions of the mesh, where the cells can be very elongated
radially ($\Delta r > r\Delta \phi$).  Omitting for simplicity the
multiplication by the Courant number, the shear time limit will be
important only if it is smaller than other time limits (see section
\ref{subsect:time-step}), in particular the time limit imposed by
sound waves, which reads here $\Delta t_\mathrm{sound} =
r\Delta\phi/C_s$. Specializing to Keplerian shear, we have here
$\Delta t_\mathrm{shear}=r\Delta\phi/(3/2\Omega \Delta r)$.  The shear
limit will therefore supersede the usual time step limit if $\Delta r
> 2H/3$, that is to say if the pressure scale length is unresolved, a
situation which should be avoided in any case.  We note that if the
cells keep an aspect ratio close to unity near the inner boundary, the
shear time step limit is of the order of the local orbital time, way
larger than other limits.

\subsection{Magnetohydrodynamics}
\label{sec:magnetohydrodynamic}
To evolve numerically the magnetic field, we follow the {\it Method of
  Characteristics + Constrained transport} (MOCCT) presented in detail
by \citet{Hawley.Stone.1995}.  The CT method, developed by
\citet{Evans.Hawley.1988}, ensures a constant (here, null) divergence
of the magnetic field to machine accuracy.  This method requires the
values at half time step of the velocity and magnetic field components
at the middle of a zone edge.  The MOC is used to calculate these
values. Instead of considering a unique characteristic velocity as is
done for the HD quantities (see section~\ref{subsect:transport}), a
more sophisticated approach is adopted in which incompressible
transverse Alfv\'en waves are considered.  Leftward and rightward
propagating Alfv\'en waves carry different eigenvalues which are
evaluated at the foot of their respective characteristics, at the
beginning of the time step, and combined at half time step to yield
the velocity and magnetic field values at the zone edge
\citep{Stone.Norman.1992.b}. The discussion in the following sections
cover the substeps of cells~8, 10 and~11 of Fig.~\ref{fig:flowchart}.

\subsubsection{Constrained transport method}
\label{subsubsect:CT}

The main idea of the constrained transport (CT) method, formulated by
\citet{Evans.Hawley.1988}, is to ensure that the divergence of the
magnetic field is maintained constant (here, naturally zero) for all
time, at machine accuracy, using a conservative formulation for its
flux based on the evaluation of electromotive forces.  Insomuch as the
sharing of fluxes on the zone faces satisfies a discrete version of
the divergence theorem, sharing electromotive forces (electric fields)
at the zone edges satisfies a discrete version of Stoke's theorem.
The integral form of Eq.~\eqref{eq:induction} reads:
\begin{equation}
  \frac{\partial \mathcal{F}^B}{\partial t} = \int_{\partial S} \left(\vec{v}\times \vec{B}-\eta\nabla\times\vec{B}\right).d\vec{l}
  \label{integrated-induction}
\end{equation}
where $\mathcal{F}^B$ is the magnetic flux through a surface S
(bounded by $\partial S$).  An approximation for the flux for the
$X$-component of the magnetic field across a zone face is:
\begin{eqnarray}
  {\mathcal{F}^B_X}_{i+1/2 } &=& {B_X}_{i+1/2 }\mathcal{S}_{i+1/2 },
\end{eqnarray}
with straightforward generalization to the $Y$ and $Z$ components. We
recall that the magnetic field is face centered like the velocity
field, as depicted in Fig.~\ref{fig:cell}.  The method to satisfy the
flux conservation is to derive the rate of change of the magnetic
fluxes by evaluating the circulation of the electromotive force (EMF)
defined by $\vec{\varepsilon}=\vec{v}\times\vec{B}$. The $X$ ($Y$,
$Z$)-component of the EMF is defined at the middle of the edge along
the $X$ ($Y$, $Z$) direction. The flux evolution is given by:
\begin{eqnarray}
\label{eq:39}
\frac{{\mathcal{F}^B_X}_{i+1/2 ,j,k}^{n+1} - {\mathcal{F}^B_X}_{i+1/2 ,j,k}^{n} } {\Delta t} = &&  \varepsilon^Y_{i+1/2 ,j,k+1/2 } \Delta Y_{i+1/2 ,j,k+1/2 }   \\
&-& \varepsilon^Y_{i+1/2 ,j,k-1/2 } \Delta Y_{i+1/2 ,j,k-1/2 } \nonumber \\
&-& \varepsilon^Z_{i+1/2 ,j+1/2 ,k}\Delta Z_{i+1/2 ,j+1/2 ,k} \nonumber \\
&+& \varepsilon^Z_{i+1/2 ,j-1/2 ,k}\Delta Z_{i+1/2 ,j-1/2 ,k}, \nonumber 
\end{eqnarray}
and similar relations for the $Y$ and $Z$ components of the flux are
obtained with a circular permutation $X\rightarrow Y\rightarrow
Z\rightarrow X$ and $i\rightarrow j\rightarrow k\rightarrow i$.  By
summing the variations of ${\mathcal{F}^B_X}_{i\pm 1/2 ,j,k}$,
${\mathcal{F}^B_Y}_{i,j\pm 1/2 ,k}$ and ${\mathcal{F}^B_Z}_{i,j,k\pm
  1/2 }$ (six terms in total), we can see that each edge is swept
twice in opposite directions, yielding a null net flux variation of
the magnetic field across the zone boundary.  If it is initially
divergence free, its divergence vanishes at all times.

\subsubsection{Method of characteristics}
\label{sec:meth-char}
\begin{figure}
  \centering
  \includegraphics[width=\columnwidth]{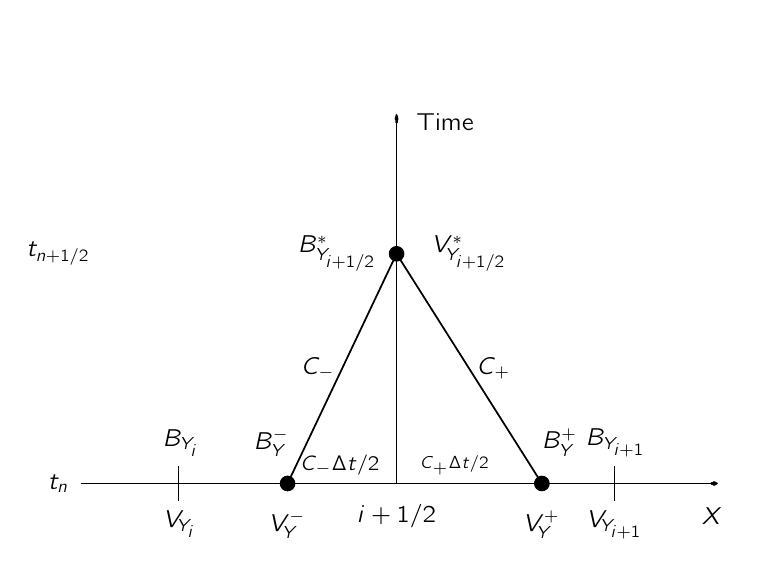}
  \caption{\label{fig:moc} Space-time diagram showing Alfv\'en's
    characteristics used to the extrapolate the EMF components, for
    the case $B_X>0$.}
\end{figure}

The integrand of Eq.~\eqref{integrated-induction} features two terms.
The first one corresponds to the field induction in ideal MHD, and the
second one to resistive effects.  We focus hereafter on the first
term.  The resistive term is added afterwards, prior to the
constrained transport substep.  As seen in
section~\ref{subsubsect:CT}, the components of the electromotive force
are defined along the edges. In order for the constraint transport
step to be second order accurate in time, they must be specified at
time $t+\Delta t/2$.  We consider specifically hereafter the case of
the $Z$-component of the electromotive force.  Other components can
trivially be deduced by the circular permutations given in
section~\ref{subsubsect:CT}.  We have:
\begin{equation}
  \label{eq:40}
  \varepsilon_Z = v_XB_Y-v_YB_X,
\end{equation}
and $\varepsilon^Z$ is defined at $(i+1/2 ,j+1/2 ,k)$, hence we need
the interpolated values ${v_{X,Y}}^*_{i+1/2 ,j+1/2 ,k}$ and
${B_{X,Y}}^*_{i+1/2 ,j+1/2 ,k}$ at time $t+\Delta t/2$.  These are
obtained using a method of characteristics which considers a
restricted problem for each pair $(v_X^*, B_X^*)$ and $(v_Y^*,
B_Y^*)$. The $X$-components are calculated by considering a 1.5D
problem along the direction $Y$, and vice-versa\endnote{The
  calculation of the star values is performed in file
  \texttt{compute\_star.c}. The functions in this file admits an
  integer argument \texttt{index} which specifies whether the flow
  speed is to be included in the characteristics speed. It must be
  included for the calculation of the electric field in
  Eq.~\eqref{eq:45} whereas it must not be included for the evaluation
  of the Lorentz force in Eq.~\eqref{eq:47}.}. We consider below the
case of the $Y$-components:
\begin{eqnarray}
  \frac{\partial v_Y}{\partial t} + v_X \frac{\partial v_Y}{\partial X} &=& \frac{B_X}{\mu_0\rho}\frac{\partial B_Y}{\partial X} \label{moc1}, \\
  \frac{\partial B_Y}{\partial t} + v_X \frac{\partial B_Y}{\partial X} &=& B_X\frac{\partial v_Y}{\partial X}. \label{moc2}
\end{eqnarray}
Eq.~\eqref{moc1} corresponds to the $Y$ projection of
Eq.~\eqref{eq:euler} in which we neglect the derivatives along the $Y$
and $Z$ coordinates, and all source terms other than the magnetic
tension.  Eq.~\eqref{moc2} corresponds similarly to the $Y$ projection
of Eq.~\eqref{eq:induction} in which we neglect resistive effects, and
derivatives along the $Y$ and $Z$ coordinates. This procedure only
retains the non-compressive, Alv\'en waves that may exist if $B_X$
does not vanish. By adding $\pm (\mu_0\rho)^{-1/2 }$ times
Eq.~\eqref{moc2} to Eq.~\eqref{moc1}, we find the characteristic
equations:
\begin{equation}
  \label{eq:41}
  \partial_t\psi_\pm+C_\pm\partial_X\psi_\pm = 0,
\end{equation}
where the characteristic speed is
\begin{equation}
  \label{eq:42}
C_\pm=v_X\mp\frac{B_X}{\sqrt{\mu_0\rho}},
\end{equation}
and where the eigenvalue $\psi_\pm$ is:
\begin{equation}
  \label{eq:43}
  \psi_\pm=v_Y\pm\frac{B_Y}{\sqrt{\mu_0\rho}}.
\end{equation}
The eigenvalues $\psi_+$ and $\psi_-$ are therefore conserved along
the characteristics with respective speeds $C_+$ and $C_-$.  Their
value is sought at the foot of these characteristics, using a
piecewise linear reconstruction of $v_Y$ and $B_Y$ with van Leer's
slope, exactly as done in the transport step (section
\ref{subsect:transport}) for the characteristic with the flow speed.
The method is depicted in Fig.~\ref{fig:moc}.
In our implementation, the values of $v_X$ and $B_X$ used to evaluate
the characteristic speeds in Eq.~\eqref{eq:42} are obtained themselves
with an upwind interpolation of these fields in $Y$ (again making use
of a zone-wise reconstruction with van Leer's slopes) with the flow
speed $\overline{v_Y}$, defined as the arithmetic average of
${v_Y}_{ij-1/2 k}$ and ${v_Y}_{i-1j-1/2 k}$.  We note these values
$v_x^u$ and $B_x^u$, where the $u$ superscript, which conveys the
upwind method, is used instead of $*$ to avoid confusion with the
values obtained through the MOC.  The value of $\rho$ used in
Eqs.~\eqref{eq:42} and~\eqref{eq:43} is the arithmetic average of the
four values $\rho_{ijk}$, $\rho_{i-1jk}$, $\rho_{ij-1k}$ and
$\rho_{i-1j-1k}$. Finally, the star value $v^*_Y$ and $B^*_Y$ are
given by:
\begin{eqnarray}
  \label{eq:44}
  v_Y^*&=&\frac 12(\psi_++\psi_-)=\frac 12\left(v_Y^++v_Y^-+\frac{B_Y^+-B_Y^-}{\sqrt{\mu_0\rho}}\right)\\\nonumber
  B_Y^*&=&\frac{\sqrt{\mu_0\rho}}{2}(\psi_+-\psi_-)=\frac 12\left[B_Y^++B_Y^-+\sqrt{\mu_0\rho}(v_Y^+-v_Y^-)\right].
\end{eqnarray}
where $v^\pm$ and $B^\pm$ are the field values at time $t$ at the foot
of the corresponding characteristics.  This corresponds to cell~8a of
Fig.~\ref{fig:flowchart}. The values given by Eq.~\eqref{eq:44} are
used to evaluate the electromotive force by use of
Eq.~\eqref{eq:40}. Nonetheless, in order to avoid anomalous EMFs near
rotational discontinuities of the magnetic field, we use the star
values only with the terms they are consistent with, which are the
upwind estimates of the transverse fields used to infer the
characteristics speeds \citep{Hawley.Stone.1995}. Eq.~\eqref{eq:40}
then takes the form:
\begin{equation}
  \label{eq:45}
	\varepsilon_Z = \frac{1}{2}\left(v^*_X B^u_Y + B^*_Y v^u_X - v^*_Y B^u_X - B^*_X v^u_Y\right).
\end{equation}
This corresponds to cell~8b of Fig.~\ref{fig:flowchart}.
Prior to applying the constrained transport through Eq.~\eqref{eq:39}
(and those obtained through circular permutations), which corresponds
to cell~10 of Fig.~\ref{fig:flowchart}, we add the
resistive term to each component, that is to say the second term of
the right hand side of Eq.~\eqref{integrated-induction}.  No special
care has to be observed to evaluate the curl of the magnetic field at
this stage: we use a simple finite difference estimate. This is done
in substep~8c of Fig.~\ref{fig:flowchart}.

Lastly, prior to the update of the field, the source term
corresponding to the Lorentz force is applied to each component of the
velocity.  Consider the update of the $X$ component. In Cartesian
coordinates, the partial differential equation corresponding to this
substep is:
\begin{equation}
  \label{eq:46}
  \partial_tv_X=\frac{1}{\mu_0\rho}(B_Y\partial_YB_X+B_Z\partial_ZB_X-B_Y\partial_XB_Y-B_Z\partial_XB_Z).
\end{equation}
We note in this equation that the term $B_X\partial_XB_X$ arising from
$\vec{B}.\nabla B_X$ cancels out with the corresponding term of the
magnetic pressure gradient $-\partial_XB^2/2$.  We deal with the first
two terms of the right hand side of Eq.~\eqref{eq:46} in a specific
substep, which we detail below.  The remaining two terms are dealt
with in the substep~1 described in section~\ref{sec:source-step}: the
gradient of the zone centered two-point average of the transverse
terms is applied in much the same way as the thermal pressure
gradient.  This is done at the beginning of the time step (in cell~7a
of Fig.~\ref{fig:flowchart}), with the value of the magnetic field at
time~$t$.

The magnetic tension source term, corresponding to the first two terms
of Eq.~\eqref{eq:46}, use the star value of the $B_X$ field on the
edge of the face at $i-1/2 $ to evaluate the gradients. The
corresponding substep reads therefore:
\begin{eqnarray}
  \label{eq:47}
  \frac{{v^c_X}_{i-1/2 jk}-{v^b_X}_{i-1/2 jk}}{\Delta t}&=&\frac{1}{\mu_0}\frac{2}{\rho_{ijk}+\rho_{i-1jk}}\times\\\nonumber
&&\left(\overline{B_Y}\frac{{B^*_X}_{i-1/2 j+1/2 k}-{B^*_X}_{i-1/2 j-1/2 k}}{\Delta Y_{i-1/2 k}}+\right.\\\nonumber
&&\left.\overline{B_Z}\frac{{B^*_X}_{i-1/2 jk+1/2 }-{B^*_X}_{i-1/2 jk-1/2 }}{\Delta Z_{i-1/2 j}}\right).
\end{eqnarray}
The momentum transport is dealt with during the transport step of the
full update (in cells~12a to~12d of Fig.~\ref{fig:flowchart}), and
similarly, the transport of the magnetic field is dealt with during
the constrained transport update (cell~10 of
Fig.~\ref{fig:flowchart}).  There is therefore no need to use an
upwind method to seek the values of the magnetic field to be used in
Eq.~\eqref{eq:47}.  The fluid element sitting at a face edge at the
beginning of the time step, and the associated magnetic field, can be
regarded as drifting together during the substep in which magnetic
tension is applied.  This has sometime been described as a Lagrangian
substep \citep{Hawley.Stone.1995}.  As a consequence, the starred
values in Eq.~\eqref{eq:47} are obtained with a method of
characteristics in which the characteristic speed is only
$\pm{B_X}/{\sqrt{\mu_0\rho}}$, instead of the speed given in
Eq.~\eqref{eq:42}.  Similarly, we use for the bar values in
Eq.~\eqref{eq:47} the suitable four-point averages rather than
advection upwind estimates as in Eq.~\eqref{eq:45}.  The different
notation (superscript $u$ versus bar) is meant to convey this
difference. The evaluation of the Lorentz force, and the corresponding
update of velocities, are performed in cell~8d of
Fig.~\ref{fig:flowchart}.

Finally, when the mesh is non-Cartesian, geometric source terms arise
from the $\vec{B}\cdot\nabla\vec{B}$ term. Those are also included in
substep~1, \emph{i.e.}  in cell~7a of Fig.~\ref{fig:flowchart}.  For
the sake of completeness we hereafter indicate the partial
differential equations corresponding to this update.

\begin{description}
\item[Cylindrical coordinates] 
  \begin{eqnarray}
    \label{eq:48}
    \partial_tv_\phi&=&\frac{B_rB_\phi}{\mu_0 \rho r}\\
    \partial_tv_r&=&-\frac{B_\phi^2}{\mu_0\rho r}
  \end{eqnarray}
\item[Spherical coordinates] 
  \begin{eqnarray}
    \label{eq:49}
    \partial_tv_\phi&=&\frac{B_rB_\phi+B_\theta B_\phi\cot\theta}{\mu_0\rho r}\\
    \partial_tv_r&=&-\frac{B_\phi^2+B_\theta^2}{\mu_0\rho r}\\
    \partial_tv_\theta&=&\frac{B_rB_\theta}{\mu_0\rho r}-\frac{B_\phi^2\cot\theta}{\mu_0\rho r}
  \end{eqnarray}
\end{description}

\subsection{FARGO algorithm in MHD - Orbital Advection}
\label{subsect:fargo-mhd}

The generalization of the orbital advection (aka FARGO algorithm) to
the MHD equations has first been considered by
\citet{2008ApJS..177..373J}.  Subsequently, a method that generalizes
the CT method and therefore automatically ensures the preservation of
the divergence free property of the magnetic field was devised by
\citet{2010ApJS..189..142S}.  To the best of our knowledge, our
implementation is the first implementation of this method in a staggered
mesh code.  Other implementations of the orbital advection of the
magnetic field were with centered methods
\citep{2010ApJS..189..142S,Mignone.2012}.  Our implementation does not
require any adaptation, though: the centering of the magnetic field is
the same in both kinds of codes, and they only differ by how the
orbital advection of hydrodynamical quantities is implemented.

The following discussion sums up the principles of the method of
\citet{2010ApJS..189..142S}, and shows that it amounts to an azimuthal
shift of the components of the magnetic field, similar to those of the
hydrodynamics variables (see section~\ref{subsect:fargo}), when there
is no shear.

As for the hydrodynamics, orbital advection consists in splitting the
set of governing equations into a set based on the residual velocity
$\delta v$, and a set based on the orbital velocity, which is
exclusively azimuthal, and axisymmetric.  The first set is solved as
detailed in the previous section.  Orbital advection has been
discussed in section~\ref{subsect:fargo}. Its magnetic supplement
corresponds to solving the following equation:
\begin{equation}
  \label{eq:50}
  \partial_t\vec{B}-\nabla\times(\vec{v_0}\times\vec{B})=\vec{0},
\end{equation}
Component wise, Eq.~\eqref{eq:50} reads (the discussion hereafter
specializes to the two-dimensional shearing sheet, but generalization
to three dimensions in cylindrical or spherical coordinates is
immediate):
\begin{eqnarray}
  \label{eq:51}
  \partial_tB_X+v_0\partial_XB_X&=&qB_Y\\\nonumber
  \partial_tB_Y+v_0\partial_XB_Y&=&0,
\end{eqnarray}
where $q=\partial_Yv_0$ may depend on $Y$ but neither on $X$ nor on
time, and where the constraint $\nabla\cdot\vec{B}=0$ has been used in
the first of these equations, which state that the transverse field
simply obeys an advection equation at the uniform speed $v_0$, and the
shear of the transverse field acts as a source term on the
longitudinal component, in addition to its advection.  The system
above trivially admits the following exact solution:
\begin{eqnarray}
  \label{eq:52}
  B_X(x,t)&=&B^0_X(x-v_0t)+tqB^0_Y(x-v_0t)\\
\label{eq:53}
  B_Y(x,t)&=&B^0_Y(x-v_0t),
\end{eqnarray}

where $B_{X/Y}^0(x)$ are the $X$ and $Y$ components of the magnetic
field at $t=0$.  For any finite time interval $\Delta t$, we define
the upstream average of the vertical electric field as:
\begin{equation}
  \label{eq:54}
  \overline{E_z}(x,t,\Delta t) = \frac{1}{\Delta x}\int_{x-\Delta x}^xv_0B_y(x',t)dx',
\end{equation}
where $\Delta x= v_0\Delta t$. Using Eqs.~\eqref{eq:53} and~\eqref{eq:54}, we have:
\begin{equation}
  \label{eq:55}
  -\partial_x\overline{E_z}(x,t,\Delta t)=\frac{1}{\Delta t}\left[B_y(x,t+\Delta t)-B_y(x,t)\right].
\end{equation}
Similarly, we have:
\begin{equation}
  \label{eq:56}
  \partial_y\overline{E_z}(x,t,\Delta t)=qB_y(x-\Delta x,t)-\frac{1}{\Delta t}\left[B_x(x,t)-B_x(x-\Delta x,t)\right],
\end{equation}
where we have used the property $\nabla\cdot\vec{B}=0$. Using Eq.~\eqref{eq:52}, this result can be recast as:
\begin{equation}
  \label{eq:57}
  \partial_y\overline{E_z}(x,t,\Delta t)=\frac{1}{\Delta t}\left[B_x(x,t+\Delta t)-B_x(x,t)\right].
\end{equation}
Eqs.~\eqref{eq:55} and~\eqref{eq:57} state that the upstream average
of the electric field can be used to get the exact variation of the
magnetic field, regardless of the magnitude $\Delta t$ of the time
step.  This is because the flow velocity is uniform along the
streamlines and constant in time, so that the upstream average
coincides with the Eulerian time average over the time step.  With
discretization in space, we perform the integral of Eq.~\eqref{eq:54}
over as many zones as necessary to cover the arcs of length $\Delta
x$.  Only the $Y$ and $Z$ components of the electric field do not
vanish.  They are respectively staggered in $Z$ and $Y$.  The velocity
$v_0$ used in their evaluation is the two-point arithmetic average of
the azimuthal velocities in the zones adjacent to the face considered.
It may be enlightening to consider the special case in which there is
no shear, and the arcs span an integer amount $N$ of cells in
azimuth. Under these special circumstances, we have:
\begin{equation}
  \label{eq:58}
  \overline{E_z}_{i-1/2 j-1/2 }=\frac{v^0_{j-1/2}}{N}\sum_{m=1}^{N}{B_Y}_{i-mj-1/2}.
\end{equation}
The update of the $Y$ component reads (all indices for the $Y$
direction have implicitly value $j-1/2$):
\begin{eqnarray}
  \label{eq:59}
  \Delta X\frac{B^{n+1}_{Yi}-{B^{n}_Y}_i}{\Delta t}&=&-\overline{E_z}_{i+1/2}+\overline{E_z}_{i-1/2}\\
  &=&\frac{v^0_{j-1/2}}{N}\left(-{B^n_Y}_i+{B^n_Y}_{i-N}\right).
\end{eqnarray}
Since by assumption $v^0_{j-1/2}\Delta t=N\Delta X$, we are simply
left with:
\begin{equation}
  \label{eq:60}
  B^{n+1}_{Yi} = {B^n_Y}_{i-N}.
\end{equation}
Similarly, the update of the $X$ component reads:
\begin{eqnarray}
  \label{eq:61}
  \Delta Y\frac{B^{n+1}_{Xi-1/2}-{B^{n}_X}_{i-1/2}}{\Delta t}&=&\overline{E_z}_{i-1/2j+1/2}-\overline{E_z}_{i-1/2j-1/2}\\  \nonumber
&=&\frac{v^0_{j+1/2}}{N}\sum_{m=1}^N{B_Y}_{i-mj+1/2}\\\nonumber
&&-\frac{v^0_{j-1/2}}{N}\sum_{m=1}^N{B_Y}_{i-mj-1/2}
\end{eqnarray}
Using the no shear assumption, we can write $v^0_{j+1/2}/\Delta
X_{j+1/2}=v^0_{j-1/2}/\Delta X_{j-1/2}$, where $\Delta X_{j\pm 1/2}$
is the length of the upper (lower) edge in $X$ of the cell $(i,j)$. We
therefore have:
\begin{eqnarray}
  \label{eq:62}
  \Delta Y\frac{B^{n+1}_{Xi-1/2}-{B^{n}_X}_{i-1/2}}{\Delta t}=\\\nonumber
\frac{v^0_{j+1/2}}{\Delta X_{j+1/2}N}\sum_{m=1}^N\Delta X_{j+1/2}{B_Y}_{i-mj+1/2}-\Delta X_{j-1/2}{B_Y}_{i-mj-1/2}.
\end{eqnarray}
Using the divergence-free property of the magnetic field, the generic
term of the sum of Eq.~\eqref{eq:62} can be recast as $\Delta
Y({B_x}_{i-m-1/2}-{B_x}_{i-m+1/2})$. We therefore have:
\begin{equation}
  \label{eq:63}
  \frac{B^{n+1}_{Xi-1/2}-{B^{n}_X}_{i-1/2}}{\Delta t}=\frac{v^0_{j+1/2}}{\Delta X_{j+1/2}N}({B^n_X}_{i-N-1/2}-{B^n_X}_{i-1/2}),
\end{equation}
which yields, since $v^0_{j+1/2}\Delta t=N\Delta X_{j+1/2}$:
\begin{equation}
  \label{eq:64}
  B^{n+1}_{Xi-1/2} = {B^n_X}_{i-N-1/2}.
\end{equation}
Eqs.~\eqref{eq:60} and~\eqref{eq:64} show that in the special case
under consideration the constrained transport using the upstream
average electric field has characteristics similar to those of orbital
advection in the hydrodynamical case: it merely amounts to a circular
permutation of the cell values, with no numerical diffusivity, and no
impact on the Courant condition.  Instead of being hard coded as a
shift as in the hydrodynamical case, however, here we require the
evaluation of the upstream averages.  Most of their terms cancel out,
which \emph{de facto} amounts to a shift to machine accuracy.  In the
more general case with shear and when the upstream arc does not
represent an integer number of cells, the electromotive force on the
residual arc is evaluated using a piecewise parabolic interpolation of
the electric field\endnote{Orbital advection is implemented in the
  file \texttt{fargo\_mhd.c}, which contains the invocation to the
  different sub steps, and in the file
  \texttt{integrate\_emf.c}.}. The orbital advection of the magnetic
field corresponds to cell~11 of Fig.~\ref{fig:flowchart}.

\subsection{Orbital integrator for planets}
\label{sec:orbit-integr-plan}
Like its ancestor FARGO, the FARGO3D code features the possibility to
simulate an arbitrary number of point-like masses around a central
mass, which interact with the gas.  To evolve planetary systems, we
use the fifth order Cash-Karp method
\citep{Cash:1990:VOR:79505.79507}, a Runge-Kutta method with a fixed
time step, here governed by the CFL condition.  As the time step is
decoupled from the interaction between bodies, this method does not
ensure a good solution for close encounters.  However, it is good
enough for planetary systems in which planets do not suffer close
encounters.  Should more accuracy be needed at some point, it is easy
to sub-cycle the planetary integration with an adaptive time step
decoupled from the CFL condition to advance the planetary system
with a high level of precision.  We also mention that our N-body
module is, like in the FARGO code, well decoupled from the rest of the
code and easy to substitute, should one need a more sophisticated
integrator.  This has been for instance the approach of
\citet{2012A&A...546A..18M}, who implemented the SYMBA solver
\citep{1998AJ....116.2067D} in the FARGO code in order to detect close
encounters and collisions.

For the sake of completeness, we lay down here the algorithm used.
The differential equation reads formally:
\begin{equation}
  \frac{d \vec{y}}{dt} = \vec{f}(t,\vec{y}),
\end{equation}
where $\vec{y}$ is the $6N$ vector component of the positions and
velocities of the $N$ planets.  The approximate solution given by the
Cash-Karp method is:
\begin{equation}
  \vec{y}^{n+1} = \vec{y}^n + h\sum_i^6 b_i \vec{k_i}
\end{equation}
with $\displaystyle{\vec{k_i} =
  \vec{f}\left(t_n+c_ih,\vec{y}^n+h\sum_{j=1}^6a_{ij}\vec{k_j}\right)}$,
where the coefficients are given by the standard Butcher tableau
(Tab.~\ref{rk}).

The force exerted by the gas onto the planets is evaluated only once
per hydrodynamical time step\endnote{the force exerted by the disk on
  the planets is evaluated in \texttt{compute\_force.c}.}, and used to
update the planetary velocities. It is evaluated in either of two
manners:
\begin{itemize}
\item by direct summation of the force exerted by all the cells,
\item or by removing, prior to this summation, the axisymmetric part
  of the gas density to the content of all cells\endnote{This option
    is activated by the use of the compilation flag
    \texttt{BM08}.}. Since an axisymmetric disk cannot exert a torque
  on an embedded, coplanar planet, this does not alter the tidal
  torque exerted on the planet. An axisymmetric disk does however
  exert a radial force, on the planets and on itself. Since we do not
  take self-gravitation into account, the disk and the planets orbit
  in effectively different potentials, which leads to a spurious shift
  of resonances in the disk \citep{2008ApJ...678..483B}, which can
  severely bias the rate of change of orbital elements.  A workaround
  to this issue without resorting to self-gravitational, expensive
  calculations, can consist in removing the axisymmetric component of
  the disk mass prior to the force evaluation.
\end{itemize}
The update of the planetary positions and velocities is performed in
cell~5a of Fig.~\ref{fig:flowchart}.

\begin{table*}
\caption{Butcher tableau for the Cash-Karp method.}
\label{rk}
\centering
\def\arraystretch{.8}
{\footnotesize
\begin{tabular}{c|cccccc}
$0$ & & & & &\\
$1/5$ & $1/5$ & & & & &\\
$3/10$ & $3/40$ & $9/40$ & & &\\
$3/5$ & $3/10$ & $-9/10$ & $6/5$ & & &\\
$1$ & $-11/54$ & $5/2$ & $-70/27$ & $35/27$ & &\\
$7/8$ & $1631/55296$ & $175/512$ & $575/13824$ & $44275/110592$ & $253/4096$ &\\
\hline\\
&$37/378$ & $0$ & $250/621$ & $125/594$ & $0$ & $512/1771$
\end{tabular}}
\end{table*}

\section{Implementation considerations}
\label{sec:impl-cons}
\subsection{General considerations}
\label{sec:gener-cons}
FARGO3D is entirely written in C\endnote{While some functions have
  been adapted from the former FARGO code, most of them have been
  written from scratch.}, and parallelized using MPI (\emph{Message
  Passing Interface}) and a slab domain decomposition in the $Y$ and
$Z$ directions only (we have chosen to have the full extent of the $X$
direction, along which orbital advection is performed, on a unique
processing element).  As can be seen in Tab.~\ref{tab:corresp}, the
mesh is therefore split radially and vertically in cylindrical
coordinates, and in radius and colatitude in spherical coordinates. As
in all grid based codes, the meshes are surrounded by several extra
layers of cells, usually called buffer zones or ghost zones, which are
used either to specify the boundary conditions, or to receive the
values of neighboring meshes upon an MPI communication. The number of
extra layers depends both on the problem at hand and the frequency
with which communications are performed (\emph{i.e.} how many times
per time step).  There is a trade off between the number of ghost
layers and the frequency: the more frequent the communications, the
thinner the ghost layers.  In our implementation we perform
communications only twice per time step (corresponding to cells~2
and~9 of Fig.~\ref{fig:flowchart}), and we have three layers of ghost
zones\endnote{Defined by the preprocessor variables \texttt{NGHY} and
  \texttt{NGHZ} in \texttt{define.h}.}.  As this number also depends
on the complexity of the problem, it may need an increase when new
physics is included (such as thermal diffusivity, or any other new
source sub step in which the new value of a zone depends on its
neighbors).  The correct number of ghost layers should be determined
via an indiscernibility test: it is the minimum number of ghost layers
that are required to yield an outcome strictly independent of the
number of processors on which the test is spawn.

\subsection{Building for GPUs}
\label{sec:building-gpus}

In order to avoid programming manually in CUDA (which is tedious and
error prone) and in order to keep the development as simple as that of
a code only meant to run on CPU, we have chosen to convert
automatically expensive CPU routines to their CUDA counterpart.  The
computational cost comes from a particular kind of routines, that we
dub {\it mesh functions}. A mesh function is a function that takes
certain inputs such as scalar values, geometrical quantities and
meshes (data cubes), then processes them using nested loops over the
mesh in order to update a data cube. They correspond to the various
substeps of the numerical algorithm. For instance, a mesh function is
used to fill the pressure array, from the internal energy
array. Another one is used to update the array of velocity in the
$X$-direction from the pressure and potential gradients. Still another
one is used to divide an array by another array, cell by cell (this is
required by the consistent transport mentioned at the end of
section~\ref{subsect:transport}).

 The general form of this kind of
function is:
\begin{verbatim}
function mesh_function_cpu(arguments):
  global variables 
  local variables 
  do something (initializations, etc.)
  loop k 
    loop j 
      loop i 
        q(i,j,k) = f(neighbor cells) 
        
  do something (e.g. reductions, etc.) 
end function 
\end{verbatim}
It is possible to build a few set of rules to transform this
pseudo-language to any computational language, when the transformation
rule for each specific part is defined.  In order to transform this
piece of pseudo code to CUDA, which generally has another memory
space, it is also necessary to deal with transfers between host (CPU)
and device (GPU) memories.  Dealing manually with such transfers can
quickly turn out to be a logistic nightmare, especially during the
early phases of development, when some mesh functions are already
running on the device while others are kept on the host.  For this
purpose, we have developed two commands, \texttt{INPUT()} and
\texttt{OUTPUT()}, which automatically deal with data transfers
between host and device: each mesh structure has two flags which
indicate whether the data is up-to-date on the host and device
memories.  \texttt{INPUT()} uses this information to trigger a data
transfer whenever required, while \texttt{OUTPUT()} resets both flags,
depending on were the mesh function is run.  The portrait of a general
mesh function must therefore be amended as:
\begin{verbatim}
function mesh_function_cpu(arguments):
  INPUT(fields required by function f) 
  OUTPUT(q) 
    etc. 
\end{verbatim}
We note that recent versions of CUDA (from 6.0 onward) allow the same
array to be used on the host and device, dealing automatically with
data transfers under the hood. These versions were not available when
we began the development of FARGO3D.  One could therefore conceive a
simpler conversion of the CPU code to the GPU code, in which explicit
data transfers are not necessary, removing the need for the
\texttt{INPUT/OUTPUT} declarations at the beginning of mesh functions.
However, the development of the \texttt{INPUT/OUTPUT} commands allowed
us to include more logistical analysis in them.  In particular, we can
use the same storage area for two different data cubes which are used
in different parts of a time step\endnote{Through the use of the
  \texttt{CreateField()} and \texttt{CreateFieldAlias()} commands in
  \texttt{LowTasks.c}.}, without sacrificing the legibility of the
code.  Another task performed by the \texttt{INPUT} command is to
check whether the storage area of its argument array is not presently
occupied by other data: this allows us to reuse as much as possible
previously allocated arrays, yielding the smallest possible memory
footprint. Also, dealing semi-automatically with data transfer through
the use of the \texttt{INPUT/OUTPUT} commands allows us to perform all
the transfers needed and \emph{only} when they are needed, thereby
ensuring the best possible performance.  Typically, transfers from the
normal RAM to the GPU (host to device transfers) are performed over
the first timestep, because the initial data are initialized on the
CPU, and the \texttt{INPUT} directives trigger their upload to the
GPU. In the subsequent timesteps, no large data transfers are required
between the host and device: all three-dimensional HD or MHD arrays
are updated directly on the GPU. Large data transfer from the GPU to
the host are subsequently triggered only when the execution flow
encounters an \texttt{INPUT} directive for a function that runs on the
CPU. Under normal circumstances, this only happens when data is
written to the disk: all mesh functions do not actually require data
transfers, because the fields required by the functions are already
stored on the GPU. The GPU to host communications are therefore
sparse, and occur only during the data output to the disk. They do not
constitute a bottleneck in our implementation. An exception to this is
naturally when one of the mesh functions of a full update is run on
the CPU rather than the GPU. The \texttt{INPUT/OUTPUT} machinery
triggers the data transfers required to execute this function on the
CPU and the subsequent functions on the GPU. In this case large
data transfers occur for every timestep, with a significant impact on
performance. For this reason the user is encouraged to adhere to the
strict syntax rules that we have developed, in order to get the mesh
function automatically converted to CUDA to run on GPU. All mesh
functions of the public version of FARGO3D do follow these rules and
can run on the GPU.

Finally, although we have developed a parser from the C code to CUDA,
we could easily conceive a parser from C to OpenCL in order to run
FARGO3D on GPUs of any brand. In this case, the \texttt{INPUT/OUTPUT}
nomenclature throughout the code is necessary.

The parsing of a general mesh function must, in addition to writing
the CUDA kernel (the GPU counterpart of the CPU routine), write a
wrapper function, which is a C++ function that acts as intermediary
between the C code and the kernel.  The final result after the parsing
process is:
\begin{verbatim}
function kernel_mesh_function(extended arguments):
  local variables
  get i,j,k from thread and block indices  
  q(i,j,k) = f(neighbor cells)        
end function

function mesh_function_gpu(some arguments)
  INPUT(fields required by function f) 
  OUTPUT(q) 
  do something (initializations, etc.)
  execute kernel_mesh_function (some arguments,
                               +arguments,
                               +globals)
  do something (reductions, etc.)
end function
\end{verbatim}
The parsing process takes care of converting the loop limits of the C
code into tests on \texttt{i}, \texttt{j} and \texttt{k}, obtained
from the thread and block indices. The kernel invocation passes all
the variables needed by the kernel and, whenever required, small size
one-dimensional arrays or scalar global variables are copied into
either the so-called constant memory (which is a 64k cached memory
available on all devices), or directly to the device's global memory
if they do not fit in the constant memory, before the kernel
invocation. The corresponding \texttt{cudaMemcpyToSymbol} or
\texttt{cudaMemcpy} statements are automatically issued by the
parser. We do not manually use the shared memory available on the
streaming multiprocessors. Rather, we rely on the cache that first
appeared on platforms with compute capabilities 2.0. We therefore
issue, prior to the kernel invocation, a statement:
\begin{verbatim}
cudaFuncSetCacheConfig (kernelname,
             cudaFuncCachePreferL1);
\end{verbatim}
in order to get the maximum available amount of cache. Also, prior to
the kernel invocation, we perform some algebra to determine the number
of CUDA blocks and their size in threads. The spawning of a large
number of CUDA threads obeys a two-level hierarchy: at the top level
is a grid of CUDA blocks. Subsequently, each block is itself an array
of threads. A given block runs within one of the many streaming
multiprocessors that exist on the GPU. The total number of threads
thus spawned must be at least equal to the number of cells in our
computational mesh, so that one thread updates the content of one
cell. The division of the grid in blocks, and of the blocks in
threads, must be done in an efficient way: if one chooses a small
number of large blocks, the register pressure in each block may
significantly degrade the performance. If, on the other hand, one
chooses a large number of small blocks, the streaming multiprocessors
may be underused when running these small blocks. For each setup we
choose a characteristic block size, so that all CUDA kernels are
launched with blocks of this size by default. However, we have also
implemented an option in our parser which produces a code that wraps a
given kernel invocation in a loop that explores all the possible block
sizes and remember the size that yields the shortest execution
time. By executing this search prior to a large run, we ensure that
each kernel is launched with the best possible block size. This
feature allows us to gain a further $10-20$~\% speed up with respect to
the unique default size. We emphasize that each of our mesh functions
performs simple operations, and are designed to avoid race
conditions. Should a race condition appear in one of them, we would
have to split it into more elementary mesh functions. Therefore, the
core part of each mesh function has exactly same structure in the CPU
version (in that case it is the part executed within the nested loops)
and in the GPU version (in that case it is the action executed by each
thread upon completion of the indices' algebra). The only difference
between these two versions is the algebra on array offsets: data are
contiguous in the CPU memory, while they have some extra padding on
the GPU to preserve the alignment of each row. Our conversion process
is therefore simple and systematic, hence the performance achieved on
the GPU (measured in terms of speed up factor with respect to the CPU
version) does not vary significantly from kernel to kernel.

The parsing and production of CUDA code from the C code is done on the
fly during the compilation of the code.  The conversion process is
guided by comments in the C code that follow strict syntax rules,
which allow to separate the different parts of the mesh function
(variables with a scope limited to the function, or \emph{internal}
variables; variables that must be passed as argument to the kernel, or
\emph{external} variables; core code of the mesh function, pre- and
post-operations, etc.) Note that not all
computationally expensive functions are \emph{mesh functions}, in
which one or several data cubes are updated as function of local data
(values of data cubes in same cell or neighboring cells). This, in
particular, is the case of reductions, in which one single value is
obtained as a function of all cell values. Diagnostic functions, which
integrate a value over the whole mesh (such as the mass or angular
momentum), or the routine that uses the Courant condition (see
section~\ref{subsect:time-step}) to evaluate the maximal affordable
time step, correspond to reductions (respectively with the \emph{sum}
and \emph{min} operations). Reductions have been implemented once for
all in the code\endnote{In the files \texttt{reduction\_generic.c} and
  \texttt{reduction\_generic\_gpu.cu}, in which the reduction
  operation is a macrocommand \texttt{macro(x,y)}, which can be
  arbitrary. These files perform the reduction in the $X$ direction,
  thereby returning an array of dimension $n-1$ for a run in $n$
  dimensions. The final reduction of that intermediate array is always
  performed on the CPU. The generality of the generic reductions is
  exploited in files \texttt{reduction\_sum.c},
  \texttt{reduction\_min.c}, \texttt{reduction\_min\_device.cu} and
  \texttt{reduction\_sum\_device.cu}.}, following the efficient scheme
of \citet{MH07}.

We typically achieve a $40\times$ speed up in the execution on one GPU
(with ECC activated) compared to one CPU of similar
generation \endnote{In addition to code execution timing, one can test
  the speed up on individual routines using the
  \texttt{FARGO\_SPEEDUP()} macrocommand defined in
  \texttt{define.h}.} (e.g. between an Intel$^\circledR$
Core\texttrademark i7 950 at 3.07 GHz and a Tesla C2050, or between
an Intel$^\circledR$ Xeon$^\circledR$ E5-2609v2 at 2.5 GHz and a
Kepler 20), although this ratio depends on the problem and on the mesh
size (larger speedups are obtained with larger meshes).  The error
control is naturally activated on the GPUs in these tests.

\begin{table*}
  \centering
  \begin{tabular}{lcc}
   Setup & Computational throughput (Mcell/s) & Memory footprint (bytes/cell)\\
\hline
    $a$  & $36.2$ & 138\\
\hline
    $b_1$ & $3.88$ & 280\\
    $b_2$ & $2.44$ & 2060\\
\hline
    $c$ & $20.4$ & 168\\
\hline
    $d$ & $22.6$ & 200\\
  \end{tabular}
  \caption{\label{tab:compchar}Computational throughput and memory
    footprint for different setups. Setup $a$: HD only, locally
    isothermal, 2D (azimuth and radial) in cylindrical geometry
    (therefore strictly similar to the former FARGO code). Setup
    $b_1$: MHD, Cartesian mesh in YZ, equation of energy (corresponds
    to the Orszag and Tang test, see
    section~\ref{sec:orzag-tang-vortex}). Setup $b_2$: same as $b_1$,
    but the mesh is in XY. Setup $c$: HD, spherical mesh, locally
    isothermal. Setup $d$: MHD, cylindrical mesh, locally isothermal
    (corresponds to the unstratified MRI test presented in
    section~\ref{sec:unstr-mri-test}). The low performance for the
    setup $b_1$ is due to the fact that reductions in YZ are performed
    on the CPU (only reductions along the X axis are performed on the
    GPU). The low performance for the setup $b_2$ is due to the fact
    that the XY mesh is sliced in between three layers of ghost cells
    in Z on each side (MHD setups require all three coordinates,
    regardless of the mesh size). All tests have been performed on one
    NVIDIA's K20 graphics card in double precision. \emph{Note: all
      throughputs on the GPU have been measured with ECC on.}}
\end{table*}

Table~\ref{tab:compchar} presents the computational throughput and
memory footprint of different setups. Tests $a$, $b_1$ and $b_2$
correspond to two-dimensional setups, while tests $c$ and $d$
correspond to three-dimensional setups. When MHD is included, all
three directions are calculated and ghost zones are always added to
the mesh in $Y$ and $Z$. The 2D Cartesian MHD tests $b_1$ and $b_2$,
performed respectively in $YZ$ and $XY$, have very different memory
footprints. The $YZ$ test has an active mesh size (respectively in
$X$, $Y$, and $Z$) of $1\times 1024\times 1024$, and the augmented mesh
(including the ghost zones) has size $1\times 1030 \times 1030$.
Similarly, the $XY$ set has an active mesh of size
$1024\times1024\times 1$, the augmented mesh has size
$1024\times1030\times 7$, hence an approximately 7-fold larger memory
footprint. Since calculations occur in the ghost zones, and since our
computational throughput is evaluated using exclusively the cells of
the \emph{active} mesh, there is an apparent drop, typically by a
factor of $7$, of the computational throughput of test $b_2$. The low
throughput measured for test $b_1$ has a different origin: here the
active and augmented meshes have sensibly same size. In our
implementation, reductions are firstly performed on the GPU in the
$X$-direction, and completed in the $Y$ and $Z$ direction on the CPU
(the cost of these last two steps being negligible compared to the
first one if there are many cells in $X$). Here, however, we only have
one cell in the $X$ direction: the reduction (which corresponds to the
search of the minimal timestep limit over the mesh) can be regarded as
performed only on the CPU, which constitutes a bottleneck. This is the
only bottleneck in our present implementation, and it occurs for
setups which are not among the primary goals of the code. 

Tab.~\ref{tab:compchar} can also be used to estimate the maximal size
of a simulation for a given platform. The memory footprint does depend
on the problem dimensionality, on whether the equation of energy is
solved, and on whether MHD is included, but it does not depend on the
mesh geometry. We can for instance estimate that on a K20x device with
6~Gb of RAM, we could run a Cartesian three-dimensional isothermal
HD simulation on a cube of size $\sim (6\cdot
10^9/168)^{1/3}\approx 329$~cells. This is naturally an upper limit, since
some memory space must be reserved for auxiliary two- and
one-dimensional arrays. Assuming arbitrarily that about $10$~\% of the
card memory should be reserved for overhead, we find that a simulation
of size $320^3$ could fit on one K20x.

\subsection{MPI and GPUs}
\label{sec:mpi-gpus}
\begin{figure}
  \centering
  \includegraphics[width=\columnwidth]{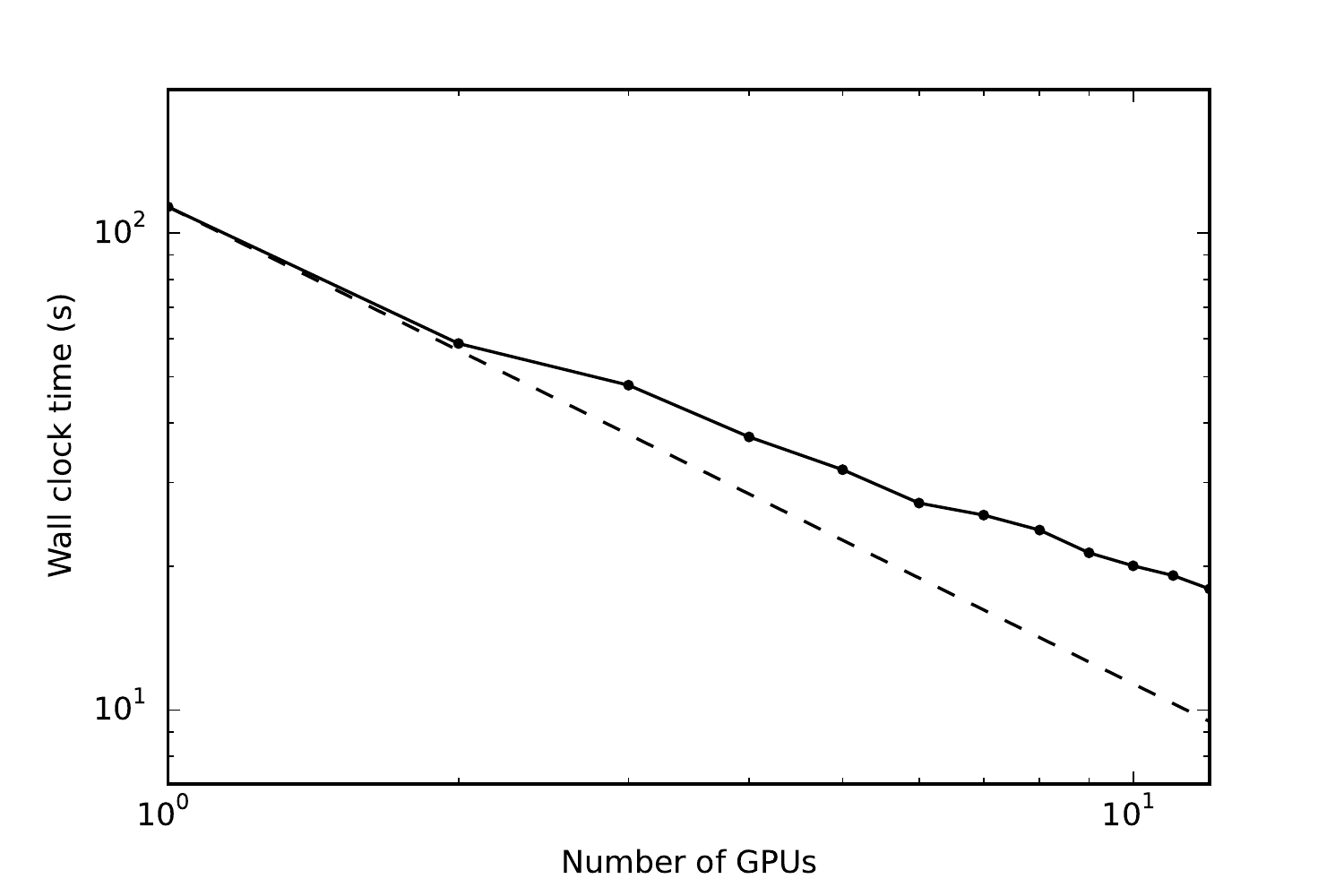}
  \caption{\label{fig:strongscaling} Wall clock time as a function of
    the number of GPUs, on a cluster of K20 GPUs, for a problem of
    given size (strong scaling test). The dots represent the results
    of measurements. The dashed line shows the ideal case obtained by
    extrapolating the time measured when the test is run on one GPU
    only.}
\end{figure}

Our implementation makes use of the peer-to-peer and \emph{universal
  virtual addressing} features of CUDA, available since version 4.0,
which allow to send data directly from one GPU to another one, thereby
reducing the cost of inter-GPU MPI communications. This feature is
requested at build time\endnote{By the use of the compilation option
  \texttt{MPICUDA=1}.}, and can be disabled to fall back on
traditional MPI communications between CPUs (accompanied with CPU-GPU
data transfer), should peer-to-peer communications be not supported on
a given platform.  We note that the typical speed up ratio between a
CPU core and a GPU is about 40, which is of the same order of
magnitude as the bandwidth ratio between a gigabit and InfiniBand
network.  The scaling performance of a parallel GPU code on an
InfiniBand cluster should therefore be similar to the scaling
efficiency of its CPU version on a gigabit cluster \emph{with one core
  per node}\footnote{Although a GPU cluster can have several GPUs per
  node, the bandwidth of the bus transited by data in intranode
  communications has same order of magnitude than the bandwidth of
  InfiniBand networks.}.  We show in Fig.~\ref{fig:strongscaling} the
results of a strong scaling test of FARGO3D on a cluster of Kepler~20
graphics card. The cluster has four such cards per nodes, and the
nodes are interconnected with an FDR InfiniBand network at 56
Gb/s. The setup considered in this test is irrelevant for the purpose
of this section, and it suffices to know that the mesh considered has
size $N_x=1300$, $N_y=435$ and $N_z=43$ (so that the problem just fits
in the RAM of one graphics card). 

\section{Tests}
\label{sec:tests}
We present below a series of tests which all have been published
elsewhere, for comparison
purposes. Section~\ref{sec:hydrodynamical-tests} presents the purely
hydrodynamical tests, while section~\ref{sec:magn-tests} presents MHD
tests. Within each section, the tests are broadly organized by
increasing dimensionality of the setup.

\subsection{Hydrodynamical tests}
\label{sec:hydrodynamical-tests}
\subsubsection{Sod shock-tube test}
\label{sec:sod-shock-tube}
\begin{figure}
  \centering
  \includegraphics[width=\columnwidth]{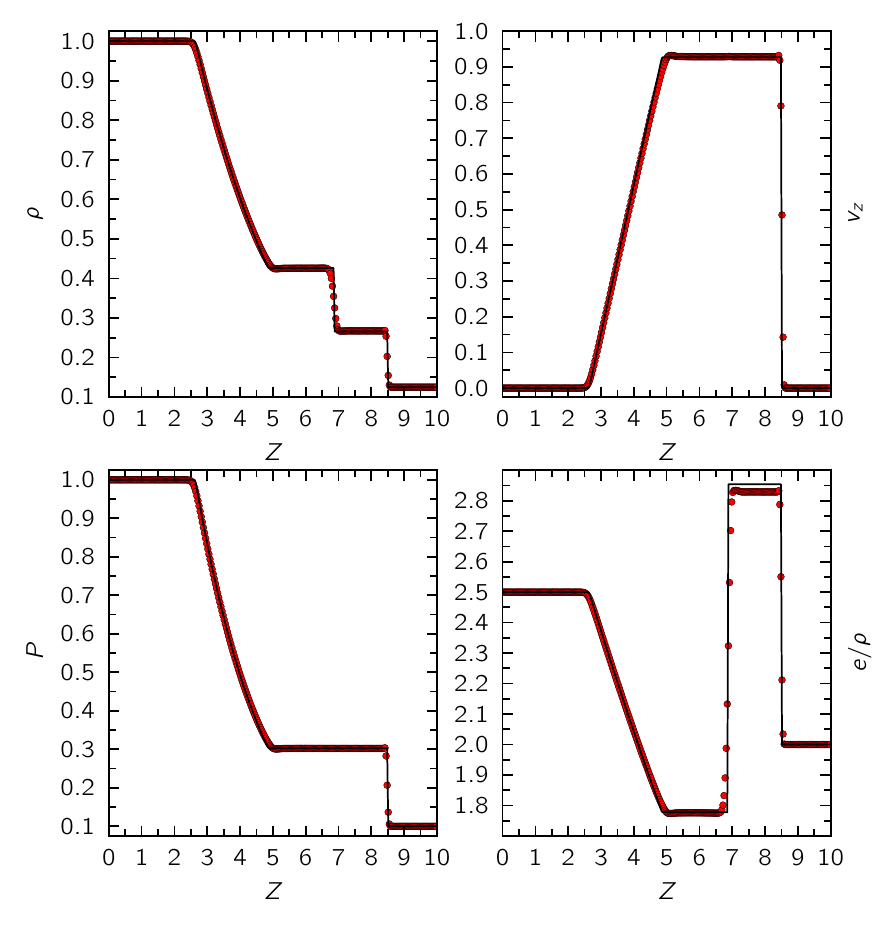}
  \caption{\label{fig:sod1d} Sod shock tube test, with 300 grid cells,
    at $t=2$. The shock (at $z\approx 8.5$) is spread over $2-3$
    zones, while the contact discontinuity (at $z\approx 6.9$) is
    spread over $7-8$ zones. There is a hardly noticeable offset of
    $\delta \rho/\rho \sim 10^{-2}$ between the numerical solution
    (red dots) and the exact one (solid black curve), between the
    contact discontinuity ($z\sim 7$) and the shock ($z \sim
    8.5$). The impact of this offset is readily apparent on the right
    bottom plot, because of the reduced range of the $y$-axis.}
\end{figure}

The Sod shock tube test consists of a one-dimensional Riemann problem,
in which a discontinuity (between two uniform states) is set up in the
initial conditions.  This standard hydrodynamical test has an analytic
solution.  Our domain is $0\leq z \leq 10$, and the boundary between
the two states ($-$ \& $+$) is at $z=5$.  We set $\rho^-=P^-=1$, while
the right state has $\rho^+=0.125$ and $P^+=0.1$.  The initial
velocity is set everywhere to $0$.  We use an adiabatic equation of
state with a specific heats ratio $\gamma=1.4$.  Figure
(\ref{fig:sod1d}) shows the result of this test, compared to the
analytical solution.  For this test we use 300 cells and van Leer's
slope limiter.  Despite a small deviation in the specific energy
value, the results are in good agreement with the exact solution.

\subsubsection{Viscous spread of a gaseous ring}
\label{sec:gas-ring-diffusion}
One test to verify the validity of the viscous module is the gas ring
diffusion problem \citep[see][]{Speith.Riffert.1999}.  An axisymmetric
thin disk around a central star experiences a radial drift as a
consequence of viscous stresses.  Integrating the hydrodynamical
equations in $Z$ and neglecting pressure forces, the evolution for the
density is governed by:
\begin{equation}
  \frac{\partial \Sigma}{\partial t} = \frac{3}{r} \frac{\partial}{\partial r}\left[\sqrt{r}\frac{\partial}{\partial r}\left(\nu\sqrt{r}\Sigma\right) \right].
\end{equation}
Under this assumption, the azimuthal velocity is Keplerian, while the
radial velocity is given by:
\begin{equation}
  r\Sigma v_r = -3\sqrt{R}\frac{\partial}{\partial r}\left(\nu \sqrt{r} \Sigma \right).
\end{equation}
It can be proven \citep{1974MNRAS.168..603L} that if the initial
condition is that of an infinitely narrow axisymmetric ring of mass
$M$ and radius $R_0$:
\begin{equation}
  \Sigma(r,t=0) = \frac{M}{2\pi R_0}\delta(r-R_0),
\end{equation}
and if the viscosity is a constant $\nu_0$, then the surface density
evolves as:
\begin{equation}
  \Sigma(r,t) = \frac{M}{\pi R_0^2}\frac{1}{\tau
    u^{1/4}}I_n\left(\frac{2u}{\tau}
  \right)\exp{\left(-\frac{1+u^2}{\tau}\right)},
\label{gasring_dens}
\end{equation}
where $I_n$ is the modified Bessel function of the first kind, and the
radial velocity evolves as:
\begin{equation}
  v_r(r,t) =
  \frac{6\nu}{\tau}\left[r-\frac{I_{n-1}(2u/\tau)}{I_{n}(2u/\tau)}\right],
\label{gasring_vr}
\end{equation}
where $u=r/R_0$ and $\tau = 12\nu_0t/R_0^2$.

\begin{figure}
  \centering
  \includegraphics[width=\columnwidth]{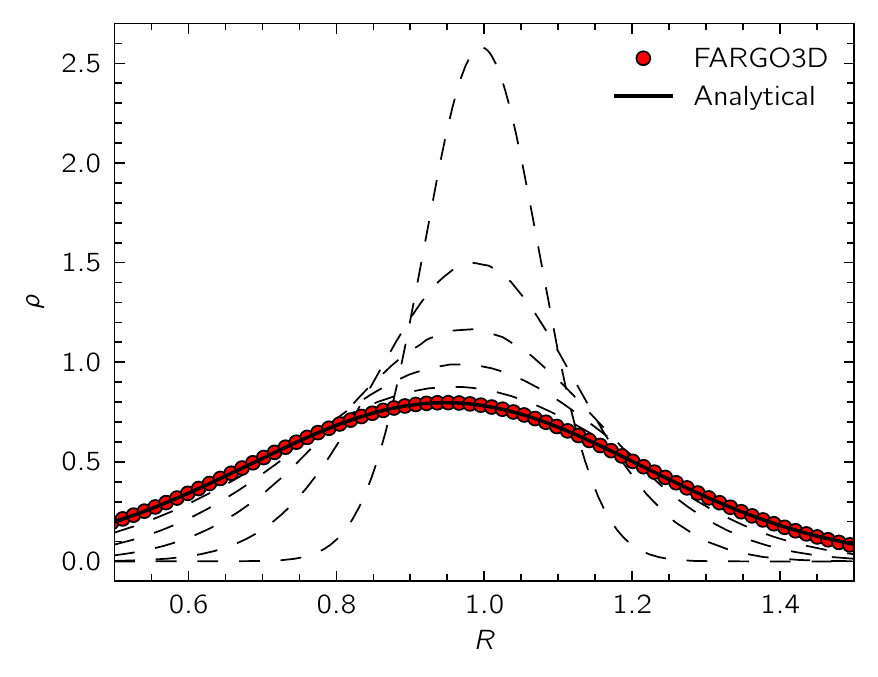}
  \caption{\label{fig:gasring} Viscous ring test. Dashed lines
    correspond to the numerical results for times
    $t=100,300,500,700,900$.  The solid line corresponds to the
    analytical solution at time $t=1100$, while the red points show
    the numerical results at the same time.}
\end{figure}

In order to test the viscous module, we initialize an axisymmetric
Keplerian ring, with an initial surface density profile and radial
velocity given respectively by Eqs.~\eqref{gasring_dens}
and~\eqref{gasring_vr}, evaluated at $t=100$.  Our constant viscosity
is $\nu_0=10^{-5}$.  The radial extent is $0.1\leq r \leq 1.6$, and we
use 512 cells evenly spaced, our mesh being cylindrical.  Boundaries
are set to the zero gradient condition for each primitive variable.
In figure~\ref{fig:gasring} we show the density at different
times. There is a good agreement between the results and analytic
expectations.

\subsection{Magnetohydrodynamics tests}
\label{sec:magn-tests}
\subsubsection{MHD Riemann problem}
\label{sec:mhd-riemann-problem}
\begin{figure}
  \centering
  \includegraphics[width=1.0\columnwidth]{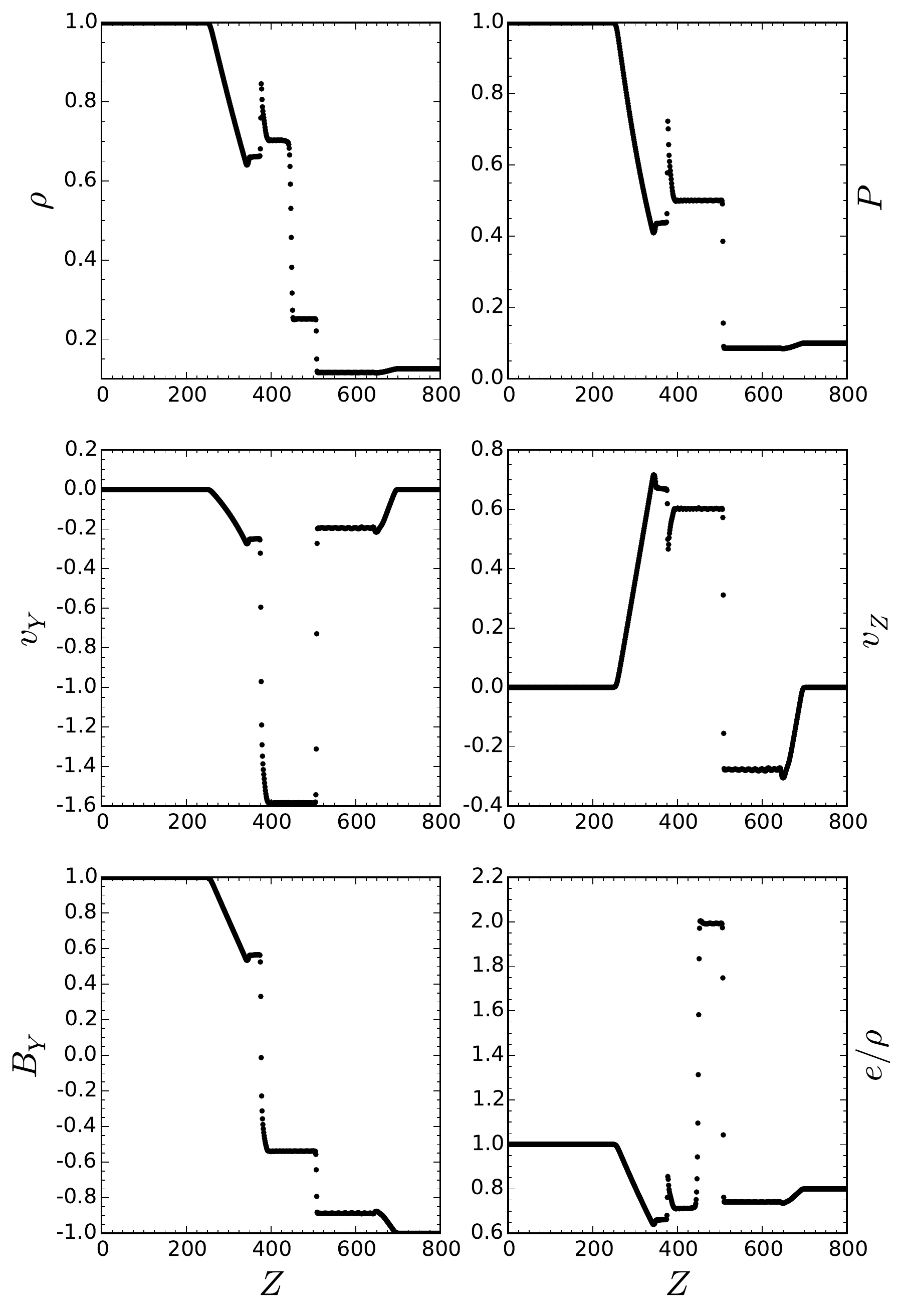}
  \caption{\label{fig:briowu} MHD Riemann problem, also known as the
    Brio \& Wu test. For the initial condition see the text.These
    results are to be compared with Fig. 6 of
    \citet{Stone.Norman.1992.b} and Fig 2. of \citet{Brio.Wu.1988}.}
\end{figure}

The MHD Riemann problem, also known as the \citet{Brio.Wu.1988} test,
is a standard Sod shock tube test where the effects of a jump in the
transverse magnetic field are considered.  We perform this 1.5D test
along the $z$ direction, using the same initial conditions and
resolution as described by \citet{Stone.Norman.1992.b}.  The test
consists in two initially uniform states ($-$ \& $+$), separated by a
discontinuity.  The domain is $0 \leq z \leq 800$, where the
discontinuity is located at $z=400$. At $t=0$ we initialize $P^- = 1$,
$P^+=0.1$, $\rho^-=1.0$, $\rho^+=0.1$, $B_z^{\pm}=0.75$ while
$B^-_y=1.0$ and $B_y^+=-1.0$.  We use a ratio of specific heats
$\gamma=2$ and $800$ cells equally spaced over the domain.  We use
reflecting boundary conditions.  Figure~\ref{fig:briowu} shows
different quantities at $t=80$.

Despite small oscillations in the velocity around $500 \leq z\leq
650$, we observe an overall good behavior for this test.

\subsubsection{Current Sheet Diffusion}
\label{sec:curr-sheet-diff}
In order to test our implementation of the physical resistivity, we
perform the following simple 1.5D test, in which a current sheet
diffuses by Ohmic resistivity.  The problem is formally defined as:
\begin{equation}
B_y(z) = \left\{
\begin{array}{cl}
  \displaystyle{B_0} & z > 0 \\
  -B_0 & z < 0, 
\end{array}\right. \nonumber 
\end{equation}
and $B_z, v_y, v_z$ vanish over the whole domain.  The resistivity is
a constant $\eta_0$.  We integrate the magnetic field passively, so
that we discard the effect of the Lorentz force on the fluid
evolution.  We initially set a uniform pressure $P=1$, hence the
velocity vanishes at all times.  Since the velocity is zero, the
induction equation reduces to a diffusion equation, and it can easily
be shown that the solution is given by:
\begin{equation}
\label{eq:currdiff}
B_y(z,t) = B_0 \text{erf}\left(\frac{x}{\sqrt{4\eta t}}\right) .
\end{equation}
Our domain is $-1\leq z \leq 1$ and we use $512$ cells evenly
spaced. We have $B_0=1$ and $\eta=0.25$.  In order to have a smooth
behavior of the magnetic field, we use the profile given by
Eq.~\eqref{eq:currdiff} at $t_0=(5\Delta z/2)^2$ as initial condition
for $B_y$, where $\Delta z$ is our the Z-size of a cell. This amounts
to smoothing the current sheet over $\sim 10$ cells.

\begin{figure}
  \centering
  \includegraphics[width=\columnwidth]{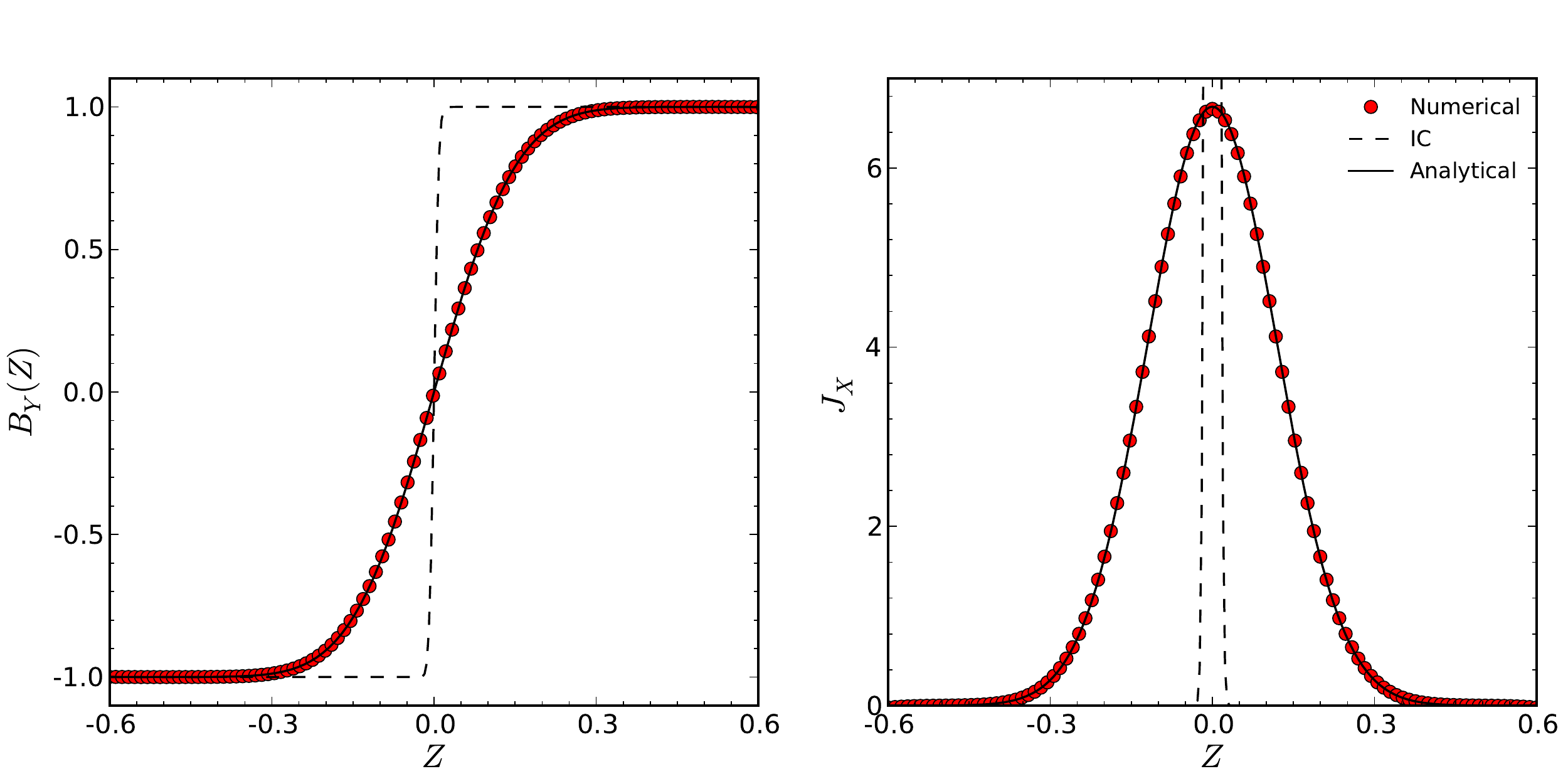}
  \caption{\label{fig:curdiff} Current sheet diffusion test.  The left
    panel shows the evolution of the magnetic field while the right
    panel shows the current.  Dashed lines correspond to the
    initial conditions.  Red points are the numerical solution at time
    $t=30t_0$. The expected analytic solution, given by
    eq.~\eqref{eq:currdiff}, is shown with a solid line.}
\end{figure}

We plot the results in Figure~\ref{fig:curdiff}, which shows a good
agreement between analytics and numerics.

\subsubsection{MHD Rotor}
\label{sec:mhd-rotor}
We perform the test suggested by \citet{Balsara.1999}, known as the
MHD rotor test.  It was designed for testing the propagation of strong
torsional Alfv\'en's waves, which have implications in star formation
problems.  This test consists in having a dense, rapidly spinning
cylinder in a light background fluid. The MHD Rotor test, by its
geometry, is an excellent test to validate the implementation of the
MHD solver in cylindrical coordinates.  We have performed this test
both in Cartesian and cylindrical coordinates.

In Cartesian coordinates, the test is defined in the $YZ$ plane, with
an initial state given by:
\begin{equation}
\rho(y,z) = \left\{
\begin{array}{cl}
  1  & r < r_0 \\
  10 & r_0 \leq r \leq r_1 \\
  1+9f(r) & r_1 < r < r_2 \\
  1 & r \geq r_2 \\
\end{array}\right. \nonumber
\end{equation}
\begin{equation}
v_y(y,z) = \left\{
\begin{array}{cl}
  0 & r < r_0\\
  -v_0\left(z-0.5\right)/r_1 & r_0 \leq r \leq r_1 \\
  -f(r)v_0\left(z-0.5\right)/r_1 & r_1 < r < r_2 \\
  0 & r \geq r_2\\
\end{array}\right. \nonumber
\end{equation}
\begin{equation}
v_z(y,z) = \left\{
\begin{array}{cl}
  0 & r < r_0\\
 v_0\left(y-0.5\right)/r_1 & r \leq r_1 \\
 f(r)v_0\left(y-0.5\right)/r_1 & r_1 < r < r_2 \\
 0 & r \geq r_2\\
\end{array}\right.,
\nonumber
\end{equation}
where $r=\sqrt{(y-1/2)^2+(z-1/2)^2}$ and the smoothing function $f$ is
defined as $f(r)=(r_2-r)/(r_2-r_1)$.  We set $r_0 = 0.05$, $r_1=0.1$,
$r_2=0.115$, $v_0=2$, $\gamma=1.4$, a uniform pressure $P=1.0$, and an
initially uniform magnetic field with $B_y=5/\sqrt{4\pi}$ and $B_z=0$.
We use $512^2$ cells uniformly distributed over a domain covering from
$Y\in [-0.5,0.5]$ and $Z\in [-0.5,0.5]$.

For the cylindrical version of this test, the initial condition we
adopt is the same for the density, we set $v_r=0$ everywhere and $v_\phi$
is given by:
\begin{equation}
v_\phi(r) = \left\{
\begin{array}{cl}
  0 & r < r_0\\
 v_0r/r_1 & r \leq r_1 \\
 f(r)v_0r/r_1 & r_1 < r < r_2 \\
 0 & r \geq r_2\\
\end{array}\right.,
\nonumber
\end{equation}
We use a grid with ($n_r,n_\phi$)=(256,1024), the cells being
uniformly distributed over a domain $r\in[0.01,0.5]$,
$\phi\in[0,2\pi]$. Boundaries condition are reflective in both
implementations and an adiabatic equation of state is used.

\begin{figure*}
  \centering
  \includegraphics[width=\textwidth]{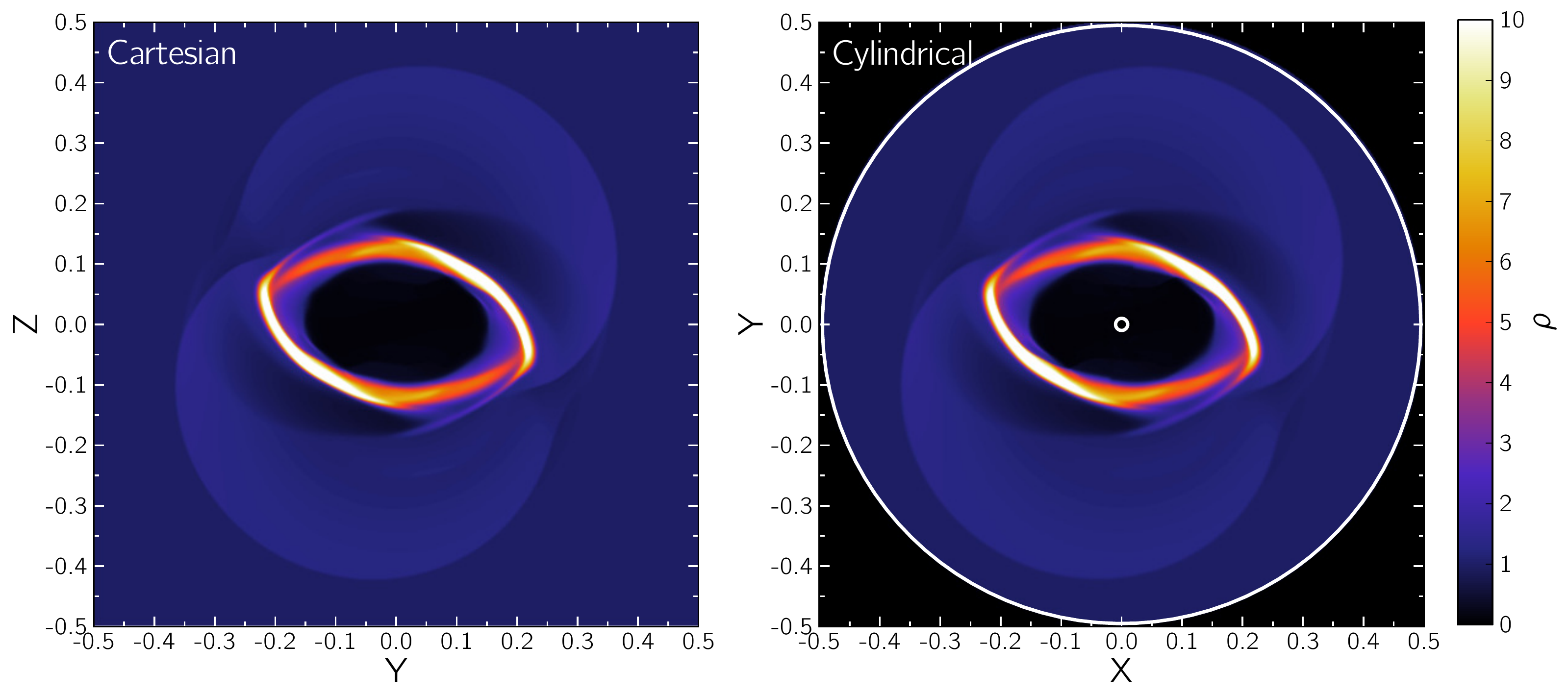}
  \caption{\label{fig:mhdrotor} MHD ROTOR test results at $t=0.15$. In
    the left panel the Cartesian version of the test is shown. In the
    right panel, the cylindrical version is plotted, showing with
    white lines the boundaries of the cylindrical mesh. As long as the
    torsional waves launched by the central rotor do not reach the
    domain boundaries, as it is the case here, the agreement between
    both panels is remarkable.}
\end{figure*}
  
Results are shown in Figure~\ref{fig:mhdrotor}. This figure can be
compared to Fig.~2 of \citet{Balsara.1999}. Note that we have
introduced a difference with respect to the standard MHD Rotor test:
we preferred to put a hole in the inner region of the cylinder to
alleviate possible issues arising from the inner boundary in the
cylindrical setup\footnote{Note that the results presented in this
  paper do not correspond to the initial conditions advertised by
  \citet{1999JCoPh.153..671B}.}.  Our results can be closely compared
with the results displayed on the webpage of the FLASH code, which
have the exact same conditions as ours, except that they consider a
whole cylinder instead one with a hole in the center
(\url{http://flash.uchicago.edu}, manual page of version~4.2,
chapter~VII, section 25.2.5).

\subsubsection{Orzag-Tang Vortex}
\label{sec:orzag-tang-vortex}
We perform the well known \citet{Orszag.Tang.1979} magnetic vortex
test. This test has become a standard test for MHD codes, and there is
a lot of comparison material for it.

\begin{figure*}
  \centering
  \includegraphics[width=\textwidth]{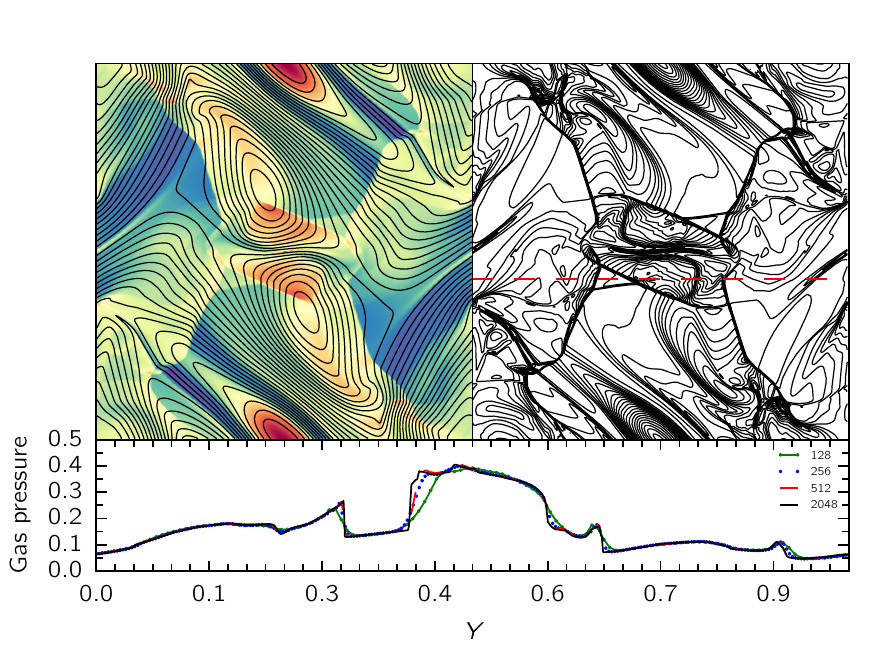}
  \caption{\label{fig:otvortex} Orzag-Tang vortex test at t=0.5.  The
    upper left panel shows in gray scale the density map, with the
    magnetic field lines superimposed.  The upper right panel shows
    the gas pressure contours.  In the bottom panel we show a cut
    along the dashed line shown in the right panel (y=0.4277).  This
    cut is common in literature, and can be used for comparison
    purposes.}
\end{figure*}

The problem is defined on the domain $0 \leq y \leq 1$, $0 \leq z \leq
1$, and the initial conditions are given by:
\begin{eqnarray}
  v_y(z) &=& -\sin{\left(2\pi z\right)} \nonumber \\
  v_z(z) &=& \sin{\left(2\pi y\right)} \nonumber \\
  B_y    &=& -B_0 \sin{\left(2\pi z\right)} \\
  B_z    &=& B_0 \sin{\left(4\pi y\right)} \nonumber \\
  P     &=& 5/(12\pi), \nonumber
\end{eqnarray}
with $B_0=(4\pi)^{-1/2}$, the density and pressure being initially
uniform with values respectively $25/(36\pi)$ and $5/(12\pi)$.  We use
a ratio of specific heats $\gamma=5/3$. The figure \ref{fig:otvortex}
can be directly compared with the figures 10 and 11 of
\citet{Londrillo.DelZanna.2000}.

\subsubsection{MHD Loop}
\label{sec:mhd-loop}
\begin{figure}
  \centering
  \includegraphics[width=\columnwidth]{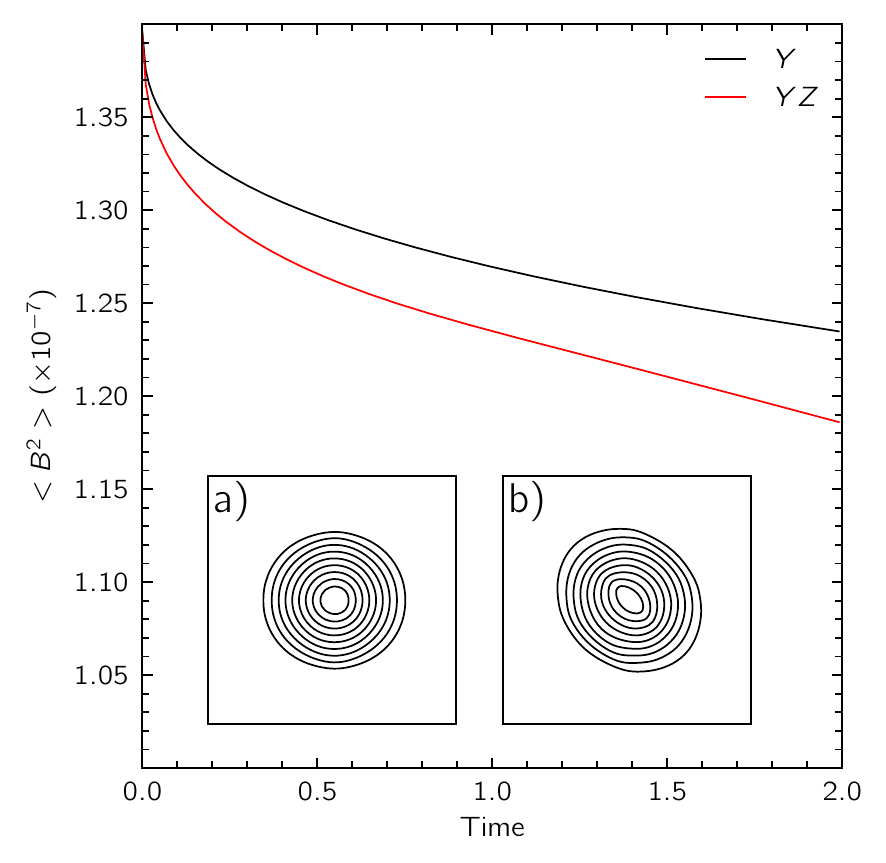}
  \caption{\label{fig:mhdloop} The main panel shows the time evolution
    of the mean magnetic energy in the magnetic loop advection test.
    We show in the inset panels the shape of the loops after a first
    travel over the mesh, for the horizontal advection case in panel
    \emph{a} and for the diagonal advection case in panel \emph{b}.}
\end{figure}

This two dimensional test is based on the advection of a magnetic
closed loop.  It consists in a weak magnetic loop located in a uniform
density medium with a uniform velocity field.  The test is performed
in the limit of $\beta \gg 1$, where the magnetic pressure is
negligible compared with thermal pressure.  The initial conditions are
similar to those of \citet{Gardiner.Stone.2005} and
\citet{Fromang.Hennebelle.Teyssier.2006}.  In order to have initially
a divergence free magnetic field, we initialize the loop using the
potential vector $\vec{A}$:
\begin{equation}
A_x(r) = \left\{
\begin{array}{cl}
  A_0\left(R-r\right) & \mbox{if } r<R \\
  0 & \mbox{otherwise }
\end{array}\right. \nonumber
\end{equation}
with $A_0=10^{-3}$, $R=0.3$ and $r=\sqrt{y^2+z^2}$.  We use a
Cartesian mesh over the domain $-1 \leq y \leq 1$, $-0.5 \leq z \leq
0.5$.  We use 128 cells in $y$ and 64 in $z$. Periodic boundary
conditions are used in $y$ and $z$.  We perform two simulations for
$v_y = \sqrt(5)/2$, $v_z=0$ and $v_y =1.0$, $v_z=0.5$, corresponding
respectively to horizontal and diagonal advection.  We show in
Figure~\ref{fig:mhdloop} the evolution of the mean magnetic energy of
the mesh and the shape of the loop upon an advection over one full
spatial period for the two cases.  The decay of magnetic energy for
the horizontal advection case should be compared to the curves given
\citet{Fromang.Hennebelle.Teyssier.2006} in their Fig.~5.
As these authors have twice a larger advection speed than ours, one
should compare the magnetic energy at $t=2$ in our case ($E_m\approx
1.235\cdot 10^{-7}$) to theirs at $t=1$ (respectively $E_m\approx
1.225\cdot 10^{-7} $ and $E_m\approx 1.19\cdot 10^{-7}$ for the Roe
Riemann solver and for the Lax-Friedrichs solver, so that on this
specific problem the decay of magnetic energy that we get is similar
to the one obtained by these authors with the Roe Riemann solver).

\subsubsection{Current Sheet}
\label{sec:current-sheet}
In order to assess the numerical diffusivity of the MHD solver, we
perform the current sheet test. It consists in a two dimensional
problem in which discontinuities of magnetic field create current
sheets.  The details of the test have been designed by
\citet{Gardiner.Stone.2005}, and this test has also been performed by
\citet{Fromang.Hennebelle.Teyssier.2006}.  We do not include physical
resistivity in this test, so that the evolution of the sheet is
exclusively governed by numerical resistivity.

\begin{figure*}
  \centering
  \includegraphics[width=\textwidth]{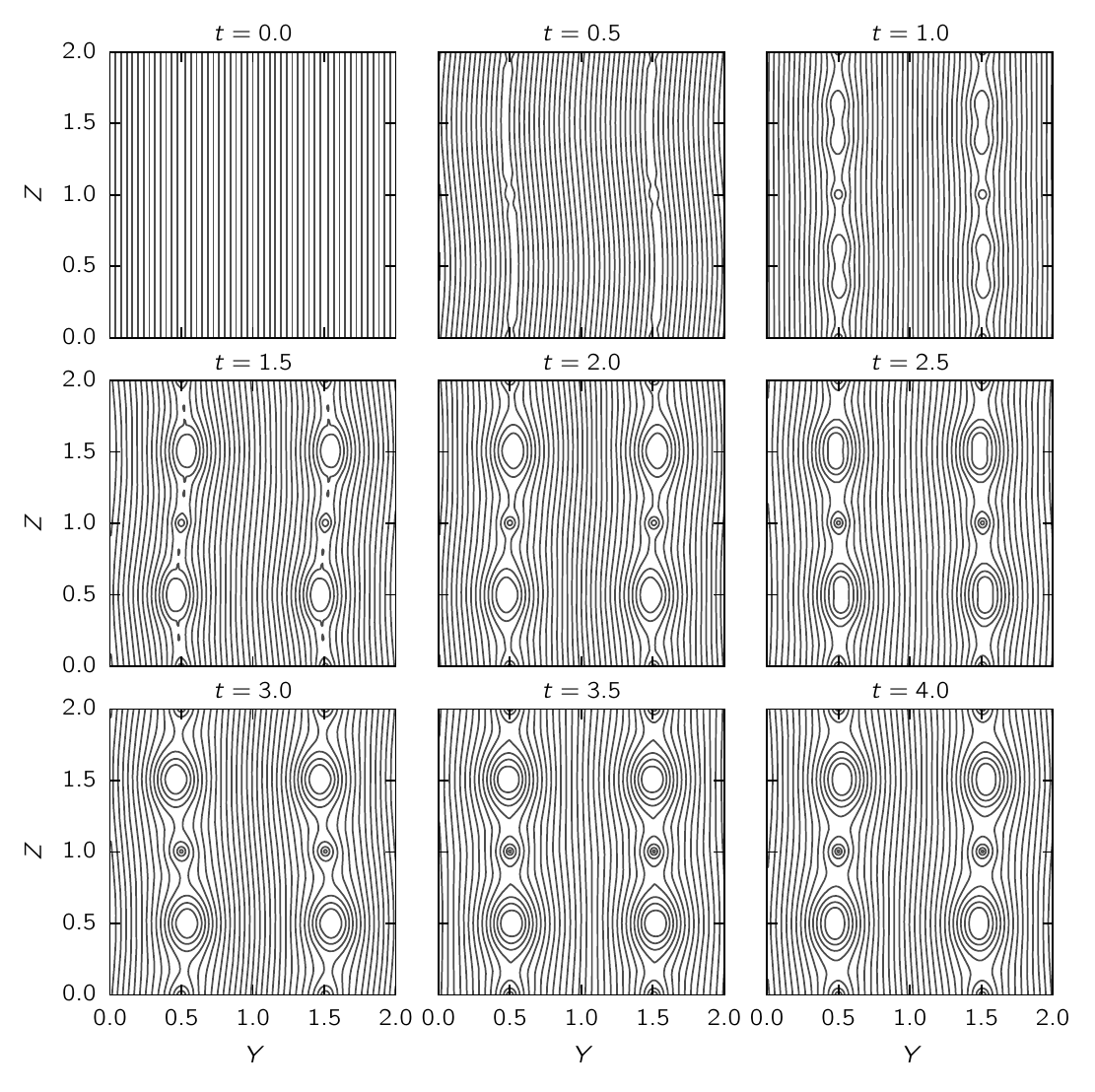}
  \caption{\label{fig:cursheet} Time evolution of the magnetic field
    lines in the current sheet test of
    \citet{Gardiner.Stone.2005}. The box has $256^2$ cells. The times
    of the snapshots are shown on the top of each panel.}
\end{figure*}

We use the exact same setup as \citet{Gardiner.Stone.2005} and
\citet{Fromang.Hennebelle.Teyssier.2006}: our computational domain is
$0\leq y \leq 2$, $0\leq z\leq 2$ and the resolution is $256$ in each
direction.  Initially, the density is $\rho=1$, the pressure $P=0.1$
and the magnetic field components $B_x=0$ and $B_y=0$ everywhere,
while $B_z$ is given by:
\begin{equation}
B_z(y) = \left\{
\begin{array}{cl}
 +1 & \mbox{if } |y-1|<0.5 \\
 -1 & \mbox{otherwise }
\end{array}\right. \nonumber
\end{equation}
The velocity is initialized as $v_z=0$ everywhere and
$v_y=v_0\sin{(\pi z)}$.  We use $v_0=0.1$ and $\gamma=5/3$.  In figure
\ref{fig:cursheet} we show the evolution of the magnetic field lines.
We see a behavior very similar to that of
\citet{Fromang.Hennebelle.Teyssier.2006}.
At larger time ($t > 2.5$),
four islands remain along the location of each discontinuity, instead
of two in the case of \citet{Gardiner.Stone.2005}.  We also note that
our scheme conserves the central symmetry with respect to the velocity
nodes, as does the RAMSES code on this test
\citep{Fromang.Hennebelle.Teyssier.2006}.

\section{Discussion}
The tests presented in the previous section allow us to validate the
implementation of the different part of the core algorithm of the
FARGO3D code, in different geometries. We now turn to a discussion
about some aspects of the code. Namely, in a first part, we assess the
impact of orbital advection on the code properties for two setups
which are of high interest for the scope of the code. Secondly, we
address briefly the question of single versus double precision
calculations, which is a recurrent question in high performance
computing on GPUs.

\label{sec:discussion}
\subsection{Impact of orbital advection on the code properties}
\subsubsection{Vortex in a two-dimensional Keplerian disk}
\label{sec:vort-two-dimens}
We reproduce here the test described in section~3.1 of
\citet{Mignone.2012}.  Our setup and Courant number are exactly the
same as those used by \citet{Mignone.2012} and our runs are performed
for the same three resolutions as in that work. The setup consists of
a two-dimensional disk orbiting a central potential $\Phi=-1/r$, with
uniform surface density $\Sigma\equiv 1$ and pressure
$p\equiv 1/(\gamma{\cal M})$, ${\cal M}$ being the Mach number of the
Keplerian flow at $r_0=1$, chosen to be ${\cal M}=10$. The unperturbed
flow is exactly Keplerian, since there is no pressure
gradient. Superimposed on this flow, we impose a perturbation
$(\delta v_\phi, \delta v_r)$ of velocity that has the form:
\begin{equation*}
\begin{pmatrix} \delta v_\phi\\\delta
  v_r\end{pmatrix}=K\exp\left(-\frac{x^2+y^2}{h^2}\right)
\begin{pmatrix}
-\sin\phi&\cos\phi\\
\cos\phi&\sin\phi\\
\end{pmatrix}
\begin{pmatrix}
  -y\\x
\end{pmatrix},
\end{equation*}
where $h=c_s(r_0)/2\Omega(r_0) $ is the vortex size,
$K=-1$ is the vortex amplitude,
$x=r\cos\varphi-r_0\cos\varphi_0$
and $y=r\sin\varphi-r_0\sin\varphi_0$
are the Cartesian coordinates measured in the frame centered on the
vortex, which is located initially at $r_0$
and $\varphi_0=\pi/4$.
\begin{figure}
  \centering
  \includegraphics[width=\columnwidth]{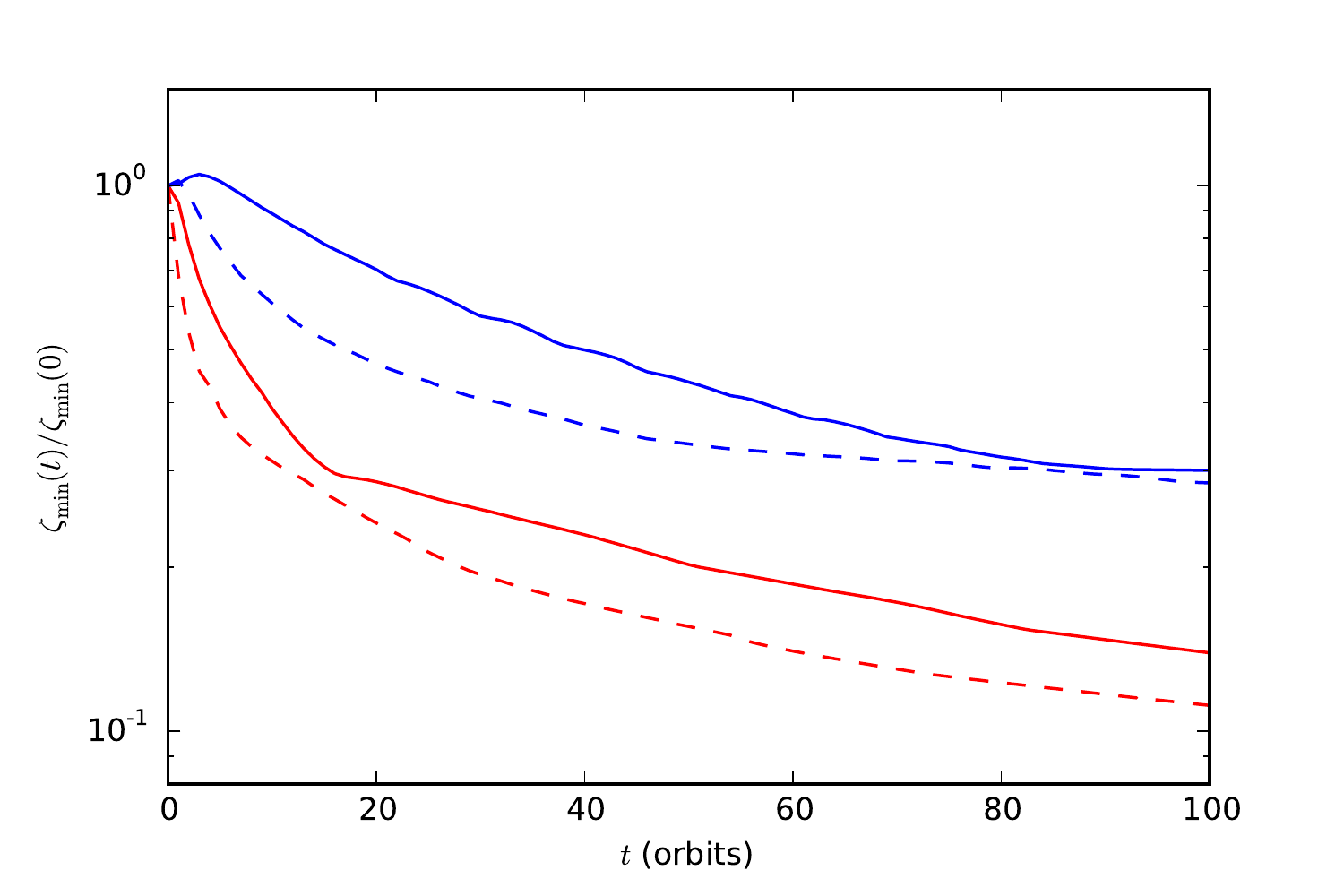}
  \includegraphics[width=\columnwidth]{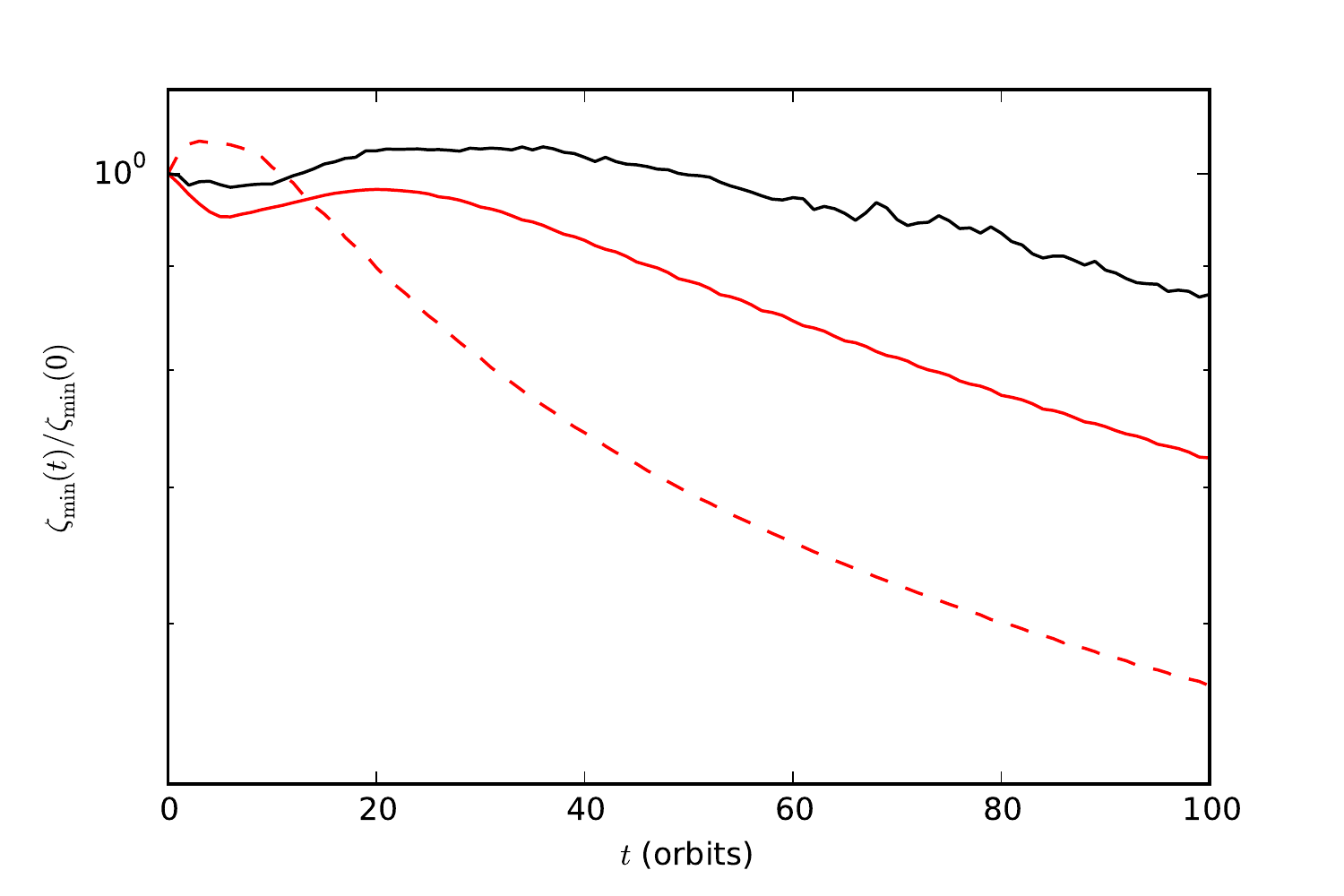}
  \caption{\label{fig:vortex12} Left: decay of vortensity minimum as a
    function of time for resolution $1024\times 256$ (red) and
    $2048\times 512$ (blue).  Solid lines show the results with
    orbital advection, and dashed lines without it.  Right: results
    for the highest resolution case ($4096\times 1024$).  The red
    curves show results for FARGO3D (with and without orbital
    advection) and the black curve shows the results for PLUTO with
    orbital advection (which we have reproduced using the version 4.0
    of that code)}
\end{figure}
The results in Fig.~\ref{fig:vortex12} show that FARGO3D, like the
PLUTO code, benefits significantly from orbital advection, as this
dramatically improves the conservation of vortensity at all
resolutions.  Qualitatively, without orbital advection, vortensity
conservation is slightly better, at a given resolution, with FARGO3D
than with PLUTO, whereas the opposite holds when orbital advection is
turned on.

\subsubsection{MRI in unstratified disks}
\label{sec:unstr-mri-test}
We present a calculation of the development of the magneto-rotational
instability (MRI) in cylindrical coordinates with no vertical external
force, with a setup similar to that of \citet{2011AA...533A..84B}.
The mesh has a radial extent $1<r<8$, a vertical extent $-0.3<z<+0.3$,
and covers half a disk in azimuth: $0 < \varphi < \pi$.  A locally
isothermal equation of state is used (see eq. \ref{eq:eos1}), in which
we have everywhere $c_s(r)= 0.1v_k(r)$, where $c_s$ is the sound speed
and $v_k(r)=(GM_\star/r)^{1/2}$ the Keplerian speed at radius $r$.
Periodic boundary conditions are used in $z$.  The density follows a
power law of radius: $\rho \propto r^{-1/2}$ (the multiplicative
constant of this law does not matter in an isothermal setup).  The
magnetic field is initially exclusively toroidal, and such the thermal
to magnetic pressure ratio be uniform and equal to $\beta = 50$.  We
introduce some noise on the radial and vertical components of the
velocity, with an amplitude of $5$~\% of the sound speed.  We also
follow a standard practice that consists in introducing resistive
buffers in the inner part ($r=1$ to $r=2$) and outer part ($r=7$ and
$r=8$) of the mesh.  The resistivity scales linearly in these buffers,
from zero at the frontier with the active disk to a maximal value at
the mesh edge.  This maximal value is $10^{-4}$ at the outer edge, and
$2\cdot 10^{-5}$ at the inner edge.  Similarly to
\citet{2011AA...533A..84B}, our mesh is rotating with an angular
velocity $\Omega_f=r_0^{-3/2}$, which sets the corotation at $r_0=3$.
Symmetric boundary conditions are applied in radius for the radial
component of the EMF, whereas antisymmetric conditions are applied on
the azimuthal and vertical components in order to conserve the magnetic
flux to machine accuracy.  We run two calculations, each over
$300$~orbits at the inner edge.  One calculation uses orbital
advection (see section~\ref{subsect:fargo-mhd}), whereas the other
does not.  We refer to the latter as the standard calculation.
\begin{figure}
  \centering
  \includegraphics[width=\columnwidth]{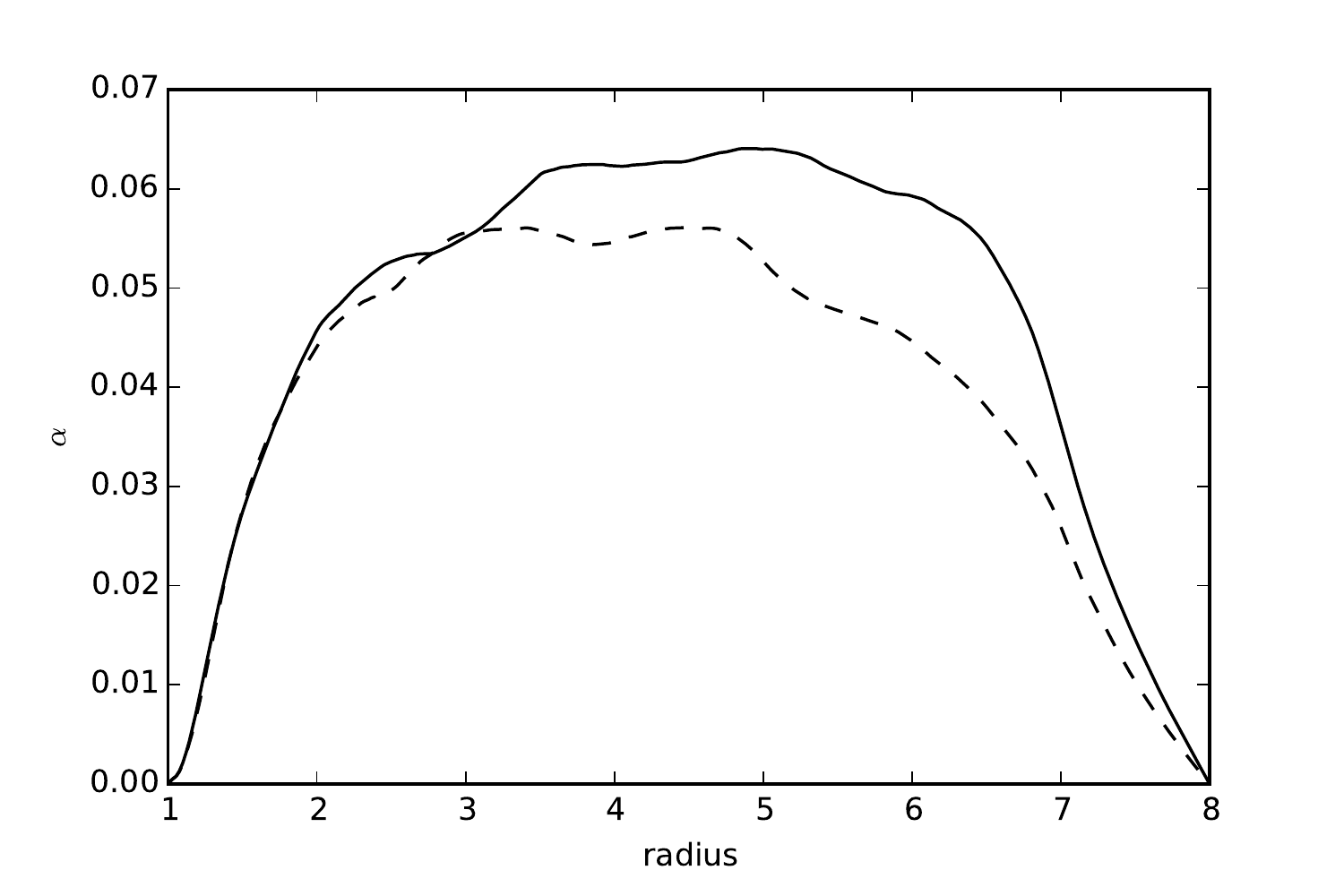}
  \caption{\label{fig:avr} Value of $\alpha$ as a function of radius
    for the FARGO-MHD case (solid line) and standard case (dashed
    line). The drop near the edges is attributable to the resistive
    buffers.}
\end{figure}

Fig.~\ref{fig:avr} shows the value of the $\alpha$ coefficient as a
function of radius, time averaged between 120 and 300 orbits at the
inner edge (over this time frame the turbulence appears to have
reached a saturated state at all radii in both calculations), where
$\alpha$ is defined as:
\begin{equation}
  \label{eq:65}
  \alpha(r,t) = \alpha_\mathrm{Reynolds}(r,t)+\alpha_\mathrm{Maxwell}(r,t),
\end{equation}
with:
\begin{equation}
  \label{eq:66}
  \alpha_\mathrm{Reynolds}(r,t) =\frac{\int\!\int\rho\delta v_r\delta v_\phi d\phi dz}{c_s^2(r)\int\!\int\rho d\phi dz},
\end{equation} 
where $\delta v_r$ and $\delta v_\phi$ are the velocity fluctuations
with respect to the mean flow at a given radius, and:
\begin{equation}
  \label{eq:67}
  \alpha_\mathrm{Maxwell}(r,t) =-\frac{\int\!\int B_r B_\phi d\phi dz}{\mu_0 c_s^2(r)\int\!\int\rho d\phi dz}.
\end{equation} 
We recover the fact that the standard advection scheme is more
diffusive away from corotation \citep{2008ApJS..177..373J} and
consequently that the $\alpha$ coefficient is lower for the standard
calculation than for the calculation with orbital advection, except at
corotation ($r=3$) where their values coincide.
\begin{figure}
  \centering
  \includegraphics[width=\columnwidth]{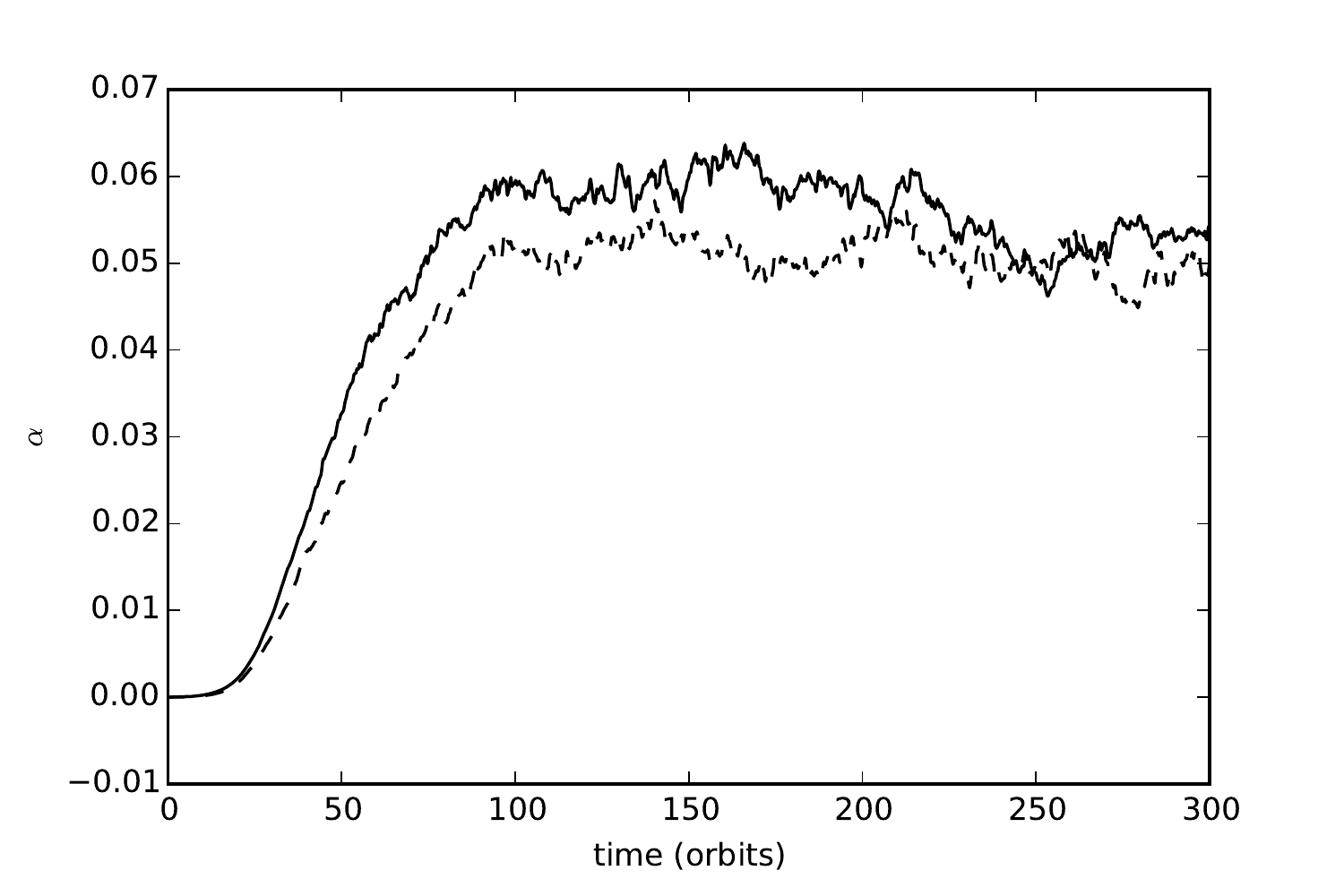}
  \caption{\label{fig:avt} Value of $\alpha$ as a function of time for
    the FARGO-MHD (solid line) and standard (dashed line) cases.}
\end{figure}

Fig.~\ref{fig:avt} shows the temporal behavior of the $\alpha$
coefficient in both calculations, averaged from radius $r=2.1$ to
$r=6.5$.  As anticipated, its value levels off at a slightly smaller
level in the standard case, which also reaches saturation a bit later
than the calculation with orbital advection.  Fig.~\ref{fig:avr}
and~\ref{fig:avt} should be compared to Fig.~6 of
\citet{2011AA...533A..84B}, in particular with the curve displaying
NIRVANA's results, which were obtained for the same initial value of
$\beta$.  Our runs display a significantly larger plateau value of
$\alpha$, but saturation is reached in a longer time.  This might be
due to the fact that no resistive buffers were used in the simulation
with NIRVANA.

\begin{figure}
  \centering
  \includegraphics[width=\columnwidth]{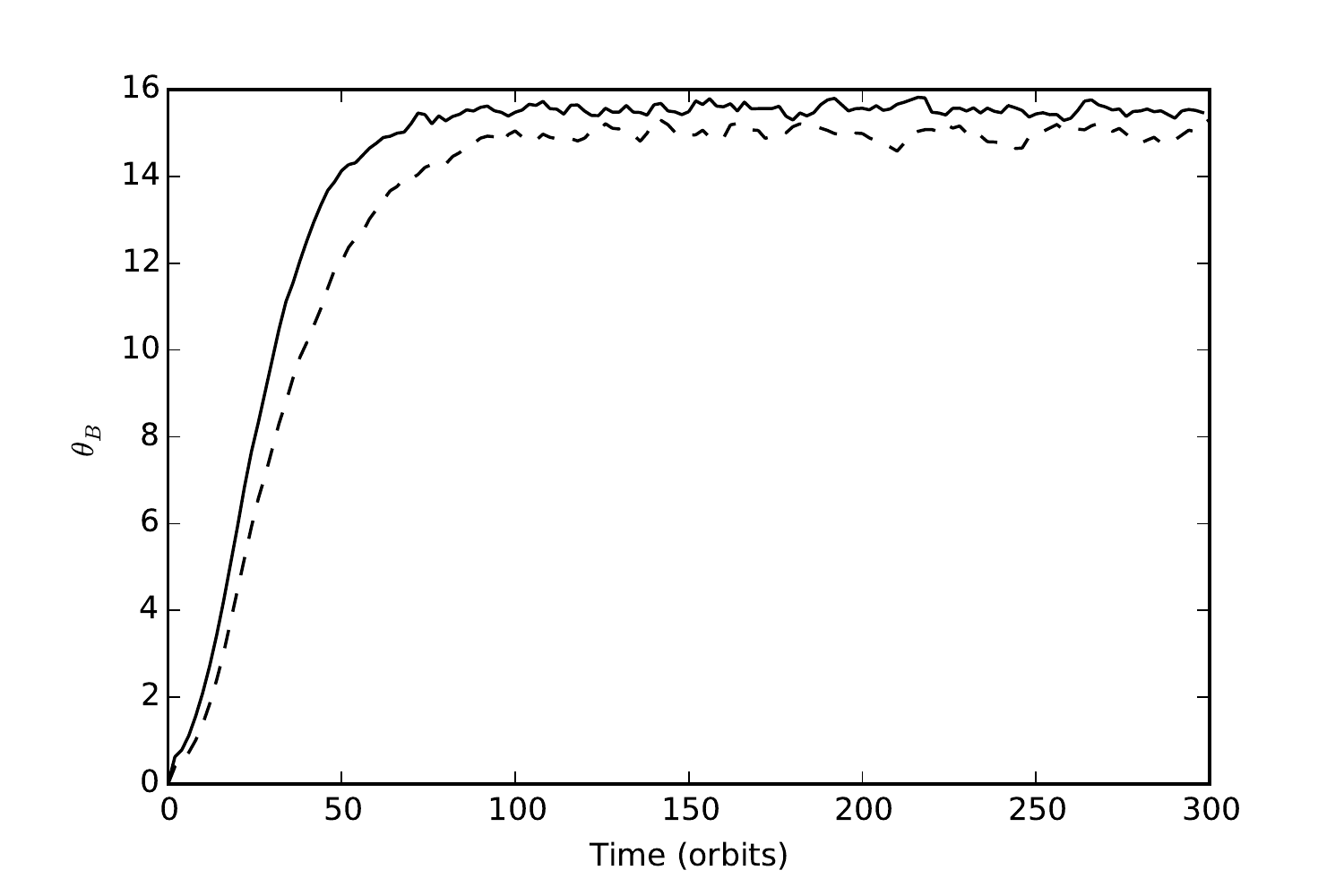}
  \caption{\label{fig:tiltb} Value of the tilt angle of the magnetic
    field (in degrees) as a function of time for the FARGO-MHD (solid
    line) and for the standard (dashed line) cases.}
\end{figure}
Fig.~\ref{fig:tiltb} shows the averaged value $\overline \theta_B$ of
the tilt angle of the magnetic field, defined by
\citet{2012ApJ...749..189S} or \citet{Mignone.2012} as:
\begin{equation}
  \label{eq:68}
  \sin 2\theta_B=\frac{|B_\phi B_r|}{P_{\text{mag}}}.
\end{equation}
It converges towards a value comprised between $15^\circ$ and
$16^\circ$ in the FARGO-MHD case, and towards a marginally lower value
in the standard case.  Again, the leveling off indicative of the
saturation of turbulence is reached later in the standard case.

\begin{figure}
  \centering
  \includegraphics[width=\columnwidth]{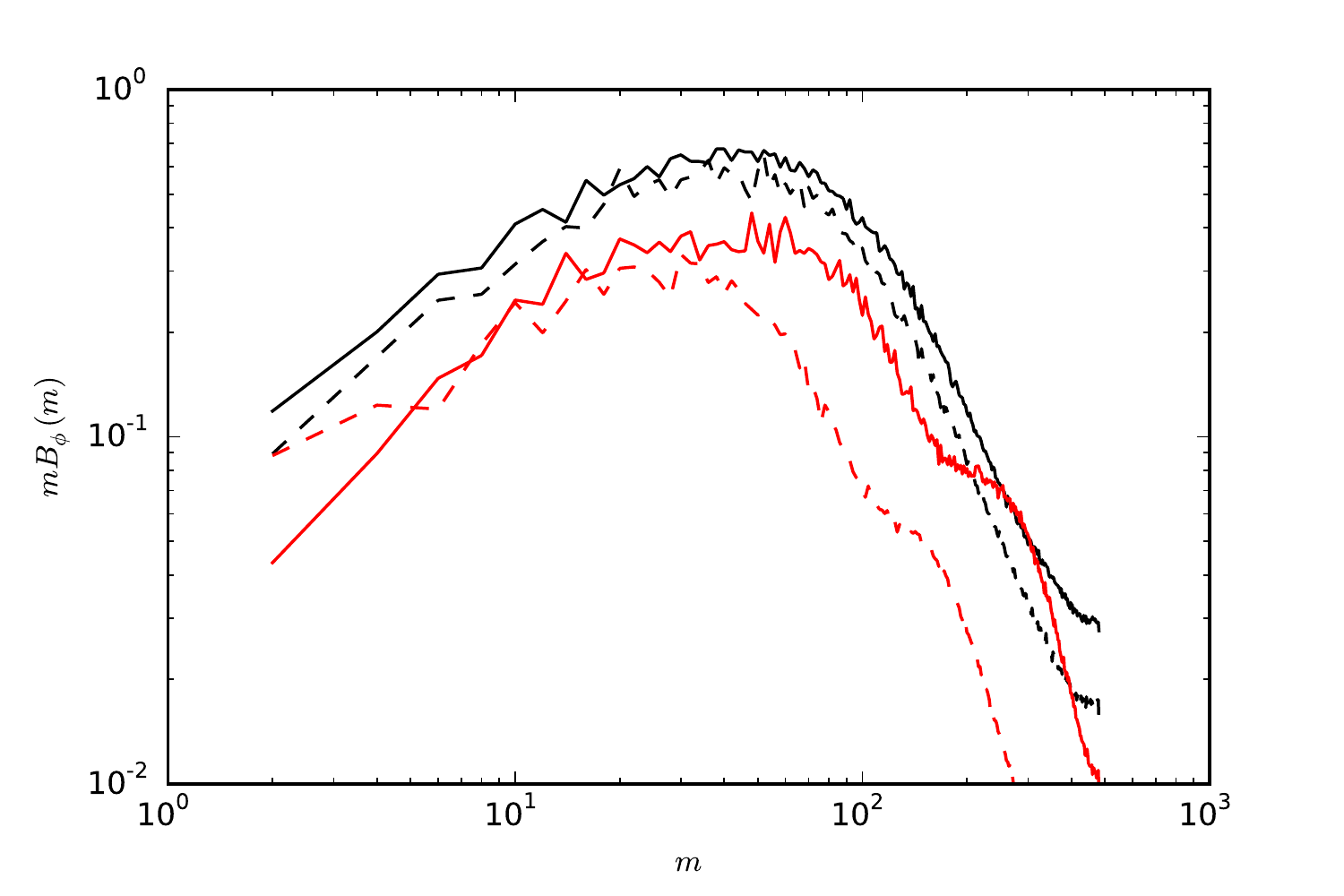}
  \caption{\label{fig:specb} Spectrum of the azimuthal component of the
    magnetic field as a function of the azimuthal wave number. Solid
    lines refer to the FARGO-MHD case, dashed lines to the standard
    case. The black curves correspond to a narrow ring encompassing
    corotation ($r=2.76$ to $r=3.24$) whereas the red curves
    correspond to the outer ring ($r=6.00$ to $r=6.48$). Both spectra
    are obtained at $t=100$~orbits at the inner edge. Note that our
    spectra are defined only for even values of $m$ since our
    azimuthal domain spans only $\pi$ radians.}
\end{figure}

Fig.~\ref{fig:specb} shows the spectrum of the azimuthal component of
the magnetic field, for the standard and orbital advection cases, in
two rings: one sitting above corotation, and one located further out
in the disk. These are obtained as:
\begin{equation}
  \label{eq:69}
  mB_\phi(m)=m\sum_{\gamma=0}^{N_z-1}\sum_{\beta=\beta_\mathrm{min}}^{\beta_\mathrm{max}}\left|
    \sum_{\alpha=0}^{N_\phi-1}
    B_{\alpha\beta\gamma}e^{i\frac{2\pi}{N_\phi}\alpha m}\right|,
\end{equation}
where in the above expression we use $\alpha\beta\gamma$ indices
instead of $ijk$ in order to avoid confusion with $\sqrt{-1}$, and
where $\beta_\mathrm{min}$ and $\beta_\mathrm{max}$ are the indices of
the edges of the ring over which the spectra are calculated.  The
spectra of both calculations are sensibly the same for the first ring
(the orbital advection case showing nevertheless systematically more
signal at higher azimuthal wave numbers), whereas the spectra of the
outer ring show smaller resolved scales in the FARGO-MHD case, roughly
by a factor of~2, as noted by \citet{Mignone.2012}.

\subsection{Single or double precision ?}
While calculations done on CPUs are most of the time done in double
precision, the question of whether a given calculation should be
performed in single or double precision arises frequently in the GPU
HPC community. This, in part, is due to the limited amount of memory
available on GPUs compared to their host counterparts. It is also due
to the fact that the endeavor of coding for GPUs is reserved for very
computationally demanding tasks, and working in single precision
typically means an extra factor of two in computational
throughput. There is no definite answer to that question: the choice
of single versus double precision should be made on a case by case
basis. We hereafter entertain two cases in the context of
astrophysical disks, which we think represent two extreme cases. In
one of them, the use of single precision floating point is obviously
to be discouraged, while for the other case the use of single
precision yields essentially same results as double point precision.

\subsubsection{Two dimensional laminar disk at high resolution}
\begin{figure}
  \centering
  \includegraphics[width=\columnwidth]{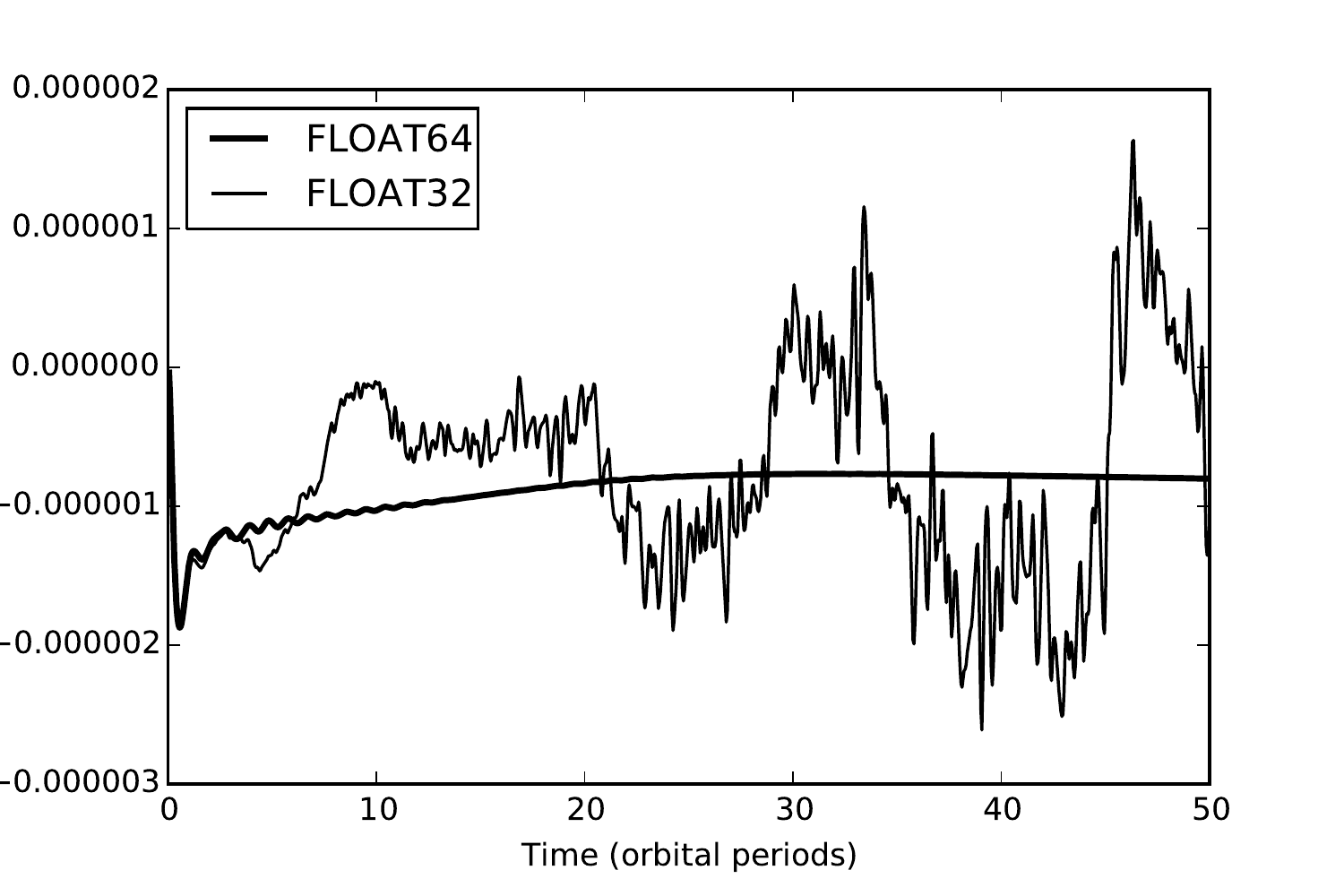}
  \caption{\label{fig:single1} Specific torque exerted by the disk on
    the planet, as a function of the time elapsed after the planet
    insertion in the disk. The nearly horizontal, thick curve
    represents the torque obtained in the double precision
    calculation, while the thin, very time variable curve represents
    the torque in the single precision case. Apart from the floating
    point accuracy, the two setups are strictly identical. The torques
    begin to diverge after $3-4$~orbits.}
\end{figure}

In this case, we consider an Earth-mass planet embedded in a locally
isothermal, inviscid disk with an initially uniform surface density,
and a constant aspect ratio $h=0.05$. The mesh extends from $r=0.6a$
to $1.4a$, where $a$ is the planetary orbital radius. It has
$N_\phi=4000$~cells in azimuth, and $N_r=2000$~cells in radius, which
implies a radial resolution of $\Delta r=4\cdot 10^{-4}a$. The
calculation is performed in the frame corotating with the
planet. Fig~\ref{fig:single1} shows how the torque behaves in the
single and double precision cases.

Some insight into the reasons for the behavior observed in the single
precision case can be gained by evaluating the number of significant
digits over which the vorticity is represented in the single precision
case.  The difference between the single and double precision values
of the azimuthal velocity at $t=0$ shows typical r.m.s. fluctuations
of the single precision value of $\sigma\sim 10^{-7}$.

\begin{figure}
  \centering
  \includegraphics[width=\columnwidth]{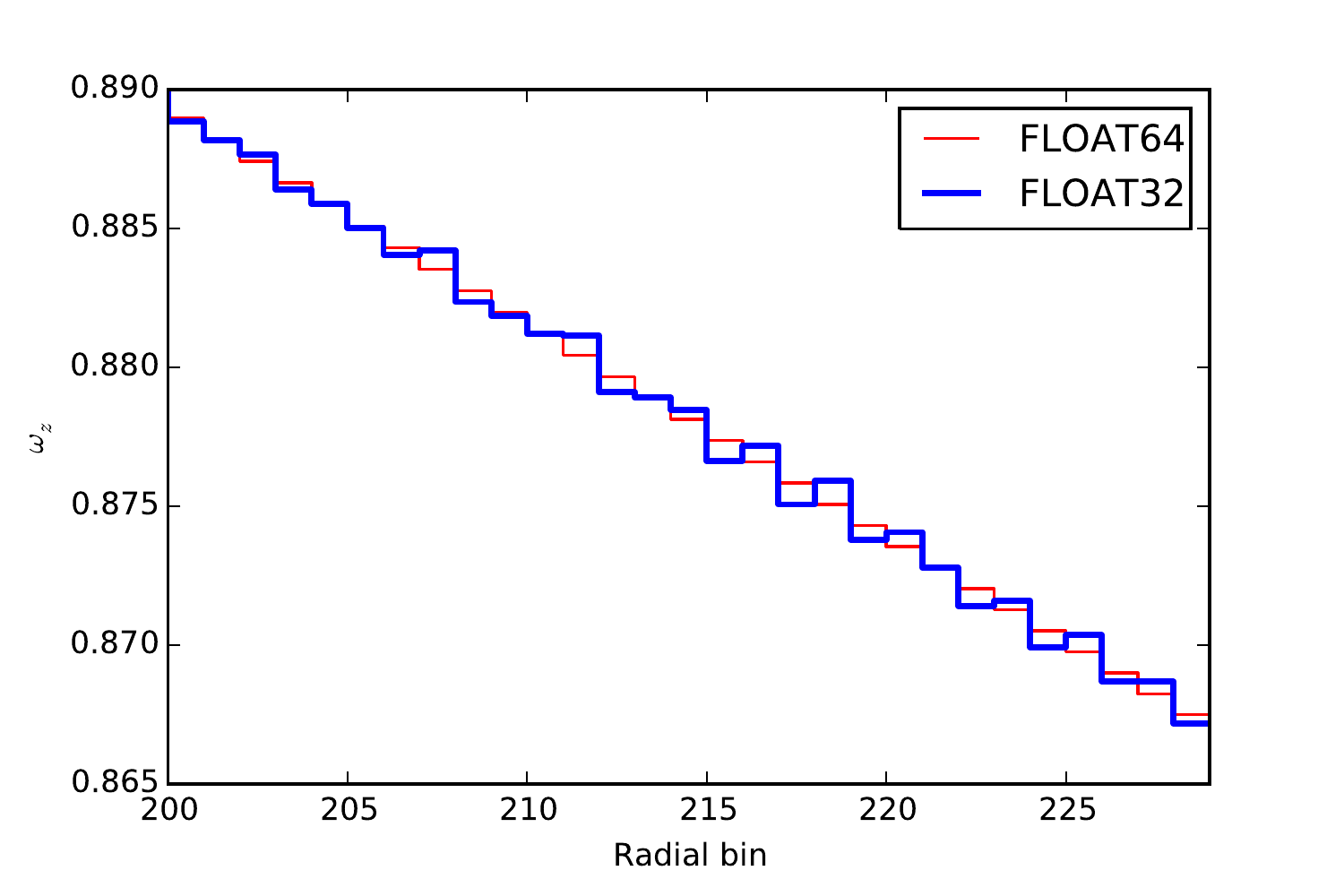}
  \caption{\label{fig:single4} Close up of the vertical component of
    the vorticity as a function of radius, near the inner disk
    edge. The thin red line shows the double precision data, which is
    well behaved (the vorticity decays monotonically), and the thick
    blue line shows the single precision data, which shows many local
    maxima and minima.}
\end{figure}

\begin{figure}
  \centering
  \includegraphics[width=\columnwidth]{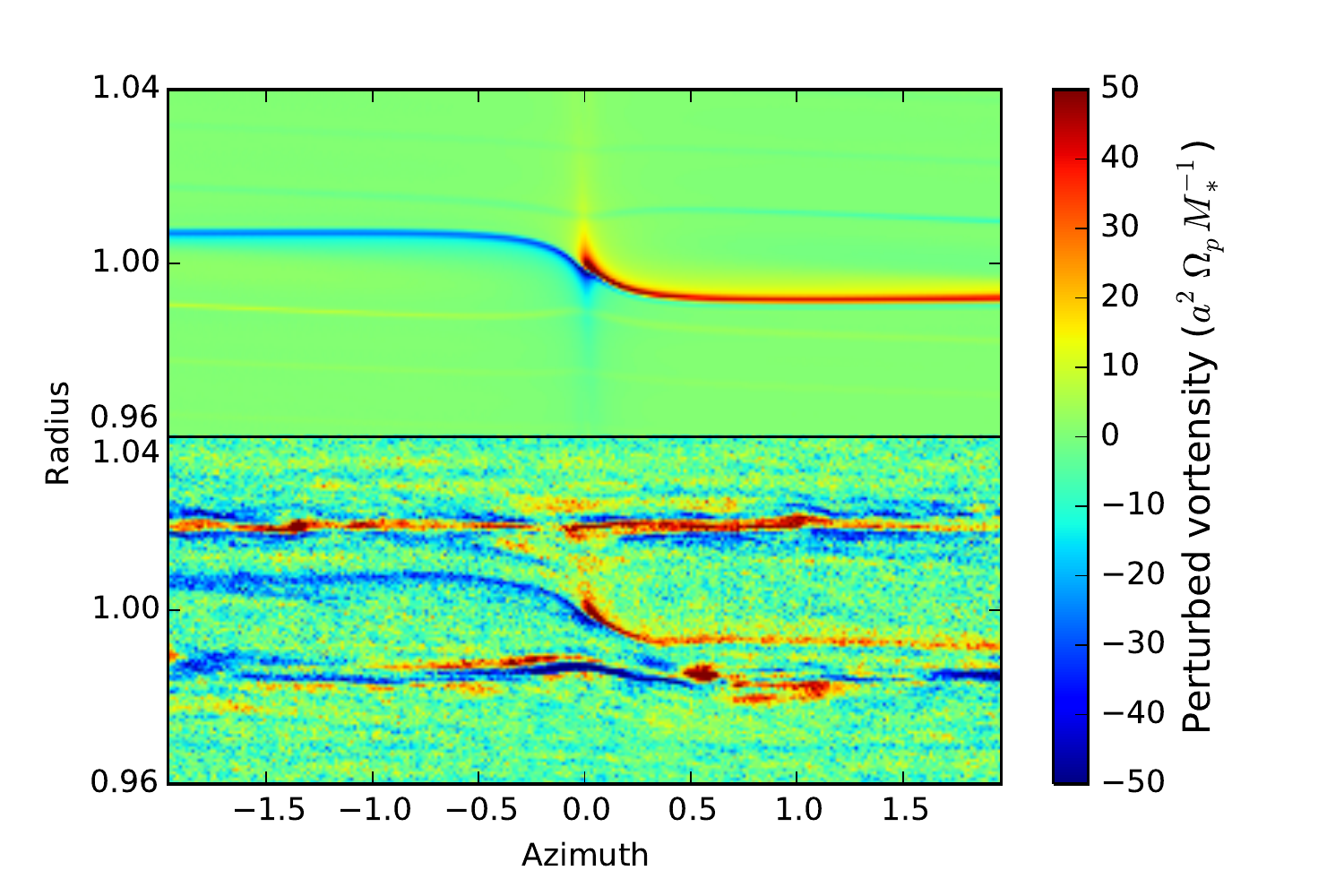}
  \caption{\label{fig:single5} Perturbed vortensity distribution after
    $50$~orbital periods, in the double precision case (top) and
    single precision case (bottom). The blue and red stripes
    originating near the planet are physical, and come from
    vortensity's advection in the planetary horseshoe region on the
    one hand, and the driving of vortensity arising from the radial
    temperature gradient on the other hand \citep{cm09}. The single
    precision case shows significant noise in the data, owing to the
    small number of significant digits over which the vortensity is
    represented, and many vortices absent from the double precision
    case.}
\end{figure}

The (vertical component of the) vorticity is obtained by deriving
radially the azimuthal velocity (appropriately weighted to account for
the cylindrical geometry). Since the velocity and its radial
derivative are $O(1)$, we expect the vorticity, in the single
precision case, to be accurate only to within
$\sim\sqrt{2}\sigma/\Delta r\sim 3.5\cdot 10^{-4}$. This is precisely
what we find on the difference of the single and double precision
values of the vorticity. \emph{This typical error on the vorticity is
  comparable to its variation between successive radial bins} $\Delta
r|\partial_R\omega_z|\sim \Delta r$, since $\partial_R\omega_z=O(1)$.
Consequently, the vorticity is not a monotonous function of radius in
single precision, as we see in Fig.~\ref{fig:single4}, and neither is
the vortensity.
Local extrema of vortensity are unstable to the Rossby
wave instability \citep[RWI,][]{lovelace99,li2000} . We find that the
computational domain is invaded by vortices after a few dynamical
times in the single precision case (see Fig.~\ref{fig:single5}).
The acoustic wakes triggered by these vortices pass in the planet vicinity
where their combined action yield the stochastic behavior of
Fig.~\ref{fig:single1}. Our interpretation is that these vortices are
triggered by the RWI which occurs at many radii in the single
precision case, the breakdown of azimuthal symmetry being provided by
the planet. We note that the problem may be worsened by our numerical
algorithm: at each timestep, a new momentum and new density is
inferred for each cell, the division of which yields the new velocity
(this kind of procedure is commonplace in numerical hydrocodes). This
may induce a random walk of the velocity with respect to the double
precision value. We comment that adopting a significantly smaller
resolution in single precision mitigates the problem exposed here
(albeit this cannot be regarded as a solution or workaround), and that
a similar issue should be expected in double precision calculations at
much larger resolution (typically for $\Delta r\sim
\sqrt{10^{-16}}a=10^{-8}a$, largely beyond the reach of present day
platforms, at least over the time scale over which RWI sets in).

\subsubsection{Magnetorotational Instability in unstratified disks}
\begin{figure}
  \centering
  \includegraphics[width=\columnwidth]{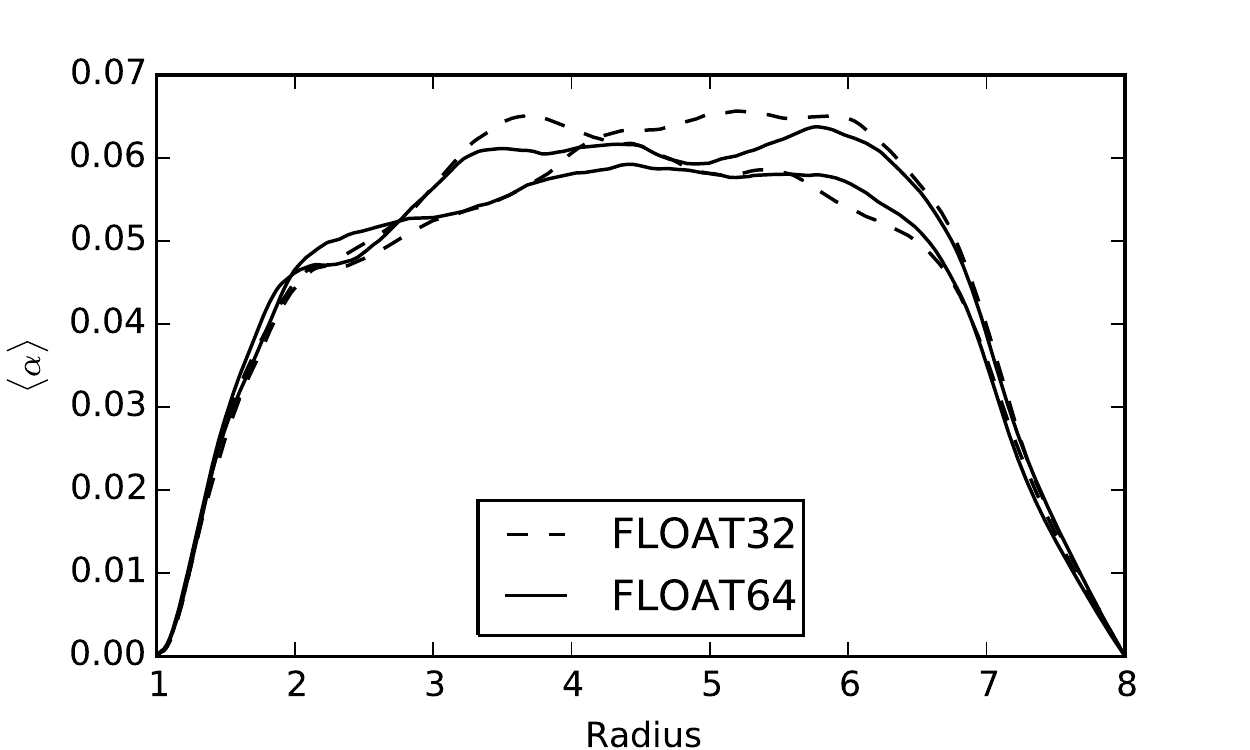}
  \caption{\label{fig:mrisd} Value of $\alpha$ time averaged between
    $100$ and $200$~orbits at the disk inner edge, as a function of
    radius, for double precision calculations (solid lines) and for
    single precision calculations (dashed lines).}
\end{figure}

We consider again the setup of section~\ref{sec:unstr-mri-test}, which
we now run both in single and double precision. We compare in
Fig.~\ref{fig:mrisd} the time averaged value of the $\alpha$
coefficient as a function of radius for the single and double
precision calculations.
In order to get a sense of the variation
expected from one realization to another, we display for each case two
calculations made with different random seeds. We see that the single
precision case yields results are perfectly comparable to those
obtained in double precision. Other diagnostics such as those shown in
section~\ref{sec:unstr-mri-test} yield similar conclusions.

\section{Perspectives}
\label{sec:perspectives}
The development of an astrophysical code for fluid dynamics is
generally an unending process in which new physics as well as minor
improvements are constantly added.  We draw below a list of some of
our recently added features and ongoing projects, not publicly
released yet, which will be presented in full detail elsewhere.
Nonetheless, the features that are presented in this paper are the
basic blocks of the code, and we intend to keep them unchanged for the
foreseeable future.  They constitute the bases that any user
interested in working on the code should know.

Recent work has highlighted the importance of radiative transfer
processes for the physics of protoplanetary disks and their
interactions with forming planets
\citep{2008A&A...478..245P,2008ApJ...679..797J,2014MNRAS.440..683L}.
For this purpose we have implemented a simple radiative transfer
module in FARGO3D, based on a gray approximation and flux limited
diffusion, and a two-temperature (gas and photons) approach
\citep{2013A&A...549A.124B}.  Similarly, for the description of
passive disks, the irradiation of the disk's photosphere by the
stellar light is crucially important
\citep{2000A&A...361L..17D,2001ApJ...560..957D,1998ApJ...500..411D,2014ApJ...790...78F},
and we have implemented a module to account for this effect by ray
tracing.  Ray tracing requires radial integration of the zones optical
depths, which can be tricky on parallel platforms, especially GPUs.
We will present in a forthcoming work our implementation strategy for
the radiative transfer and stellar irradiation modules.

The importance of using very large resolution for computations of
embedded planets is becoming increasingly evident.  Such resolution is
way beyond current computational resources for global disk
calculations with meshes with uniform resolution.  A strategy can be
to have smoothly variable cell sizes in order to achieve small cells
in the planet vicinity \citep{2003MNRAS.341..213B}.  Another strategy
can be the use of nested meshes of increasing resolution
\citep{2003ApJ...586..540D}.  This is the approach that we use in
FARGO3D. There is an ongoing project aimed at merging the nested-mesh
capability of the JUPITER code \citep{2014ApJ...782...65S} with the
FARGO3D code.

Other improvements have been the implementation of multi-fluid
capability, and the incorporation of Hall effects and ambipolar
diffusion to the MHD solver. Each of these new features will be
presented in future publications.

\begin{table*}
  \centering 
  \begin{tabular}{|l|c|c|c|}
    \hline 
    factor & Cartesian & cylindrical & spherical \\
    \hline 
    \phantom{$\displaystyle\int$}$s_x^j(j)$ & $Y_{\rm max}-Y_{\rm min}$ & $Y_{\rm max}-Y_{\rm min}$ & $\frac 12(Y_{\rm max}^2-Y_{\rm min}^2)$\\
    \hline 
    \phantom{$\displaystyle\int$}$s_x^k(k)$ & $Z_{\rm max}-Z_{\rm min}$ & $Z_{\rm max}-Z_{\rm min}$ & $Z_{\rm max}-Z_{\rm min}$\\
    \hline 
    \phantom{$\displaystyle\int$}$s_y^j(j)$ & $1$ & $Y_{\rm min}\Delta x$ & $Y_{\rm min}^2\Delta x$\\
    \hline 
    \phantom{$\displaystyle\int$}$s_y^k(k)$ & $\Delta x(Z_{\rm max}-Z_{\rm min})$ & $Z_{\rm max}-Z_{\rm min}$ & $\cos(Z_{\rm min})-\cos(Z_{\rm max})$\\
    \hline 
    \phantom{$\displaystyle\int$}$s_z^j(j)$ & $\Delta x(Y_{\rm max}-Y_{\rm min})$ & $\frac 12\Delta x(Y_{\rm max}^2-Y_{\rm min}^2)$ & $\frac 12\Delta x(Y_{\rm max}^2-Y_{\rm min}^2)$ \\
    \hline 
    \phantom{$\displaystyle\int$}$s_z^k(k)$ & $1$ & $1$ & $\sin(Z_{\rm min})$\\
    \hline
    $V_j(j)^{-1}$ & $(Y_\textrm{max}-Y_\textrm{min})^{-1}$ & $2/[(Y_\textrm{max}^2-Y_\textrm{min}^2)\Delta x]$ & $3/[(Y_\textrm{max}^3-Y_\textrm{min}^3)\Delta x]$\\
    \hline 
  \end{tabular}
\medskip
\caption{One dimensional coefficients in the three different
  geometries. The meaning of the $X$, $Y$ and $Z$ coordinates for each
  geometry is given by Tab.~\ref{tab:corresp}. In each row, the
  coefficient either depends on $j$ or $k$, never on both. The min/max
  subscript refers to the location (in $Y$ or $Z$, depending on the
  case) of the lower/upper side of the cell (at location $j,k$, the
  value of index $i$ being irrelevant).}
\label{tab:geom}
\end{table*}
\begin{acknowledgements}
  Pablo Ben\'\i tez-Llambay acknowledges financial support from
  CONICET and the computational resources provided by IATE and CCAD
  (Universidad Nacional de C\'ordoba). F. Masset acknowledges support
  from CONACyT grant~178377. \\ We thank S\'ebastien Fromang for
  discussions and guidance during the development of the MHD solver,
  and for comments on an early version of this work. We thank Gloria
  Koenigsberger for a thorough reading of this manuscript, and
  constructive comments. We thank Ulises Amaya Olvera, Reyes Garc\'\i
  a Carre\'on and J\'er\^ome Verleyen for their assistance in setting
  up the GPU cluster on which most of the calculations presented here
  have been run.
\end{acknowledgements}

\appendix

\section{Geometric coefficients}
\label{sec:geom-coeff}
During a full update on the mesh, one requires at several places the
volume of a zone, the surface of all its faces, and the length of its
edges.  In order to save memory (which is always a concern on GPUs)
without sacrificing speed, we define a number of one-dimensional
arrays, and use products of these arrays to obtain the face surfaces
or the zone volumes. Our aim is that on relatively modest setups,
these one-dimensional arrays may fit in the so-called \emph{constant
  memory}, which is cached (fast) read-only memory on board the
graphics card\endnote{The compilation variable \texttt{BIGMEM}
  controls this behavior: when set to \texttt{0} the constant memory
  is used, while when set to \texttt{1} the standard, global memory is
  used to store them.}.  In FARGO3D the cell size is uniform in $X$
(hence azimuth in cylindrical or spherical geometries). This is a
limitation imposed by the (nearly systematic) use of orbital advection
along this direction\endnote{The zone width in $X$ is accessible
  throughout the code via the global variable \texttt{Dx}.}.
\subsection{Surfaces}
\label{sec:surfaces}
We have the relationships:
\begin{eqnarray}
  \label{eq:70}
  S_x(j,k) &=& s_x^j(j)s_x^k(k)\\
  S_y(j,k) &=& s_y^j(j)s_y^k(k)\\
  S_z(j,k) &=& s_z^j(j)s_z^k(k),
\end{eqnarray}
where the coefficients $s_{x/y/z}^{j/k}$ are defined in
Tab.~\ref{tab:geom}, and where $S_A(j,k)$ is the surface of the lower
face of cell $(j,k)$ perpendicular to direction $A$ (the value of $i$
is unimportant for the reason given above, regardless of the
geometry).
\subsection{Volumes}
\label{sec:volumes}
As we more often use the inverse of the volume of a zone than its mere
value, and since divisions have a high computational cost, we prefer
to have an expression that directly gives us the inverse of a cell
volume. We use:
\begin{equation}
  \label{eq:71}
  V^{-1}=\frac{V_j(j)^{-1}}{s_y^k(k)},
\end{equation}
where $V$ is the cell volume, and where the coefficients $V_j(j)^{-1}$
are defined in Tab.~\ref{tab:geom}.  We leave as an exercise to the
reader to check that Eqs.~\eqref{eq:70} and~\eqref{eq:71} lead to the
exact surfaces and (inverse) volumes in all cases.

\section{Viscous stress tensor}
\label{sec:visc-stress-tens}
\begin{figure}
  \centering
  \includegraphics[width=0.5\columnwidth]{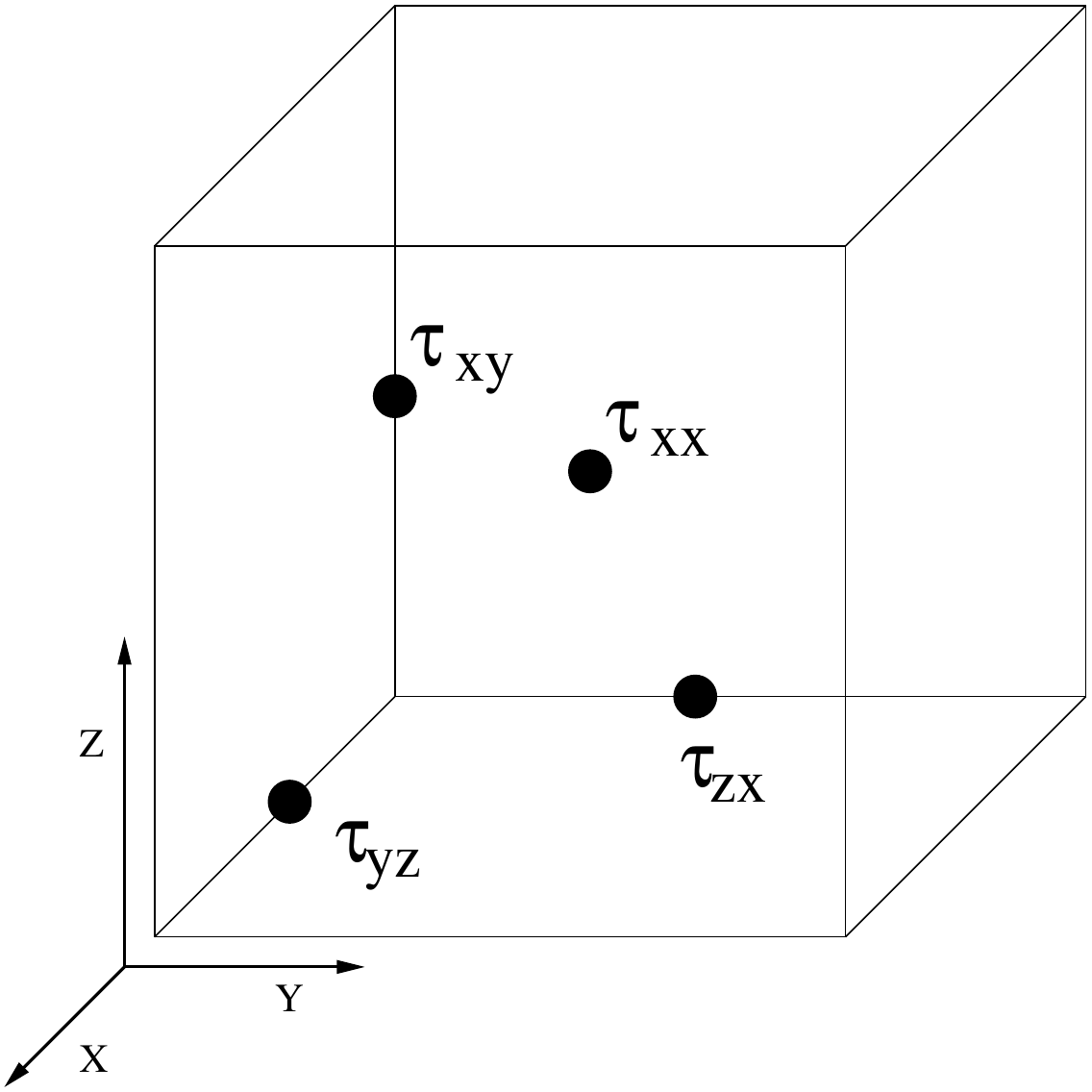}
  \caption{\label{fig:stress} Location of the different components of
    the stress tensor.}
\end{figure}

We provide hereafter the expression of its components in the three
geometries \citep[see e.g.][]{1978trs..book.....T}, the different
terms being written exactly as implemented\endnote{Respectively in
  files \texttt{visctensor\_cart.c}, \texttt{visctensor\_cyl.c} and
  \texttt{visctensor\_sph.c}.}.
\begin{description}
\item[Cartesian coordinates:] 
  \begin{equation}
    \label{eq:72}
    \tau_{ij}=-\rho\nu\left(\partial_iv_j+\partial_jv_i-\frac 23\delta_{ij}\nabla\cdot\vec{v}\right).
  \end{equation}
\item[Cylindrical coordinates:] 
  \begin{eqnarray}
    \label{eq:73}
    \tau_{\phi\phi}&=&-\rho\nu\left[2\left(\frac 1r\partial_\phi v_\phi+\frac{u_r}{r}\right)-\frac 23\nabla\cdot\vec{v}\right]\\\nonumber
    \tau_{rr}&=&-\rho\nu\left(2\partial_rv_r-\frac 23\nabla\cdot\vec{v}\right)\\\nonumber
    \tau_{zz}&=&-\rho\nu\left(2\partial_zv_z-\frac 23\nabla\cdot\vec{v}\right)\\\nonumber
    \tau_{\phi r}=\tau_{r\phi}&=&-\rho\nu\left(\partial_rv_\phi-\frac{v_\phi}{r}+\frac 1r\partial_\phi v_r\right)\\\nonumber
    \tau_{rz}=\tau_{zr}&=&-\rho\nu\left(\partial_rv_z+\partial_zv_r\right)\\\nonumber
    \tau_{\phi z}=\tau_{z\phi}&=&-\rho\nu\left(\partial_zv_\phi+\frac 1r\partial_\phi v_z\right). \nonumber
  \end{eqnarray}
\item[Spherical coordinates:] 
  \begin{eqnarray}
    \label{eq:74}
    \tau_{rr}&=&-\rho\nu\left(2\partial_rv_r-\frac 23\nabla\cdot\vec{v}\right)\\\nonumber
    \tau_{\phi\phi}&=&-\rho\nu\left[2\left(\frac{1}{r\sin\theta}\partial_\phi v_\phi+\frac{v_r}{r}+\frac{v_\theta\cot\theta}{r}\right)-\frac 23\nabla\cdot\vec{v}\right]\\\nonumber
    \tau_{\theta\theta}&=&-\rho\nu\left[2\left(\frac 1r\partial_\theta v_\theta+\frac{v_r}{r}\right)-\frac 23\nabla\cdot\vec{v}\right]\\\nonumber
    \tau_{\phi r}=\tau_{r\phi}&=&-\rho\nu\left(\frac{1}{r\sin\theta}\partial_\phi v_r+\partial_rv_\phi-\frac{v_\phi}{r}\right)\\\nonumber
    \tau_{r\theta}=\tau_{\theta r}&=&-\rho\nu\left(\partial_rv_\theta-\frac{v_\theta}{r}+\frac 1r\partial_\theta v_r\right)\\\nonumber
    \tau_{\phi\theta}=\tau_{\theta\phi}&=&-\rho\nu\left[\frac{\sin\theta}{r}\partial_\theta\left(\frac{v_\phi}{\sin\theta}\right)+\frac{1}{r\sin\theta}\partial_\phi v_\theta\right].
  \end{eqnarray}
\end{description}
The different components of the stress tensor are located at different
positions.  The diagonal terms are zone centered, whereas the cross
terms are defined at the middle of the edges.  For instance, inside a
given cell, $\tau_{xy}$ is defined at the middle of the edge along the
$z$ dimension which has the lowest value for $x$ and $y$. By circular
permutation of the indices, the centering of other components
ensues. The position of the different components is depicted in
Fig.~\ref{fig:stress}.

The different components of the stress tensor, once calculated, are
used to update directly the components of the
velocity\endnote{Respectively in files \texttt{addviscosity\_cart.c},
  \texttt{addviscosity\_cyl.c} and
  \texttt{addviscosity\_sph.c}.}. This corresponds to cell~7d
of Fig.~\ref{fig:flowchart}. We list hereafter the partial derivative
equations that correspond to this substep, in the three geometries.
\begin{description}
\item[Cartesian coordinates:] 
  \begin{equation}
    \label{eq:75}
    \partial_tv_i = -\partial_j\tau_{ij}.
  \end{equation}
\item[Cylindrical coordinates:] 
  \begin{eqnarray}
    \label{eq:76}
    \partial_tv_\phi&=&-\frac{1}{\rho}\left[\frac{1}{r^2}\partial_r(r^2\tau_{\phi r})+\frac 1r\partial_\phi\tau_{\phi\phi}+\partial_z\tau_{\phi z}\right]\\\nonumber
    \partial_tv_r&=&-\frac{1}{\rho}\left[\frac 1r\partial_r(r\tau_{rr})+\frac 1r\partial_\phi\tau_{r\phi}-\frac{\tau_{\phi\phi}}{r}+\partial_z\tau_{rz}\right]\\\nonumber
    \partial_tv_z&=&-\frac{1}{\rho}\left[\frac 1r\partial_r(r\tau_{rz})+\frac 1r\partial_\phi\tau_{\phi z}+\partial_z\tau_{zz}\right].\\\nonumber
  \end{eqnarray}
\item[Spherical coordinates:] 
  \begin{eqnarray}
    \label{eq:77}
    \partial_tv_\phi&=&-\frac{1}{\rho}\left[\frac{1}{r^2}\partial_r(r^2\tau_{\phi r})+\frac 1r\partial_\theta\tau_{\phi\theta}\right.\\\nonumber
&&\left.+\frac{1}{r\sin\theta}\partial_\phi\tau_{\phi\phi}+\frac{\tau_{r\phi}}{r}+\frac{2\tau_{\theta\phi}\cot\theta}{r}\right]\\\nonumber
\partial_tv_r&=&-\frac{1}{\rho}\left[\frac{1}{r^2}\partial_r(r^2\tau_{rr})+\frac{1}{r\sin\theta}\partial_\theta(\tau_{r\theta}\sin\theta)\right.\\\nonumber
&&+\left.\frac{1}{r\sin\theta}\partial_\phi\tau_{r\phi}-\frac{\tau_{\theta\theta}+\tau_{\phi\phi}}{r}\right]\\\nonumber
    \partial_tv_\theta&=&-\frac{1}{\rho}\left[\frac{1}{r^3}\partial_r(r^3\tau_{r\theta})+\frac{1}{r\sin\theta}\partial_\theta(\tau_{\theta\theta}\sin\theta)\right.\\\nonumber 
&&+\left.\frac{1}{r\sin\theta}\partial_\phi\tau_{\theta\phi}-\frac{\tau_{\phi\phi}\cot\theta}{r}\right],\\\nonumber 
  \end{eqnarray}
\end{description}
where, again, the writing of the different terms reflects the actual
implementation.

\theendnotes 

%\bibliography{fargo3d}

\end{document}